\documentclass[%preprint,
superscriptaddress,
amsmath,
amssymb,
aps,
pra,
10pt, twocolumn
]{revtex4-2}

\usepackage{graphicx}
\usepackage{dcolumn}
\usepackage{bm}
\usepackage{physics}
\usepackage{color}
\begin{document}

\title{Non-equilibrium generalized Langevin equation for multi-dimensional observables}

\author{Benjamin J. A. Héry}
\affiliation{Department of Physics, Freie Universit\"at Berlin, 14195 Berlin, Germany.}
\author{Lucas Tepper}
\affiliation{Department of Physics, Freie Universit\"at Berlin, 14195 Berlin, Germany.}
\author{Andrea Guljas}
\affiliation{Department of Physics, Freie Universit\"at Berlin, 14195 Berlin, Germany.}
\author{Artem Pavlov}
\affiliation{Institut für Chemie und Biochemie, Freie Universit\"at Berlin, 14195 Berlin, Germany.}
\author{Beate Koksch}
\affiliation{Institut für Chemie und Biochemie, Freie Universit\"at Berlin, 14195 Berlin, Germany.}
\author{Cecilia Clementi}
\affiliation{Department of Physics, Freie Universit\"at Berlin, 14195 Berlin, Germany.}
\author{Roland R. Netz}
\email{corresponding author: rnetz@physik.fu-berlin.de}
\affiliation{Department of Physics, Freie Universit\"at Berlin, 14195 Berlin, Germany.}

\date{\today}

\begin{abstract}
    \setlength{\parindent}{0pt}
    The Mori-Zwanzig formalism is a powerful theoretical framework for deriving equations of motion for coarse-grained observables in the form of generalized Langevin equations (GLEs) involving evolution and projection operators. Using a time-dependent many-body Hamiltonian and a multi-dimensional Mori projection operator, we derive a non-equilibrium Mori GLE for a multi-dimensional observable of interest $\vec{A}$ that consists of a Markovian force, a non-Markovian friction force, and an orthogonal force that is often interpreted as a random force. We study the structure of the derived GLE in three limiting cases: when the components of $\vec{A}$ are uncorrelated, when the Hamiltonian is time-independent and thus the system is at equilibrium, and when both conditions are simultaneously satisfied. Our non-Markovian framework is a first  step towards the systematic modeling of the coupled kinetics of coarse-grained reaction coordinates in biological complex systems, for example for the coupled intra- and inter-protein  fibril dynamics  of the human islet amyloid polypeptide (IAPP), for which we present some illustrative molecular dynamics (MD)  simulation data.
\end{abstract}

\maketitle

\section{\label{s1:introduction}Introduction}

The Mori-Zwanzig formalism is a theoretical framework developed by Mori~\cite{mori_transport_1965}, Nakajima~\cite{nakajima_quantum_1958}, and Zwanzig~\cite{zwanzig_ensemble_1960}, that employs operator formalism to derive generalized Langevin equations (GLEs), i.e. equations of motion for coarse-grained observables in complex systems~\cite{izvekov_microscopic_2013, izvekov_microscopic_2017, di_pasquale_systematic_2019, izvekov_microscopic_2019, lee_multi-dimensional_2019, klippenstein_introducing_2021, glatzel_interplay_2022, jung_non-markovian_2022, klippenstein_cross-correlation_2022, vroylandt_derivation_2022, vroylandt_position-dependent_2022, jung_dynamic_2023}. In the classical limit, this formalism is used to describe both equilibrium~\cite{izvekov_microscopic_2013, izvekov_microscopic_2017, montoya-castillo_approximate_2017, ayaz_generalized_2022, vroylandt_derivation_2022, vroylandt_position-dependent_2022, ayaz_self-consistent_2022} and non-equilibrium~\cite{meyer_non-stationary_2017, izvekov_microscopic_2019, vrugt_projection_2020, izvekov_mori-zwanzig_2021, jung_non-markovian_2022, netz_derivation_2024, hery_derivation_2024, izvekov_mori-zwanzig_2025} phenomena, with various applications in biology~\cite{hsu_zwanzig-mori_2009}, chemistry~\cite{freed_excluded_1976, lindenfeld_identity_1977, wertheimer_theory_1978, zippelius_kinetic_1978, schranner_dynamic_1979, wang_brillouin_1979, bose_time_1980, lanzafame_ultrafast_1992, schilling_mode_1997, chong_mode-coupling_1998, swinburne_phonon_2015, castellano_mode-coupling_2023}, computational physics~\cite{fulde_computation_1982, chorin_optimal_2000, li_variational_2007, silbermann_coarse-grained_2008, darve_computing_2009, hijon_morizwanzig_2010, kauzlaric_markovian_2013, mukhopadhyay_numerical_2013, carof_coarse_2014, carof_two_2014, lang_tagged-particle_2014, parish_dynamic_2017, parish_non-markovian_2017, maeyama_extracting_2020, curtis_dynamic-mode_2021, klippenstein_introducing_2021, tian_data-driven_2021, wulkow_memory-based_2021, klippenstein_cross-correlation_2022, brunig_barrier-crossing_2022, brunig_pair-reaction_2022, brunig_time-dependent_2022, jung_dynamic_2023}, hydrodynamics~\cite{akcasu_theory_1970, duderstadt_calculation_1970, mashiyama_origin_1978, grossmann_correlation_1982, bixon_hard_1989, koide_transport_2008, gouasmi_priori_2017, bian_note_2018, kadam_dynamic_2022}, and statistical physics~\cite{nordholm_systematic_1975, tokuyama_statistical-mechanical_1976, zwanzig_nonlinear_1978, kawasaki_projector_1992, vojta_charge_1998, lamba_variable-range_1999, givon_existence_2005, lyubimov_first-principle_2011, venturi_mori-zwanzig_2017, zhu_generalized_2020, debets_generalized_2021, te_vrugt_understanding_2022, fiorentino_green-kubo_2023, kiefer_analysis_2025}. The resulting GLEs are derived with no approximations or ad-hoc assumptions and share a universal structure: a term that captures the instantaneous force, which usually derives from a potential, a running integral over time of a non-Markovian friction  force, and an orthogonal force that is often treated as a random noise. While the applications of scalar GLEs are well studied, multi-dimensional GLEs remain an open field of investigation~\cite{zon_mode_2001, lee_multi-dimensional_2019, vroylandt_position-dependent_2022, kiefer_uence_2025}. Moreover, since many reaction coordinates in complex system display coupled kinetics~\cite{Jangi01122012, C3CP54476A, Baldovin_2020, 10.1063/5.0010074}, extending the Mori-Zwanzig formalism to multi-dimensional observables of interest for non-equilibrium systems is a critical step to investigate the intimate interplay between such macroscopic degrees of freedom. \\

This paper presents the derivation of a GLE for a multi-dimensional observable of interest $\vec{A}(t_{0}, t, \vec{w})$ of dimension $d \geq 2$ from the microscopic dynamics of a non-autonomous  (i.e. time dependent) Hamiltonian system using evolution operators and projection techniques. In section \ref{s2:derivation}, we review the derivation procedure, where we introduce a multi-dimensional version of the Mori projection operator, obtain a non-stationary multi-dimensional Mori-like GLE, and detail its properties. In section \ref{s3:limits}, we consider three limiting cases: $i)$ when the components of $\vec{A}(t_{0}, t, \vec{w})$ are uncorrelated, $ii)$ when the Hamiltonian is time-independent and the system is at equilibrium, and $iii)$ when both conditions $i)$ and $ii)$ are simultaneously satisfied. We discuss the structure of the derived GLE in the three aforementioned cases, and show in section~\ref{s4:discussion} as an example MD simulation data of the fibril dynamics of the human islet amyloid polypeptide (IAPP), where the coupled dynamics of the inter- and intra-peptide ordering  can be described by two coupled observables.

\section{\label{s2:derivation}Derivation of a multi-dimensional GLE from a time-dependent Hamiltonian}

\subsection*{\label{s2_ss2:system}Definition of the system and its microscopic dynamics}

We consider a system of $N$ classical atoms or particles that evolve in three-dimensional space and define the 
generalized position vector $\vec{r} \equiv (r_{1}, \cdots, r_{3N})^{T}$, the generalized momentum vector 
$\vec{p} \equiv (p_{1}, \cdots, p_{3N})^{T}$, and the generalized diagonal mass matrix
\begin{align}
    \hat{m} \equiv
    \begin{pmatrix}
        m_{1} & 0 & \ldots & 0\\
        0 & \ddots & \ddots & \vdots \\
        \vdots & \ddots & \ddots & 0 \\
        0 & \ldots & 0 & m_{3N}
    \end{pmatrix}
    .
    \label{s2_ss1_def:mass_matrix}
\end{align}
We define the microscopic state $\vec{w} \equiv (r_{1}, \cdots, r_{3N}, p_{1}, \cdots, p_{3N})^{T}$ of the system that evolves in a $6N$-dimensional phase space denoted $\mathbb{W} = \mathbb{R}^{6N}$, and assume that the microscopic dynamics of the system are derived from the generic time-dependent Hamiltonian
\begin{eqnarray}
    \label{s2_ss1_def:hamiltonian}
    & H(t, \vec{w}) \equiv H_{0}(\vec{w}) - H_{1}(t, \vec{r}),
\end{eqnarray}
where $H_{0}(\vec{w}) \equiv \sum_{n = 1}^{3N} \frac{p_{n}^{2}}{2 m_{n}} + V(\vec{r})$ is a generic time-independent many-body Hamiltonian and $H_{1}(t, \vec{r})$ is a time-dependent contribution that does not depend on $\vec{p}$. Therefore, the probability density at time $t$ of a state $\vec{w}$ is determined by the Liouville equation
\begin{align}
    \label{s2_ss1_eq:liouville_equation}
    & \frac{\partial \rho(t, \vec{w})}{\partial t } = - \mathcal{L}(t) \rho(t, \vec{w}),
\end{align}
where $\mathcal{L}(t) \equiv \sum_{n=1}^{3N} \left( \frac{\partial H(t, \vec{w})}{\partial p_{n}} \frac{\partial}{\partial r_{n}} - \frac{\partial H(t, \vec{w})}{\partial r_{n}} \frac{\partial}{\partial p_{n}} \right)$ defines the time-dependent anti-self-adjoint Liouville operator associated with $H(t, \vec{w})$. Since $\mathcal{L}(t)$ is linear in $H(t, \vec{w})$, it can be split as~\cite{netz_derivation_2024, hery_derivation_2024}
\begin{align}
    \label{s2_ss1_eq:liouville_operator}
    & \mathcal{L}(t) = \mathcal{L}_{0} - \mathcal{L}_{1}(t),
\end{align}
where $\mathcal{L}_{0} \equiv \sum_{n=1}^{3N} \left( \frac{\partial H_{0}(\vec{w})}{\partial p_{n}} \frac{\partial}{\partial r_{n}} - \frac{\partial H_{0}(\vec{w})}{\partial r_{n}} \frac{\partial}{\partial p_{n}} \right)$ is the time-independent anti-self-adjoint Liouville operator associated with $H_{0}(\vec{w})$ and $\mathcal{L}_{1}(t) = - \sum_{n=1}^{3N} \frac{\partial H_{1}(t, \vec{r})}{\partial r_{n}} \frac{\partial}{\partial p_{n}}$ the time-dependent anti-self-adjoint Liouville operator associated with $H_{1}(t, \vec{r})$. We show in section~\ref{si_1:solve} that $\rho(t, \vec{w})$ reads
\begin{align}
    \label{s2_ss1_eq:probability_density}
    & \rho(t, \vec{w}) = \exp_{S} \left( - \int_{t_{0}}^{t} du \: \mathcal{L}(u) \right) \rho(t_{0}, \vec{w}),
\end{align}
where
\begin{align}
    \label{s2_ss1_def:shcrödinger_evolution_operator}
    \exp_{S}\left( - \int_{t_{0}}^{t} du \: \mathcal{L}(u) \right) \equiv & \mathcal{I} + \sum_{n \geq 1} \prod_{k = 1}^{n} (-1)^{n} \\
    & \int_{t_{0}}^{\delta_{k,1}t + (1-\delta_{k,1})t_{k-1}} dt_{k} \prod_{j = 1}^{n} \mathcal{L}(t_{j}) \nonumber
\end{align}
defines the Schrödinger-type time-ordered propagation operator, $\rho(t_{0}, \vec{w})$ is the initial probability density distribution, $\mathcal{I}$ is the identity operator,
\begin{align}
    \label{s2_ss1_def:nested_product}
    & \prod_{k = 1}^{n} \int_{t_{0}}^{\delta_{k,1}t + (1-\delta_{k,1})t_{k-1}} dt_{k} \equiv \int_{t_0}^{t} dt_{1} \times \cdots \times \int_{t_{0}}^{t_{n-1}} dt_{n},
\end{align}
defines a nested product of integrals, and
\begin{align}
    \label{s2_ss1_def:time_ordered_product}
    & \prod_{j = 1}^{n} \mathcal{L}(t_{j}) = \mathcal{L}(t_{1}) \times \cdots \times \mathcal{L}(t_{n})
\end{align}
is the left-to-right time-ordered product of $\mathcal{L}(t)$.

\subsection*{\label{s2_ss2:observable_of_interest}Definition of the observable of interest}

We define the time-dependent expectation value
\begin{align}
    \label{s2_ss2_def:ensemble_expectation}
    & o(t) \equiv \int_{\mathbb{W}} d\hat{\vec{w}} \: \rho(t, \hat{\vec{w}}) O_{S}(\hat{\vec{w}})
\end{align}
of an arbitrary time-independent Schrödinger-type scalar observable $O_{S}(\vec{w})$ and where
\begin{align}
    \label{s2_ss2_def:integral_phase_space}
    & \int_{\mathbb{W}} d\hat{\vec{w}} \equiv \prod_{n = 1}^{3N} \int_{- \infty}^{+ \infty} d\hat{r}_{n} \int_{- \infty}^{+ \infty} d\hat{p}_{n}
\end{align}
denotes the integration over phase space. We show in section~\ref{si_2:generic_observable} that eq.~\ref{s2_ss2_def:ensemble_expectation} can be rewritten as
\begin{align}
    \label{s2_ss2_eq:ensemble_expectation}
    & o(t) = \int_{\mathbb{W}} d\hat{\vec{w}} \: \rho(t_{0}, \hat{\vec{w}}) O(t_{0}, t, \hat{\vec{w}}) \equiv \langle O(t_{0}, t, \hat{\vec{w}}) \rangle, 
\end{align}
where we defined the time-dependent scalar Heisenberg-type generic observable
\begin{align}
    \label{s2_ss2_def:generic_observable}
    & O(t_{0}, t, \vec{w}) \equiv \exp_{H} \left( \int_{t_{0}}^{t} du \: \mathcal{L}(u) \right) O_{S}(\vec{w})
\end{align}
as the action of the Heisenberg-type propagator
\begin{align}
    \label{s2_ss2_def:heisenberg_propagator}
    \exp_{H}\left( \int_{t_{0}}^{t} du \: \mathcal{L}(u) \right) \equiv & \mathcal{I} + \sum_{n \geq 1} \prod_{k = 1}^{n} \int_{t_{0}}^{\delta_{k,1}t + (1-\delta_{k,1})t_{k-1}} dt_{k} \nonumber \\
    & \prod_{j = n}^{1} \mathcal{L}(t_{j})
\end{align}
on the time-independent Schrödinger-type scalar observable $O_{S}(\vec{w})$. Therefore, the $d$-dimensional observable of interest $\vec{A}(t_{0}, t, \vec{w}) \equiv ( A_{i}(t_{0}, t, \vec{w}), 1 \leq i \leq d )^{T}$ is defined by the propagation
\begin{align}
    \label{s2_ss2_def:observable_of_interest}
    & A_{i}(t_{0}, t, \vec{w}) \equiv \exp_{H}\left( \int_{t_{0}}^{t} du \: \mathcal{L}(u) \right) A_{S,i}(\vec{r})
\end{align}
of each component of the Schrödinger-type $d$-dimensional observable $\vec{A}_{S}(\vec{r}) \equiv ( A_{S, i}(\vec{r}), 1 \leq i \leq d )^{T}$. The velocity $\dot{\vec{A}}(t_{0}, t, \vec{w}) \equiv ( \dot{A}_{i}(t_{0}, t, \vec{w}), 1 \leq i \leq d )^{T}$ and the acceleration $\ddot{\vec{A}}(\vec{w}, t_{0}, t) \equiv ( \ddot{A}_{i}(t_{0}, t, \vec{w}), 1 \leq i \leq d )^{T}$ of $\vec{A}(t_{0}, t, \vec{w})$ are given by~\cite{netz_derivation_2024}
\begin{subequations}
    \label{s2_ss2_eq:observable_derivatives}
    \begin{align}
        \label{s2_ss2_eq:observable_velocity}
        & \dot{A}_{i}(t_{0}, t, \vec{w}) = \exp_{H}\left( \int_{t_{0}}^{t} du \: \mathcal{L}(u) \right) \dot{A}_{S,i}(\vec{w}) \\
        \label{s2_ss2_eq:observable_acceleration}
        & \ddot{A}_{i}(t_{0}, t, \vec{w}) \equiv \exp_{H}\left( \int_{t_{0}}^{t} du \: \mathcal{L}(u) \right) \ddot{A}_{S,i}(t, \vec{w}),
    \end{align}
\end{subequations}
the component-wise  propagation of the initial $d$-dimensional velocity $\dot{\vec{A}}_{S}(\vec{w}) \equiv ( \dot{A}_{S, i}(\vec{w}), 1 \leq i \leq d )^{T}$ and initial $d$-dimensional acceleration $\ddot{\vec{A}}_{S}(t, \vec{w}) \equiv ( \ddot{A}_{S,i}(t, \vec{w}), 1 \leq i \leq d )^{T}$, where we introduced
 the notations $\dot{A}_{S,i}(\vec{w}) \equiv \mathcal{L}_{0} A_{S,i}(\vec{r})$ and $\ddot{A}_{S,i}(t, \vec{w}) \equiv \mathcal{L}(t) \mathcal{L}_{0} A_{S,i}(\vec{r})$ for the components of the initial $d$-dimensional velocity and the initial $d$-dimensional acceleration.

\subsection*{\label{s2_ss3:decomposition}Decomposition procedure to derive a GLE}

We recall the usual procedure to derive a GLE. First, we consider the $i$-th component of $\ddot{\vec{A}}(\vec{w}, t_{0}, t) \equiv ( \ddot{A}_{i}(t_{0}, t, \vec{w}), 1 \leq i \leq d )^{T}$ and decompose it using the operator identity $\mathcal{I} = \mathcal{P} + \mathcal{Q}$,  where $\mathcal{I}$ is the identity operator, i.e.
\begin{align}
    \label{s2_ss3_eq:first_decomposition}
    \ddot{A}_{i}(t_{0}, t, \vec{w}) = & \exp_{H}\left( \int_{t_{0}}^{t} du \: \mathcal{L}(u) \right) \mathcal{P} \ddot{A}_{S,i}(t, \vec{w}) \\
    & + \exp_{H}\left( \int_{t_{0}}^{t} du \: \mathcal{L}(u) \right) \mathcal{Q} \ddot{A}_{S,i}(t, \vec{w}). \nonumber
\end{align}
Next, we use the Dyson operator decomposition identity~\cite{kawasaki_simple_1973, zwanzig_nonequilibrium_2001}
\begin{align}
    \label{s2_ss3_eq:second_operator_identity}
    \exp_{H}\left( \int_{t_{0}}^{t} du \: \mathcal{L}(u) \right) = & \exp_{H}\left( \mathcal{Q} \int_{t_{0}}^{t}du \: \mathcal{L}(u) \right) \\
    & + \int_{t_{0}}^{t} ds \: \exp_{H}\left( \int_{t_{0}}^{s} du \: \mathcal{L}(u) \right) \nonumber \\
    & \mathcal{P} \mathcal{L}(s) \exp_{H}\left( \mathcal{Q} \int_{s}^{t} du \: \mathcal{L}(u) \right), \nonumber
\end{align}
where in analogy to eq.~\ref{s2_ss2_def:heisenberg_propagator}
\begin{align}
    \label{s2_ss3_def:orthogonal_projection}
    \exp_{H}\left( \mathcal{Q} \int_{t_{0}}^{t}du \: \mathcal{L}(u) \right) \equiv & \mathcal{I} + \sum_{n \geq 1} \prod_{k = 1}^{n} \int_{t_{0}}^{\delta_{k,1}t + (1-\delta_{k,1})t_{k-1}} dt_{k} \nonumber \\
    & \prod_{j = n}^{1} \mathcal{Q} \mathcal{L}(t_{j})
\end{align}
defines a right-to-left time-ordered orthogonal propagator. We obtain from eq.~\ref{s2_ss3_eq:first_decomposition}
\begin{align}
    \label{s2_ss3_eq:second_decomposition}
    \ddot{A}_{i}(t_{0}, t, \vec{w}) = & F_{\text{eff}, i}(t_{0}, t, \vec{w}) + F_{\text{nM,i}}(t_{0}, t, \vec{w}) \\
    & + F_{\mathcal{Q}, i}(t_{0}, t, \vec{w}), \nonumber
\end{align}
where we defined the  $i$-th component of the $d$-dimensional effective force
\begin{align}
    \label{s2_ss3_def:effective_force}
    & F_{\text{eff}, i}(t_{0}, t, \vec{w}) \equiv \exp_{H}\left( \int_{t_{0}}^{t} du \: \mathcal{L}(u) \right) \mathcal{P} \ddot{A}_{S, i}(t, \vec{w}),
\end{align}
the $i$-th component of the orthogonal force
\begin{align}
    \label{s2_ss3_def:orthogonal_force}
    F_{\mathcal{Q}, i}(t_{0}, t, \vec{w}) = & \exp_{H}\left( \mathcal{Q} \int_{t_{0}}^{t} du \: \mathcal{L}(u) \right) \\
    & \mathcal{Q} \ddot{A}_{S, i}(t, \vec{w}), \nonumber
\end{align}
and the $i$-th component of the non-Markovian force
\begin{align}
    \label{s2_ss3_def:non_markovian_force}
    F_{\text{nM,i}}(t_{0}, t, \vec{w}) = & \int_{t_{0}}^t ds \: \exp_{H}\left( \int_{t_{0}}^{s} du \: \mathcal{L}(u) \right) \\
    & \mathcal{P} \mathcal{L}(s) F_{\mathcal{Q}, i}(s, t, \vec{w}). \nonumber
\end{align}

\subsection*{\label{s2_ss4:projection_operator}Definition of the projection operator}

We list in section~\ref{si_3:projection_formalism} the general properties a projection operator must satisfy. In agreement with the
Mori-Zwanzig formalism ~\cite{mori_transport_1965,nakajima_quantum_1958,zwanzig_ensemble_1960}, 
we choose the projection distribution $\rho(t_{0}, \vec{w})$, used to calculate  expectation values  at the initial time,   to be the Boltzmann distribution
\begin{align}
    \label{s2_ss4_eq:initial_distribution}
    & \rho(t_{0}, \vec{w}) = \frac{\exp \left( - \beta H_{0}(\vec{w}) \right)}{Z(\beta)}
\end{align}
determined by $H_{0}(\vec{w})$ and where $\beta \equiv 1/(k_{B}T)$ is the inverse temperature. 
We construct in section~\ref{si_4:mori_projection} the multi-dimensional Mori projection operator according to
\begin{align}
    \label{s2_ss4_def:mori_projection_operator}
    \mathcal{P}_{M} O(t_{0}, t, \vec{w}) \equiv & \langle O(t', t_{0}, t, \hat{\vec{w}}) \rangle \\
    & + \sum_{i = 1}^{d} \langle O(t_{0}, t, \hat{\vec{w}}) (A_{S, k}(\hat{\vec{r}}) - \langle A_{S, k}(\tilde{\vec{r}}) \rangle ) \rangle \nonumber \\
    & (\hat{G}_{A}^{-1})_{k k'} ( A_{S, k'}(\vec{r}) - \langle A_{S, k'}(\hat{\vec{r}}) \rangle ) \nonumber \\
    & + \sum_{i = 1}^{d} \langle O(t_{0}, t, \hat{\vec{w}}) \dot{A}_{S, k}(\hat{\vec{w}}) \rangle (\hat{G}_{\dot{A}}^{-1})_{k k'} \nonumber \\
    & \dot{A}_{S, k'}(w) \nonumber ,
\end{align}
which  acts on the generic observable $O(t_{0}, t, \vec{w})$ and  where $\hat{G}_{A}$ and $\hat{G}_{\dot{A}}$ are two $d \times d$ matrices defined in sec.~\ref{si_4:mori_projection}.

\subsection*{\label{s2_ss5:md_neq_mori_gle}The multi-dimensional non-equilibrium Mori GLE}

We compute the  expressions of $F_{\text{eff}, i}(t_{0}, t, \vec{w})$ and $F_{\text{nM,i}}(t_{0}, t, \vec{w})$ in secs.~\ref{si_6:relevant_force} and~\ref{si_7:non_markovian_force} by substituting $\mathcal{P}_{M}$ in eq.~\ref{s2_ss4_def:mori_projection_operator} for $\mathcal{P}$ in eq.~\ref{s2_ss3_def:effective_force}, eq.~\ref{s2_ss3_def:orthogonal_force} and eq.~\ref{s2_ss3_def:non_markovian_force}. 
The final expression of the $i$-th component of the multi-dimensional non-equilibrium Mori GLE (MD-NEQ Mori GLE) reads
\begin{widetext}
    \begin{align}
        \label{s2_ss5_eq:mori_gle}
        \ddot{A}_{i}(t_{0}, t, \vec{w}) = & D_{i}(t) - \sum_{j = 1}^{d} (\hat{K}(t))_{ij} ( A_{j}(t_{0}, t, \vec{w}) - \langle A_{S, j}(\hat{\vec{r}}) \rangle ) + \int_{t_{0}}^{t} ds \: (\vec{\Gamma}_{1}(s, t))_{i} - \int_{t_{0}}^{t} ds \: \sum_{j = 1}^{d} (\hat{\Gamma}_{\dot{A}}(s, t))_{ij} \dot{A}_{j}(t_{0}, s, \vec{w}) \nonumber \\
        & + \int_{t_{0}}^{t} ds \: \sum_{j = 1}^{d} (\hat{\Gamma}_{A}(s, t))_{ij} ( A_{j}(t_{0}, s, \vec{w}) - \langle A_{S, j}(\hat{\vec{r}}) \rangle ) + F_{M, i}(t_{0}, t, \vec{w}). 
    \end{align}
\end{widetext}
It displays an effective force
\begin{align}
    \label{s2_ss5_def:neq_force}
    & D_{i}(t) \equiv \beta \langle \lbrack \mathcal{L}_{0} H_{1}(t, \hat{\vec{r}}) \rbrack \dot{A}_{S, i}(\hat{\vec{w}}) \rangle,
\end{align}
which only acts in the presence of a perturbation Hamiltonian $H_{1}(t, \hat{\vec{r}})$,
and a linear force proportional to the time-dependent stiffness matrix
\begin{align}
    \label{s2_ss5_def:stiffness_matrix}
    (\hat{K}(t))_{ij} \equiv & \sum_{k = 1}^{d} \left\lbrack \langle \dot{A}_{S, i}(\hat{\vec{w}}) \dot{A}_{S, k}(\hat{\vec{w}}) \rangle \right. \\
    & - \beta \langle \dot{A}_{S, i}(\hat{\vec{w}}) \lbrack \mathcal{L}_{0} H_{1}(t, \hat{\vec{r}}) \rbrack \nonumber \\
    & \left. ( A_{S, k}(\hat{\vec{r}}) - \langle A_{S, k}(\tilde{\vec{r}}) \rangle ) \rangle \right\rbrack ( \hat{G}_{A}^{-1} )_{kj}. \nonumber
\end{align}
The orthogonal force $F_{M, i}(t_{0}, t, \vec{w})$ satisfies the properties
\begin{align}
    \label{s2_ss5_def:properties_orthogonal_force}
    \left\{
    \begin{array}{c}
        \langle F_{M, i}(t_{0}, t, \hat{\vec{w}}) \rangle = 0 \\
        \langle F_{M, i}(t_{0}, t, \hat{\vec{w}}) A_{S, k}(\hat{\vec{r}}) \rangle = 0 \\
        \langle F_{M, i}(t_{0}, t, \hat{\vec{w}}) \dot{A}_{S, k}(\hat{\vec{w}}) \rangle = 0
    \end{array}
    \right.
\end{align}
$\forall i, k$ such that $1 \leq i, k \leq d$, and the non-Markovian force displays a first contribution
\begin{align}
    \label{s2_ss5_def:constant_kernel}
    & (\vec{\Gamma}_{1}(s, t))_{i} = \beta \langle \lbrack \mathcal{L}_{0} H_{1}(s, \hat{\vec{r}}) \rbrack F_{M, i}(s, t, \hat{\vec{w}}) \rangle
\end{align}
 that does not couple to $\vec{A}(t_{0}, s, \vec{w})$, a second contribution that couples to $\vec{A}(t_{0}, s, \vec{w})$ via the positional memory kernel matrix
\begin{align}
    \label{s2_ss5_def:positional_matrix_kernel}
    ( \hat{\Gamma}_{A}(s, t) )_{ij} \equiv & \sum_{k = 1}^{d} \left\lbrack \beta \langle \lbrack \mathcal{L}_{0} H_{1}(s, \hat{\vec{r}}) \rbrack F_{M, i}(s, t, \hat{\vec{w}}) \right. \\
    & \left. ( A_{S, k}(\hat{\vec{r}}) - \langle A_{S, k}(\tilde{\vec{r}}) ) \rangle \right\rbrack ( \hat{G}_{A}^{-1} )_{kj}, \nonumber
\end{align}
and a third contribution that couples to the velocity $\dot{\vec{A}}(t_{0}, s, \vec{w})$ via the friction memory kernel matrix
\begin{align}
    \label{s2_ss5_def:friction_matrix_kernel}
    ( \Gamma_{\dot{A}}(s, t) )_{ij} \equiv & \sum_{k = 1}^{d} \left\lbrack \langle F_{M, i}(s, t, \hat{\vec{w}}) F_{M, k}(s, s, \hat{\vec{w}}) \right. \\
    & \left. - \beta \langle \lbrack \mathcal{L}_{0} H_{1}(s, \hat{\vec{r}}) \rbrack F_{M, i}(s, t, \hat{\vec{w}}) \dot{A}_{S, k}(\hat{\vec{w}}) \rangle \right\rbrack \nonumber\\
    & ( \hat{G}_{\dot{A}}^{-1} )_{kj}, \nonumber
\end{align}
which constitutes a non-Markovian friction force.

\section{\label{s3:limits}Limiting cases}

Now that we have derived the MD-NEQ Mori GLE in eq.~\ref{s2_ss5_eq:mori_gle} and have given all terms explicitly, we investigate three limiting cases by defining the covariance matrix
\begin{align}
    \label{s3_ss1_def:correlation_matrix}
    ( \hat{C}(t_{0}, t, t') )_{i j} \equiv & \langle ( A_{i}(t_{0}, t, \hat{\vec{w}}) - \langle A_{i}(t_{0}, t, \vec{\tilde{w}}) \rangle ) \\
    & ( A_{j}(t_{0}, t', \hat{\vec{w}}) - \langle A_{j}(t_{0}, t', \vec{\tilde{w}}) \rangle ) \rangle: \nonumber
\end{align}

\begin{itemize}
    \item[$i)$] Uncorrelated limit: for
    \begin{align}
        \label{s3_ss1_eq:uncorrelated_limit}
        (\hat{C}(t_{0}, t, t'))_{i j} = \delta_{i j} (\hat{C}(t_{0}, t, t'))_{j j}
    \end{align}
    $\forall t, t' \in \mathbb{R}$ and $\forall i, j$ such that $1 \leq i, j \leq d$.
    \item[$ii)$] Equilibrium limit: for $H_{1}(t, \vec{r}) \longrightarrow 0$.
    \item[$iii)$] Uncorrelated equilibrium limit when eq.~\ref{s3_ss1_eq:uncorrelated_limit}  and $H_{1}(t, \vec{r}) \longrightarrow 0$ hold simultaneously.
\end{itemize}

\subsection*{\label{s3_ss2:uncorrelated}Uncorrelated limit}

In the uncorrelated limit, the covariance matrix $\hat{C}(t_{0}, t, t')$ becomes diagonal and we show in section~\ref{si_8:uncorrelated} that the two Gram matrices $\hat{G}_{A}$ and $\hat{G}_{\dot{A}}$ in eq.~\ref{s2_ss4_def:mori_projection_operator} become diagonal as well. In this limit, the structure of eq.~\ref{s2_ss5_eq:mori_gle} does not change, but the stiffness matrix now reads
\begin{align}
    \label{s3_ss2_eq:stiffness_matrix}
    ( \hat{K}(t) )_{i j} = & \delta_{i j} \frac{\langle \dot{A}_{S, j}^{2}(\hat{\vec{w}}) \rangle }{\langle ( A_{S, j}(\hat{\vec{r}}) - \langle A_{S, j}(\tilde{\vec{r}}) )^{2} \rangle} \\
    & - \beta \frac{\langle \dot{A}_{S, i}(\hat{\vec{w}}) \lbrack \mathcal{L}_{0} H_{1}(t, \hat{\vec{r}}) \rbrack ( A_{S, j}(\hat{\vec{r}}) - \langle A_{S, j}(\tilde{\vec{r}}) \rangle ) \rangle}{\langle ( A_{S,j}(\hat{\vec{r}}) - \langle A_{S, j}(\tilde{\vec{r}}) )^{2} \rangle}, \nonumber
\end{align}
while the positional kernel matrix reads
\begin{align}
    \label{s3_ss2_eq:positional_kernel_matrix}
    & ( \hat{\Gamma}_{A}(s t) )_{i j} = \\
    & \frac{\beta \langle \lbrack \mathcal{L}_{0} H_{1}(s, \hat{\vec{r}}) \rbrack F_{M, i}(s, t, \hat{\vec{w}}) ( A_{S, j}(\hat{\vec{r}}) - \langle A_{S, j}(\tilde{\vec{r}}) ) \rangle}{\langle ( A_{S, j}(\hat{\vec{r}}) - \langle A_{S, j}(\tilde{\vec{r}}) )^{2} \rangle} \nonumber
\end{align}
and the friction kernel matrix reads
\begin{align}
    \label{s3_ss2_eq:friction_kernel_matrix}
    ( \hat{\Gamma}_{\dot{A}}(s, t) )_{i j} = & \frac{\langle F_{M, i}(s, t, \hat{\vec{w}}) F_{M, k}(s, s, \hat{\vec{w}}) \rangle}{\langle \dot{A}_{S, j}^{2}(\hat{\vec{w}}) \rangle} \\
    & - \beta \frac{\langle \lbrack \mathcal{L}_{0} H_{1}(s, \hat{\vec{r}}) \rbrack F_{M, i}(s, t, \hat{\vec{w}}) \dot{A}_{S, j}(\hat{\vec{w}}) \rangle}{\langle \dot{A}_{S, j}^{2}(\hat{\vec{w}}) \rangle}. \nonumber
\end{align}
Hence, only the expression of the parameters of the MD-NEQ Mori GLE change in the uncorrelated limit.

\subsection*{\label{s3_ss3:equilibrium}Equilibrium limit}

In the equilibrium limit, i.e. for $H_{1}(t, \vec{r}) \longrightarrow 0$,
we show in section~\ref{si_9:equilibrium_observables} that the expressions of $\vec{A}(t_{0}, t, \vec{w})$ and $\vec{F}_{M}(t_{0}, t, \vec{w})$ read
\begin{subequations}
    \label{s3_ss3_eq:observable_transformation}
    \begin{align}
        \label{s3_ss3_eq:observable_of_interest}
        & A_{i}(t - t_{0}, \vec{w}) = \exp \left( (t - t_{0}) \mathcal{L}_{0} \right) A_{S,i}(\vec{r}) \\
        \label{s3_ss3_eq:orthogonal_force}
        & F_{M, i}(t - t_{0}, \vec{w}) = \exp \left( (t - t_{0}) \mathcal{Q}_{M} \mathcal{L}_{0} \right) \mathcal{Q}_{M}\ddot{A}_{S, i}(t, \vec{w}) \nonumber \\
    \end{align}
\end{subequations}
where $\mathcal{L}_{0}$ is the Liouville operator associated with the time-independent Hamiltonian $H_{0}$ and $\exp (\cdot)$ in these expressions denotes the standard operator exponential. Moreover, the non-equilibrium force $\vec{D}(t)$, the constant memory kernel $\vec{\Gamma}_{1}(s, t)$, and the positional kernel matrix $\hat{\Gamma}_{A}(s, t)$ vanish. The stiffness matrix $\hat{K}(t)$ becomes independent of time, i.e
\begin{align}
    \label{s3_ss3_eq:stiffness_matrix}
    & ( \hat{K} )_{i j} \equiv \sum_{k = 1}^{d} \langle \dot{A}_{S, i}(\hat{\vec{w}}) \dot{A}_{S, k}(\hat{\vec{w}}) \rangle ( \hat{G}_{A}^{-1} )_{k j},
\end{align}
the friction kernel matrix $\hat{\Gamma}_{\dot{A}}(s, t)$ becomes time-homogeneous,
\begin{align}
    \label{s3_ss3_eq:friction_kernel_matrix}
    & ( \hat{\Gamma}_{\dot{A}}(t - s) )_{i j} \equiv \sum_{k = 1}^{d} \langle F_{M, i}(t - s, \hat{\vec{w}}) F_{M, k}(0, \hat{\vec{w}}) \rangle ( \hat{G}_{\dot{A}}^{-1} )_{k j}, \nonumber \\
\end{align}
and the resulting  GLE reads
\begin{widetext}
    \begin{align}
        \label{s3_ss3_eq:mori_gle}
        & \ddot{A}_{i}(t - t_{0}, \vec{w}) = - \sum_{j = 1}^{d} (\hat{K})_{i j} ( A_{j}(t - t_{0}, \vec{w}) - \langle A_{S, j}(\hat{\vec{r}}) \rangle ) - \int_{t_{0}}^{t} ds \: \sum_{j = 1}^{d} (\hat{\Gamma}_{\dot{A}}(t - s))_{i j} \dot{A}_{j}(s - t_{0}, \vec{w}) + F_{M, i}(t - t_{0}, \vec{w}).
    \end{align}
\end{widetext}
Thus, the structure of the MD-NEQ Mori GLE  in eq.~\ref{s2_ss5_eq:mori_gle} changes significantly  in the equilibrium limit.

\subsection*{\label{s3_ss4:uncorrelation_equilibrium} Uncorrelated equilibrium limit}

In the uncorrelated equilibrium limit, both eqs.~\ref{s3_ss1_eq:uncorrelated_limit} and eq.~\ref{s3_ss3_eq:observable_transformation} hold. Hence, the structure of the GLE is  the same as  in eq.~\ref{s3_ss3_eq:mori_gle} and  the stiffness matrix reads
\begin{align}
    \label{s3_ss4_eq:stiffness_matrix}
    & ( \hat{K} )_{i j} = \delta_{i j} \frac{\langle \dot{A}_{S, j}^{2}(\hat{\vec{w}}) \rangle}{\langle ( A_{S,j}(\hat{\vec{r}}) - \langle A_{S, j}(\tilde{\vec{r}}) \rangle )^{2} \rangle} 
\end{align}
and the time-homogeneous friction kernel matrix $\hat{\Gamma}_{\dot{A}}(t)$ reads
\begin{align}
    \label{s3_ss4_eq:friction_kernel_matrix}
    & ( \hat{\Gamma}_{\dot{A}}(t - s) )_{i j} = \frac{\langle F_{M, i}(t - s, \hat{\vec{w}}) F_{M, j}(0, \hat{\vec{w}}) \rangle}{\langle \dot{A}_{S, j}^{2}(\hat{\vec{w}}) \rangle}.
\end{align}
Therefore, the only difference between the equilibrium and the uncorrelated equilibrium limits is the expression of the GLE parameters  
in eqs.~\ref{s3_ss4_eq:stiffness_matrix} and  \ref{s3_ss4_eq:friction_kernel_matrix}.

\section{\label{s4:discussion}Summary and discussion}

\begin{figure*}
    \begin{minipage}[t]{0.66\textwidth}
        \includegraphics[trim={0 0.5cm 0 0.5cm}, clip]{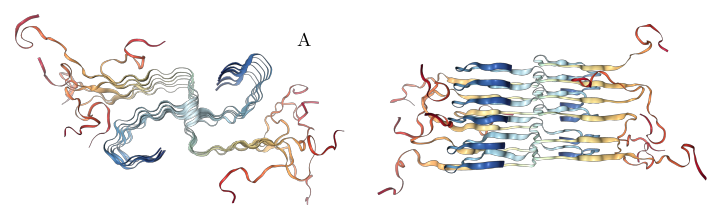}
        \includegraphics[trim={0 0.5cm 0 0.1cm}, clip]{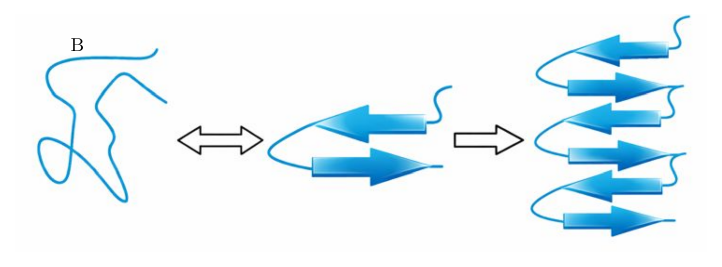}
    \end{minipage}
    \begin{minipage}[h]{0.32\textwidth}
        \vspace{1.2cm}
        \includegraphics[trim={0 0.3cm 0.2cm 0.3cm}, clip]{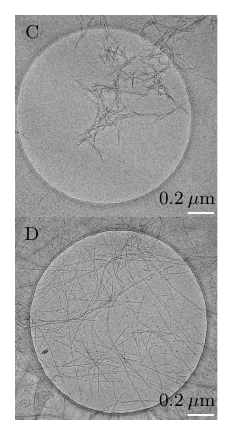}
    \end{minipage}
    \includegraphics{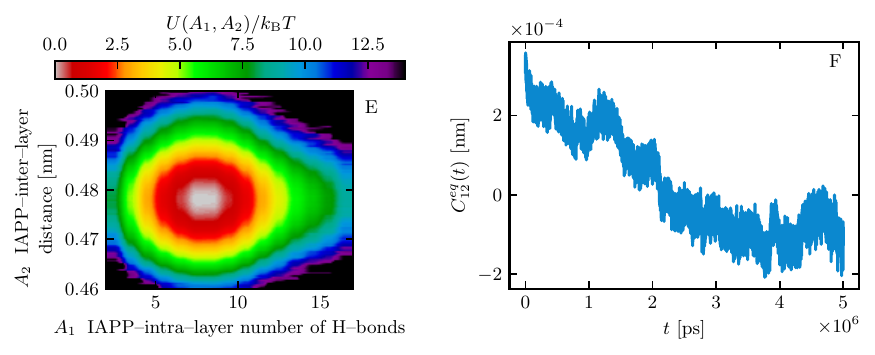}
    \caption{We show results from a 10 microsecond unconstrained molecular dynamics equilibrium simulation of a periodically replicated fibril of human islet amyloid polypeptide (IAPP) in explicit water. \textbf{A} Side and top views of a human islet amyloid polypeptide (IAPP) fibril composed of 12 IAPP monomers, shown alongside \textbf{B} a schematic representation of the IAPP fibril formation process, in which initially disordered monomers form $\beta$-sheet-rich structures and assemble into amyloid fibrils. Experimental TEM micrographs of IAPP fibrils formed at pH values of 6.8 (\textbf{C}) and 7.4 (\textbf{D}). \textbf{E} Two-dimensional free energy landscape $U(A_{1}, A_{2}) \equiv - \frac{1}{\beta} \log \langle \delta ( A_{1}(0, \vec{r}) - A_{1} ) \delta ( A_{2}(0, \vec{r}) - A_{2} ) \rangle$ as a function of the number of intra-layer hydrogen bonds $A_1$ and the distance between the centers of mass of two adjacent layers $A_2$. \textbf{F} Time correlation function ${C}_{12}^{\text{eq}}(t) \equiv \langle ( A_{1}(t, \hat{\vec{w}}) - \langle A_{1}(0, \tilde{\vec{r}}) \rangle ) ( A_{2}(0, \hat{\vec{r}}) - \langle A_{2}(0, \tilde{\vec{r}}) \rangle ) \rangle$ between the two reaction coordinates $A_1$ and $A_2$.}
    \label{fig:introduction}
\end{figure*}

In sec.~\ref{s2:derivation}, we derive the MD-NEQ Mori GLE eq.~\ref{s2_ss5_eq:mori_gle} for a multi-dimensional observable of interest $\vec{A}(t_{0}, t, \vec{w})$ of a physical system that starts in the Boltzmann distribution $\rho(t_{0}, \vec{w})$ in   eq.~\ref{s2_ss4_eq:initial_distribution} determined by the time-independent Hamiltonian $H_{0}(\vec{w})$ and is subsequently driven out of equilibrium by the time-dependent perturbation Hamiltonian $H_{1}(t, \vec{r})$. Our choice of the initial projection distribution $\rho(t_{0}, \vec{w})$  to be given by the Boltzmann distribution  is in line with the classical 
Mori-Zwanzig formalism ~\cite{mori_transport_1965,nakajima_quantum_1958,zwanzig_ensemble_1960}.
In principle it would be interesting to use for $\rho(t_{0}, \vec{w})$  distributions that differ from the  Boltzmann distribution,
but we note that the calculations would be much more involved, so we leave this for future research.\newline

In sec.~\ref{s3:limits}, we study the structure of the the MD-NEQ Mori GLE and the expression of its parameters in both the general case (i.e. the  correlated non-equilibrium case) and three limiting cases: the uncorrelated non-equilibrium limit, the correlated  equilibrium limit, and the uncorrelated equilibrium limit. On one hand, the structure of the  GLE changes significantly when comparing eq.~\ref{s2_ss5_eq:mori_gle} in the general case with eq.~\ref{s3_ss3_eq:mori_gle} in the equilibrium limit, where we observe that the non-equilibrium force $\vec{D}(t)$, the constant memory kernel $\vec{\Gamma}_{1}(s, t)$, and the positional memory kernel $\hat{\Gamma}_{A}(s, t)$ vanish. On the other hand, we notice significant  changes in the expression of the MD-NEQ Mori GLE parameters when comparing eqs.~\ref{s2_ss5_def:stiffness_matrix}, ~\ref{s2_ss5_def:positional_matrix_kernel}, and~\ref{s2_ss5_def:friction_matrix_kernel} in the general case (or eqs.~\ref{s3_ss3_eq:stiffness_matrix} and~\ref{s3_ss3_eq:friction_kernel_matrix} in the uncorrelated non-equilibrium limit) with eqs.~\ref{s3_ss2_eq:stiffness_matrix}, ~\ref{s3_ss2_eq:positional_kernel_matrix}, and~\ref{s3_ss2_eq:friction_kernel_matrix} in the equilibrium limit (or eqs.~\ref{s3_ss4_eq:stiffness_matrix} and~\ref{s3_ss4_eq:friction_kernel_matrix} in the uncorrelated equilibrium limit). These results indicate that the presence of  non-equilibrium  and the presence of correlations between observable components  both affect the global structure of the effective dynamics of $\vec{A}(t_{0}, t, \vec{w})$, which adds significantly to  the  complexity of  the analytical and numerical analysis of non-equilibrium multicomponent observables. In sec.~\ref{si_10:gle_extraction_parameters}, we explain how to  extract the parameters of the GLE for a  system that is described by a stationary perturbation Hamiltonian  $H_{1}(t, \vec{r}) \rightarrow H_{1}(\vec{r})$, these  parameters  one can in turn  use to simulate  the GLEs in eq.~\ref{s2_ss5_eq:mori_gle} or~\ref{s3_ss3_eq:mori_gle} using synthetic stochastic forces~\cite{brunig_barrier-crossing_2022, wolf_cross-correlations_2025} or orthogonal forces extracted from time series data using the methods introduced in~\cite{kiefer_analysis_2025}.\newline

Finally, we consider as an explicit example the structural dynamics of a  fibril formed from  human islet amyloid polypeptides (IAPPs). Fibril formation of this protein triggers degeneration of pancreatic cells and causes type 2 diabetes~\cite{padrickIsletAmyloidPhase2002, doi:10.1152/physrev.00042.2009, patilHeterogeneousAmylinFibril2011, buchananMechanismIAPPAmyloid2013, mooreCharacterisationStructureOligomerisation2018},
it is therefore of high interest to understand the kinetics of fibril formation and also the structural dynamics of a fibril once it has formed. According to one scenario of  fibril formation, IAPP folds and  docks onto a fibril end in a correlated fashion, meaning that the process of fibril formation is described by at least two coupled observables  or reaction coordinates. In Fig.~\ref{fig:introduction} (\textbf{A}) we show snapshots from equilibrium molecular dynamics simulations of a periodically  repeated fibril of the human IAPP in the presence of water (water molecules are not shown, see sec.~\ref{si_11:simulation_details} for  details of  the simulation setup). In Fig.~\ref{fig:introduction} (\textbf{B}) we  schematically picture the fibril formation process as the folding of a single IAPP and subsequent fibril formation, in reality though, both processes presumably occur in a time-correlated fashion. Therefore, a suitable reaction coordinate would be  the two-dimensional observable $\vec{A}(t - t_{0}, \vec{w}) = \left( A_{1}(t - t_{0}, \vec{w}), A_{2}(t - t_{0}, \vec{w}) \right)^{T}$ formed by the IAPP-fibril-intra-layer number of H bonds $A_{1}(t - t_{0}, \vec{w})$, which is a measure of the folding within one IAPP, and the IAPP-fibril-inter-layer distance $A_{2}(t - t_{0}, \vec{w})$, which is a measure of the fibril formation. The dynamics of such a two-dimensional observable can be described by the equilibrium version of the  GLE in   eq.~\ref{s3_ss3_eq:mori_gle}, for which we derive the  two-dimensional version in sec.~\ref{si_12:two-dimensional}; the actual application of  such a two-dimensional GLE to MD simulation data  is left for future work.

In Fig.~\ref{fig:introduction} (\textbf{E})  we plot the two-dimensional free energy landscape $U(A_{1}, A_{2})$, where we see that $U(A_{1}, A_{2})$ approximately factorizes into quadratic terms in A1 and A2.  This means that the multidimensional Mori GLE derived by us should be applicable to this system. In order to assess the  accuracy  of 2D Mori GLE quantitatively, one could apply the procedure introduced in sec.~\ref{si_10:gle_extraction_parameters} to extract the GLE  parameters from the MD trajectory data and then use these parameters  to simulate the GLE. The predictive usefulness of 2D GLE would then be measured by how much the free energy landscape $U(A_{1}, A_{2})$ or dynamical quantities (such as the mean-square displacement  or the mean first passage time ) extracted from MD trajectory data and GLE simulation data agree, as was done for one-dimensional observables recently~\cite{klimek_hierarchical_2026}. In Fig.~\ref{fig:introduction} (\textbf{F}) we  show the cross-correlation ${C}_{12}^{\text{eq}}(t)$ (\textbf{F}), which  decays  over a few microseconds  and becomes noisy at longer times. The noise at longer times is expected, since the total simulation time is only 10 microseconds, not much longer than the relaxation time one would estimate from the data. In fig.~\ref{fig:supplementary} in sec.~\ref{si_11:simulation_details} we present the auto-correlation function $C_{22}^{\text{eq}}(t)$, where we see  that $C_{22}^{\text{eq}}(t)$ has a  much shorter relaxation time than ${C}_{12}^{\text{eq}}(t)$. This means that the coupling between the intra-layer and inter-layer  dynamics introduces additional long  time scales that presumably are important for the dynamics of fibrils formed by  human IAPP. Such effects shall be studied in the future with extended MD simulations. In order to also understand the fibril formation and rupture dynamics, one could use the 2D-NEQ Mori GLE to explore non-equilibrium scenarios, where one might  expect even stronger couplings between intra-layer and inter-layers dynamics of  human IAPP fibrils.

To demonstrate that the fibril structures seen in the MD simulations actually occur also in experiments, we show our experimental  transmission electron microscopy (TEM) micrographs of IAPP fibrils formed at pH values of 6.8 (C) and 7.4 (D), where we  observe that the pH value sensitively controls  the formation of different structures of human IAPP fibrils (the methods we used to prepare the TEM micrographs are analogous  to previous literature studies~\cite{https://doi.org/10.1100/tsw.2010.73}). This suggests to perform matching MD simulations at different pH values as well. 

\section{\label{s5:ack}Acknowledgements}

We acknowledge support by Deutsche Forschungsgemeinschaft grant CRC 1114 ”Scaling Cascades in Complex Systems”, Project Number 235221301, Project B03.

\bibliography{sample}

\begin{thebibliography}{105}%
\makeatletter
\providecommand \@ifxundefined [1]{%
 \@ifx{#1\undefined}
}%
\providecommand \@ifnum [1]{%
 \ifnum #1\expandafter \@firstoftwo
 \else \expandafter \@secondoftwo
 \fi
}%
\providecommand \@ifx [1]{%
 \ifx #1\expandafter \@firstoftwo
 \else \expandafter \@secondoftwo
 \fi
}%
\providecommand \natexlab [1]{#1}%
\providecommand \enquote  [1]{``#1''}%
\providecommand \bibnamefont  [1]{#1}%
\providecommand \bibfnamefont [1]{#1}%
\providecommand \citenamefont [1]{#1}%
\providecommand \href@noop [0]{\@secondoftwo}%
\providecommand \href [0]{\begingroup \@sanitize@url \@href}%
\providecommand \@href[1]{\@@startlink{#1}\@@href}%
\providecommand \@@href[1]{\endgroup#1\@@endlink}%
\providecommand \@sanitize@url [0]{\catcode `\\12\catcode `\$12\catcode `\&12\catcode `\#12\catcode `\^12\catcode `\_12\catcode `\%12\relax}%
\providecommand \@@startlink[1]{}%
\providecommand \@@endlink[0]{}%
\providecommand \url  [0]{\begingroup\@sanitize@url \@url }%
\providecommand \@url [1]{\endgroup\@href {#1}{\urlprefix }}%
\providecommand \urlprefix  [0]{URL }%
\providecommand \Eprint [0]{\href }%
\providecommand \doibase [0]{https://doi.org/}%
\providecommand \selectlanguage [0]{\@gobble}%
\providecommand \bibinfo  [0]{\@secondoftwo}%
\providecommand \bibfield  [0]{\@secondoftwo}%
\providecommand \translation [1]{[#1]}%
\providecommand \BibitemOpen [0]{}%
\providecommand \bibitemStop [0]{}%
\providecommand \bibitemNoStop [0]{.\EOS\space}%
\providecommand \EOS [0]{\spacefactor3000\relax}%
\providecommand \BibitemShut  [1]{\csname bibitem#1\endcsname}%
\let\auto@bib@innerbib\@empty
%</preamble>
\bibitem [{\citenamefont {Mori}(1965)}]{mori_transport_1965}%
  \BibitemOpen
  \bibfield  {author} {\bibinfo {author} {\bibfnamefont {H.}~\bibnamefont {Mori}},\ }\bibfield  {title} {\bibinfo {title} {Transport, {Collective} {Motion}, and {Brownian} {Motion}},\ }\href {https://doi.org/10.1143/PTP.33.423} {\bibfield  {journal} {\bibinfo  {journal} {Progress of Theoretical Physics}\ }\textbf {\bibinfo {volume} {33}},\ \bibinfo {pages} {423} (\bibinfo {year} {1965})}\BibitemShut {NoStop}%
\bibitem [{\citenamefont {Nakajima}(1958)}]{nakajima_quantum_1958}%
  \BibitemOpen
  \bibfield  {author} {\bibinfo {author} {\bibfnamefont {S.}~\bibnamefont {Nakajima}},\ }\bibfield  {title} {\bibinfo {title} {On {Quantum} {Theory} of {Transport} {Phenomena}: {Steady} {Diffusion}},\ }\href {https://doi.org/10.1143/PTP.20.948} {\bibfield  {journal} {\bibinfo  {journal} {Progress of Theoretical Physics}\ }\textbf {\bibinfo {volume} {20}},\ \bibinfo {pages} {948} (\bibinfo {year} {1958})}\BibitemShut {NoStop}%
\bibitem [{\citenamefont {Zwanzig}(1960)}]{zwanzig_ensemble_1960}%
  \BibitemOpen
  \bibfield  {author} {\bibinfo {author} {\bibfnamefont {R.}~\bibnamefont {Zwanzig}},\ }\bibfield  {title} {\bibinfo {title} {Ensemble {Method} in the {Theory} of {Irreversibility}},\ }\href {https://doi.org/10.1063/1.1731409} {\bibfield  {journal} {\bibinfo  {journal} {The Journal of Chemical Physics}\ }\textbf {\bibinfo {volume} {33}},\ \bibinfo {pages} {1338} (\bibinfo {year} {1960})}\BibitemShut {NoStop}%
\bibitem [{\citenamefont {Izvekov}(2013)}]{izvekov_microscopic_2013}%
  \BibitemOpen
  \bibfield  {author} {\bibinfo {author} {\bibfnamefont {S.}~\bibnamefont {Izvekov}},\ }\bibfield  {title} {\bibinfo {title} {Microscopic derivation of particle-based coarse-grained dynamics},\ }\href {https://doi.org/10.1063/1.4795091} {\bibfield  {journal} {\bibinfo  {journal} {The Journal of Chemical Physics}\ }\textbf {\bibinfo {volume} {138}},\ \bibinfo {pages} {134106} (\bibinfo {year} {2013})}\BibitemShut {NoStop}%
\bibitem [{\citenamefont {Izvekov}(2017)}]{izvekov_microscopic_2017}%
  \BibitemOpen
  \bibfield  {author} {\bibinfo {author} {\bibfnamefont {S.}~\bibnamefont {Izvekov}},\ }\bibfield  {title} {\bibinfo {title} {Microscopic derivation of particle-based coarse-grained dynamics: {Exact} expression for memory function},\ }\href {https://doi.org/10.1063/1.4978572} {\bibfield  {journal} {\bibinfo  {journal} {The Journal of Chemical Physics}\ }\textbf {\bibinfo {volume} {146}},\ \bibinfo {pages} {124109} (\bibinfo {year} {2017})}\BibitemShut {NoStop}%
\bibitem [{\citenamefont {Di~Pasquale}\ \emph {et~al.}(2019)\citenamefont {Di~Pasquale}, \citenamefont {Hudson},\ and\ \citenamefont {Icardi}}]{di_pasquale_systematic_2019}%
  \BibitemOpen
  \bibfield  {author} {\bibinfo {author} {\bibfnamefont {N.}~\bibnamefont {Di~Pasquale}}, \bibinfo {author} {\bibfnamefont {T.}~\bibnamefont {Hudson}},\ and\ \bibinfo {author} {\bibfnamefont {M.}~\bibnamefont {Icardi}},\ }\bibfield  {title} {\bibinfo {title} {Systematic derivation of hybrid coarse-grained models},\ }\href {https://doi.org/10.1103/PhysRevE.99.013303} {\bibfield  {journal} {\bibinfo  {journal} {Physical Review E}\ }\textbf {\bibinfo {volume} {99}},\ \bibinfo {pages} {013303} (\bibinfo {year} {2019})},\ \bibinfo {note} {publisher: American Physical Society}\BibitemShut {NoStop}%
\bibitem [{\citenamefont {Izvekov}(2019)}]{izvekov_microscopic_2019}%
  \BibitemOpen
  \bibfield  {author} {\bibinfo {author} {\bibfnamefont {S.}~\bibnamefont {Izvekov}},\ }\bibfield  {title} {\bibinfo {title} {Microscopic derivation of coarse-grained, energy-conserving generalized {Langevin} dynamics},\ }\href {https://doi.org/10.1063/1.5096655} {\bibfield  {journal} {\bibinfo  {journal} {The Journal of Chemical Physics}\ }\textbf {\bibinfo {volume} {151}},\ \bibinfo {pages} {104109} (\bibinfo {year} {2019})}\BibitemShut {NoStop}%
\bibitem [{\citenamefont {Lee}\ \emph {et~al.}(2019)\citenamefont {Lee}, \citenamefont {Ahn},\ and\ \citenamefont {Darve}}]{lee_multi-dimensional_2019}%
  \BibitemOpen
  \bibfield  {author} {\bibinfo {author} {\bibfnamefont {H.~S.}\ \bibnamefont {Lee}}, \bibinfo {author} {\bibfnamefont {S.-H.}\ \bibnamefont {Ahn}},\ and\ \bibinfo {author} {\bibfnamefont {E.~F.}\ \bibnamefont {Darve}},\ }\bibfield  {title} {\bibinfo {title} {The multi-dimensional generalized {Langevin} equation for conformational motion of proteins},\ }\href {https://doi.org/10.1063/1.5055573} {\bibfield  {journal} {\bibinfo  {journal} {The Journal of Chemical Physics}\ }\textbf {\bibinfo {volume} {150}},\ \bibinfo {pages} {174113} (\bibinfo {year} {2019})}\BibitemShut {NoStop}%
\bibitem [{\citenamefont {Klippenstein}\ \emph {et~al.}(2021)\citenamefont {Klippenstein}, \citenamefont {Tripathy}, \citenamefont {Jung}, \citenamefont {Schmid},\ and\ \citenamefont {van~der Vegt}}]{klippenstein_introducing_2021}%
  \BibitemOpen
  \bibfield  {author} {\bibinfo {author} {\bibfnamefont {V.}~\bibnamefont {Klippenstein}}, \bibinfo {author} {\bibfnamefont {M.}~\bibnamefont {Tripathy}}, \bibinfo {author} {\bibfnamefont {G.}~\bibnamefont {Jung}}, \bibinfo {author} {\bibfnamefont {F.}~\bibnamefont {Schmid}},\ and\ \bibinfo {author} {\bibfnamefont {N.~F.~A.}\ \bibnamefont {van~der Vegt}},\ }\bibfield  {title} {\bibinfo {title} {Introducing {Memory} in {Coarse}-{Grained} {Molecular} {Simulations}},\ }\href {https://doi.org/10.1021/acs.jpcb.1c01120} {\bibfield  {journal} {\bibinfo  {journal} {The Journal of Physical Chemistry B}\ }\textbf {\bibinfo {volume} {125}},\ \bibinfo {pages} {4931} (\bibinfo {year} {2021})},\ \bibinfo {note} {publisher: American Chemical Society}\BibitemShut {NoStop}%
\bibitem [{\citenamefont {Glatzel}\ and\ \citenamefont {Schilling}(2022)}]{glatzel_interplay_2022}%
  \BibitemOpen
  \bibfield  {author} {\bibinfo {author} {\bibfnamefont {F.}~\bibnamefont {Glatzel}}\ and\ \bibinfo {author} {\bibfnamefont {T.}~\bibnamefont {Schilling}},\ }\bibfield  {title} {\bibinfo {title} {The interplay between memory and potentials of mean force: {A} discussion on the structure of equations of motion for coarse-grained observables},\ }\href {https://doi.org/10.1209/0295-5075/ac35ba} {\bibfield  {journal} {\bibinfo  {journal} {Europhysics Letters}\ }\textbf {\bibinfo {volume} {136}},\ \bibinfo {pages} {36001} (\bibinfo {year} {2022})},\ \bibinfo {note} {publisher: EDP Sciences, IOP Publishing and Società Italiana di Fisica}\BibitemShut {NoStop}%
\bibitem [{\citenamefont {Jung}(2022)}]{jung_non-markovian_2022}%
  \BibitemOpen
  \bibfield  {author} {\bibinfo {author} {\bibfnamefont {G.}~\bibnamefont {Jung}},\ }\bibfield  {title} {\bibinfo {title} {Non-{Markovian} systems out of equilibrium: exact results for two routes of coarse graining},\ }\href {https://doi.org/10.1088/1361-648X/ac56a7} {\bibfield  {journal} {\bibinfo  {journal} {Journal of Physics: Condensed Matter}\ }\textbf {\bibinfo {volume} {34}},\ \bibinfo {pages} {204004} (\bibinfo {year} {2022})},\ \bibinfo {note} {publisher: IOP Publishing}\BibitemShut {NoStop}%
\bibitem [{\citenamefont {Klippenstein}\ and\ \citenamefont {van~der Vegt}(2022)}]{klippenstein_cross-correlation_2022}%
  \BibitemOpen
  \bibfield  {author} {\bibinfo {author} {\bibfnamefont {V.}~\bibnamefont {Klippenstein}}\ and\ \bibinfo {author} {\bibfnamefont {N.~F.~A.}\ \bibnamefont {van~der Vegt}},\ }\bibfield  {title} {\bibinfo {title} {Cross-correlation corrected friction in generalized {Langevin} models: {Application} to the continuous {Asakura}–{Oosawa} model},\ }\href {https://doi.org/10.1063/5.0093056} {\bibfield  {journal} {\bibinfo  {journal} {The Journal of Chemical Physics}\ }\textbf {\bibinfo {volume} {157}},\ \bibinfo {pages} {044103} (\bibinfo {year} {2022})}\BibitemShut {NoStop}%
\bibitem [{\citenamefont {Vroylandt}(2022)}]{vroylandt_derivation_2022}%
  \BibitemOpen
  \bibfield  {author} {\bibinfo {author} {\bibfnamefont {H.}~\bibnamefont {Vroylandt}},\ }\bibfield  {title} {\bibinfo {title} {On the derivation of the generalized {Langevin} equation and the fluctuation-dissipation theorem},\ }\href {https://doi.org/10.1209/0295-5075/acab7d} {\bibfield  {journal} {\bibinfo  {journal} {Europhysics Letters}\ }\textbf {\bibinfo {volume} {140}},\ \bibinfo {pages} {62003} (\bibinfo {year} {2022})},\ \bibinfo {note} {publisher: EDP Sciences, IOP Publishing and Società Italiana di Fisica}\BibitemShut {NoStop}%
\bibitem [{\citenamefont {Vroylandt}\ and\ \citenamefont {Monmarché}(2022)}]{vroylandt_position-dependent_2022}%
  \BibitemOpen
  \bibfield  {author} {\bibinfo {author} {\bibfnamefont {H.}~\bibnamefont {Vroylandt}}\ and\ \bibinfo {author} {\bibfnamefont {P.}~\bibnamefont {Monmarché}},\ }\bibfield  {title} {\bibinfo {title} {Position-dependent memory kernel in generalized {Langevin} equations: theory and numerical estimation},\ }\href {https://doi.org/10.1063/5.0094566} {\bibfield  {journal} {\bibinfo  {journal} {The Journal of Chemical Physics}\ }\textbf {\bibinfo {volume} {156}},\ \bibinfo {pages} {244105} (\bibinfo {year} {2022})}\BibitemShut {NoStop}%
\bibitem [{\citenamefont {Jung}\ and\ \citenamefont {Jung}(2023)}]{jung_dynamic_2023}%
  \BibitemOpen
  \bibfield  {author} {\bibinfo {author} {\bibfnamefont {B.}~\bibnamefont {Jung}}\ and\ \bibinfo {author} {\bibfnamefont {G.}~\bibnamefont {Jung}},\ }\bibfield  {title} {\bibinfo {title} {Dynamic coarse-graining of linear and non-linear systems: {Mori}–{Zwanzig} formalism and beyond},\ }\href {https://doi.org/10.1063/5.0165541} {\bibfield  {journal} {\bibinfo  {journal} {The Journal of Chemical Physics}\ }\textbf {\bibinfo {volume} {159}},\ \bibinfo {pages} {084110} (\bibinfo {year} {2023})}\BibitemShut {NoStop}%
\bibitem [{\citenamefont {Montoya-Castillo}\ and\ \citenamefont {Reichman}(2017)}]{montoya-castillo_approximate_2017}%
  \BibitemOpen
  \bibfield  {author} {\bibinfo {author} {\bibfnamefont {A.}~\bibnamefont {Montoya-Castillo}}\ and\ \bibinfo {author} {\bibfnamefont {D.~R.}\ \bibnamefont {Reichman}},\ }\bibfield  {title} {\bibinfo {title} {Approximate but accurate quantum dynamics from the {Mori} formalism. {II}. {Equilibrium} time correlation functions},\ }\href {https://doi.org/10.1063/1.4975388} {\bibfield  {journal} {\bibinfo  {journal} {The Journal of Chemical Physics}\ }\textbf {\bibinfo {volume} {146}},\ \bibinfo {pages} {084110} (\bibinfo {year} {2017})}\BibitemShut {NoStop}%
\bibitem [{\citenamefont {Ayaz}\ \emph {et~al.}(2022{\natexlab{a}})\citenamefont {Ayaz}, \citenamefont {Scalfi}, \citenamefont {Dalton},\ and\ \citenamefont {Netz}}]{ayaz_generalized_2022}%
  \BibitemOpen
  \bibfield  {author} {\bibinfo {author} {\bibfnamefont {C.}~\bibnamefont {Ayaz}}, \bibinfo {author} {\bibfnamefont {L.}~\bibnamefont {Scalfi}}, \bibinfo {author} {\bibfnamefont {B.~A.}\ \bibnamefont {Dalton}},\ and\ \bibinfo {author} {\bibfnamefont {R.~R.}\ \bibnamefont {Netz}},\ }\bibfield  {title} {\bibinfo {title} {Generalized {Langevin} equation with a nonlinear potential of mean force and nonlinear memory friction from a hybrid projection scheme},\ }\href {https://doi.org/10.1103/PhysRevE.105.054138} {\bibfield  {journal} {\bibinfo  {journal} {Physical Review E}\ }\textbf {\bibinfo {volume} {105}},\ \bibinfo {pages} {054138} (\bibinfo {year} {2022}{\natexlab{a}})},\ \bibinfo {note} {publisher: American Physical Society}\BibitemShut {NoStop}%
\bibitem [{\citenamefont {Ayaz}\ \emph {et~al.}(2022{\natexlab{b}})\citenamefont {Ayaz}, \citenamefont {Tepper},\ and\ \citenamefont {Netz}}]{ayaz_self-consistent_2022}%
  \BibitemOpen
  \bibfield  {author} {\bibinfo {author} {\bibfnamefont {C.}~\bibnamefont {Ayaz}}, \bibinfo {author} {\bibfnamefont {L.}~\bibnamefont {Tepper}},\ and\ \bibinfo {author} {\bibfnamefont {R.}~\bibnamefont {Netz}},\ }\bibfield  {title} {\bibinfo {title} {Self-consistent {Markovian} embedding of generalized {Langevin} equations with configuration-dependent mass and a nonlinear friction kernel},\ }\href {https://doi.org/10.55730/1300-0101.2726} {\bibfield  {journal} {\bibinfo  {journal} {Turkish Journal of Physics}\ }\textbf {\bibinfo {volume} {46}},\ \bibinfo {pages} {194} (\bibinfo {year} {2022}{\natexlab{b}})}\BibitemShut {NoStop}%
\bibitem [{\citenamefont {Meyer}\ \emph {et~al.}(2017)\citenamefont {Meyer}, \citenamefont {Voigtmann},\ and\ \citenamefont {Schilling}}]{meyer_non-stationary_2017}%
  \BibitemOpen
  \bibfield  {author} {\bibinfo {author} {\bibfnamefont {H.}~\bibnamefont {Meyer}}, \bibinfo {author} {\bibfnamefont {T.}~\bibnamefont {Voigtmann}},\ and\ \bibinfo {author} {\bibfnamefont {T.}~\bibnamefont {Schilling}},\ }\bibfield  {title} {\bibinfo {title} {On the non-stationary generalized {Langevin} equation},\ }\href {https://doi.org/10.1063/1.5006980} {\bibfield  {journal} {\bibinfo  {journal} {The Journal of Chemical Physics}\ }\textbf {\bibinfo {volume} {147}},\ \bibinfo {pages} {214110} (\bibinfo {year} {2017})}\BibitemShut {NoStop}%
\bibitem [{\citenamefont {Vrugt}\ and\ \citenamefont {Wittkowski}(2020)}]{vrugt_projection_2020}%
  \BibitemOpen
  \bibfield  {author} {\bibinfo {author} {\bibfnamefont {M.~t.}\ \bibnamefont {Vrugt}}\ and\ \bibinfo {author} {\bibfnamefont {R.}~\bibnamefont {Wittkowski}},\ }\bibfield  {title} {\bibinfo {title} {Projection operators in statistical mechanics: a pedagogical approach},\ }\href {https://doi.org/10.1088/1361-6404/ab8e28} {\bibfield  {journal} {\bibinfo  {journal} {European Journal of Physics}\ }\textbf {\bibinfo {volume} {41}},\ \bibinfo {pages} {045101} (\bibinfo {year} {2020})},\ \bibinfo {note} {publisher: IOP Publishing}\BibitemShut {NoStop}%
\bibitem [{\citenamefont {Izvekov}(2021)}]{izvekov_mori-zwanzig_2021}%
  \BibitemOpen
  \bibfield  {author} {\bibinfo {author} {\bibfnamefont {S.}~\bibnamefont {Izvekov}},\ }\bibfield  {title} {\bibinfo {title} {Mori-{Zwanzig} projection operator formalism: {Particle}-based coarse-grained dynamics of open classical systems far from equilibrium},\ }\href {https://doi.org/10.1103/PhysRevE.104.024121} {\bibfield  {journal} {\bibinfo  {journal} {Physical Review E}\ }\textbf {\bibinfo {volume} {104}},\ \bibinfo {pages} {024121} (\bibinfo {year} {2021})}\BibitemShut {NoStop}%
\bibitem [{\citenamefont {Netz}(2024)}]{netz_derivation_2024}%
  \BibitemOpen
  \bibfield  {author} {\bibinfo {author} {\bibfnamefont {R.~R.}\ \bibnamefont {Netz}},\ }\bibfield  {title} {\bibinfo {title} {Derivation of the nonequilibrium generalized {Langevin} equation from a time-dependent many-body {Hamiltonian}},\ }\href {https://doi.org/10.1103/PhysRevE.110.014123} {\bibfield  {journal} {\bibinfo  {journal} {Physical Review E}\ }\textbf {\bibinfo {volume} {110}},\ \bibinfo {pages} {014123} (\bibinfo {year} {2024})},\ \bibinfo {note} {publisher: American Physical Society}\BibitemShut {NoStop}%
\bibitem [{\citenamefont {Héry}\ and\ \citenamefont {Netz}(2024)}]{hery_derivation_2024}%
  \BibitemOpen
  \bibfield  {author} {\bibinfo {author} {\bibfnamefont {B.~J.~A.}\ \bibnamefont {Héry}}\ and\ \bibinfo {author} {\bibfnamefont {R.~R.}\ \bibnamefont {Netz}},\ }\bibfield  {title} {\bibinfo {title} {Derivation of a generalized {Langevin} equation from a generic time-dependent {Hamiltonian}},\ }\href {https://doi.org/10.1088/1751-8121/ad91ff} {\bibfield  {journal} {\bibinfo  {journal} {Journal of Physics A: Mathematical and Theoretical}\ }\textbf {\bibinfo {volume} {57}},\ \bibinfo {pages} {505003} (\bibinfo {year} {2024})},\ \bibinfo {note} {publisher: IOP Publishing}\BibitemShut {NoStop}%
\bibitem [{\citenamefont {Izvekov}(2025)}]{izvekov_mori-zwanzig_2025}%
  \BibitemOpen
  \bibfield  {author} {\bibinfo {author} {\bibfnamefont {S.}~\bibnamefont {Izvekov}},\ }\bibfield  {title} {\bibinfo {title} {Mori-{Zwanzig} projection operator formalism: {Generalized} {Langevin} equation dynamics of a classical system perturbed by an external generalized potential and far from equilibrium},\ }\href {https://doi.org/10.1103/PhysRevE.111.034130} {\bibfield  {journal} {\bibinfo  {journal} {Physical Review E}\ }\textbf {\bibinfo {volume} {111}},\ \bibinfo {pages} {034130} (\bibinfo {year} {2025})},\ \bibinfo {note} {publisher: American Physical Society}\BibitemShut {NoStop}%
\bibitem [{\citenamefont {Hsu}\ and\ \citenamefont {Hsu}(2009)}]{hsu_zwanzig-mori_2009}%
  \BibitemOpen
  \bibfield  {author} {\bibinfo {author} {\bibfnamefont {D.}~\bibnamefont {Hsu}}\ and\ \bibinfo {author} {\bibfnamefont {M.}~\bibnamefont {Hsu}},\ }\bibfield  {title} {\bibinfo {title} {Zwanzig-{Mori} projection operators and {EEG} dynamics: deriving a simple equation of motion},\ }\href {https://doi.org/10.1186/1757-5036-2-6} {\bibfield  {journal} {\bibinfo  {journal} {PMC Biophysics}\ }\textbf {\bibinfo {volume} {2}},\ \bibinfo {pages} {6} (\bibinfo {year} {2009})}\BibitemShut {NoStop}%
\bibitem [{\citenamefont {Freed}\ \emph {et~al.}(1976)\citenamefont {Freed}, \citenamefont {Edwards},\ and\ \citenamefont {Warner}}]{freed_excluded_1976}%
  \BibitemOpen
  \bibfield  {author} {\bibinfo {author} {\bibfnamefont {K.~F.}\ \bibnamefont {Freed}}, \bibinfo {author} {\bibfnamefont {S.~F.}\ \bibnamefont {Edwards}},\ and\ \bibinfo {author} {\bibfnamefont {M.}~\bibnamefont {Warner}},\ }\bibfield  {title} {\bibinfo {title} {Excluded volume effect on quasielastic neutron scattering from concentrated polymer solutions},\ }\href {https://doi.org/10.1063/1.432188} {\bibfield  {journal} {\bibinfo  {journal} {The Journal of Chemical Physics}\ }\textbf {\bibinfo {volume} {64}},\ \bibinfo {pages} {5132} (\bibinfo {year} {1976})}\BibitemShut {NoStop}%
\bibitem [{\citenamefont {Lindenfeld}(1977)}]{lindenfeld_identity_1977}%
  \BibitemOpen
  \bibfield  {author} {\bibinfo {author} {\bibfnamefont {M.}~\bibnamefont {Lindenfeld}},\ }\bibfield  {title} {\bibinfo {title} {Identity for memory operators in classical kinetic theory},\ }\href {https://doi.org/10.1103/PhysRevA.15.1801} {\bibfield  {journal} {\bibinfo  {journal} {Physical Review A}\ }\textbf {\bibinfo {volume} {15}},\ \bibinfo {pages} {1801} (\bibinfo {year} {1977})},\ \bibinfo {note} {publisher: American Physical Society}\BibitemShut {NoStop}%
\bibitem [{\citenamefont {Wertheimer}(1978)}]{wertheimer_theory_1978}%
  \BibitemOpen
  \bibfield  {author} {\bibinfo {author} {\bibfnamefont {R.~K.}\ \bibnamefont {Wertheimer}},\ }\bibfield  {title} {\bibinfo {title} {On the theory of spectral line broadening : vibrational dephasing and resonance energy transfer in molecular liquids},\ }\href {https://doi.org/10.1080/00268977800100191} {\bibfield  {journal} {\bibinfo  {journal} {Molecular Physics}\ }\textbf {\bibinfo {volume} {35}},\ \bibinfo {pages} {257} (\bibinfo {year} {1978})},\ \bibinfo {note} {publisher: Taylor \& Francis}\BibitemShut {NoStop}%
\bibitem [{\citenamefont {Zippelius}\ and\ \citenamefont {Götze}(1978)}]{zippelius_kinetic_1978}%
  \BibitemOpen
  \bibfield  {author} {\bibinfo {author} {\bibfnamefont {A.}~\bibnamefont {Zippelius}}\ and\ \bibinfo {author} {\bibfnamefont {W.}~\bibnamefont {Götze}},\ }\bibfield  {title} {\bibinfo {title} {Kinetic theory for the coherent scattering function \${S}(q, {\textbackslash}ensuremath\{{\textbackslash}omega\})\$ of classical liquids},\ }\href {https://doi.org/10.1103/PhysRevA.17.414} {\bibfield  {journal} {\bibinfo  {journal} {Physical Review A}\ }\textbf {\bibinfo {volume} {17}},\ \bibinfo {pages} {414} (\bibinfo {year} {1978})},\ \bibinfo {note} {publisher: American Physical Society}\BibitemShut {NoStop}%
\bibitem [{\citenamefont {Schranner}\ \emph {et~al.}(1979)\citenamefont {Schranner}, \citenamefont {Grossmann},\ and\ \citenamefont {Richter}}]{schranner_dynamic_1979}%
  \BibitemOpen
  \bibfield  {author} {\bibinfo {author} {\bibfnamefont {R.}~\bibnamefont {Schranner}}, \bibinfo {author} {\bibfnamefont {S.}~\bibnamefont {Grossmann}},\ and\ \bibinfo {author} {\bibfnamefont {P.~H.}\ \bibnamefont {Richter}},\ }\bibfield  {title} {\bibinfo {title} {Dynamic correlation functions in the {Brusselator}},\ }\href {https://doi.org/10.1007/BF01332698} {\bibfield  {journal} {\bibinfo  {journal} {Zeitschrift für Physik B Condensed Matter}\ }\textbf {\bibinfo {volume} {35}},\ \bibinfo {pages} {363} (\bibinfo {year} {1979})}\BibitemShut {NoStop}%
\bibitem [{\citenamefont {Wang}\ \emph {et~al.}(1979)\citenamefont {Wang}, \citenamefont {~}, ,\ and\ \citenamefont {Jones}}]{wang_brillouin_1979}%
  \BibitemOpen
  \bibfield  {author} {\bibinfo {author} {\bibfnamefont {C.}~\bibnamefont {Wang}}, \bibinfo {author} {\bibfnamefont {L.}~\bibnamefont {~}, \bibfnamefont {Y.-H.}}, ,\ and\ \bibinfo {author} {\bibfnamefont {D.}~\bibnamefont {Jones}},\ }\bibfield  {title} {\bibinfo {title} {Brillouin scattering and segmental motion of a polymeric liquid, {II}},\ }\href {https://doi.org/10.1080/00268977900100241} {\bibfield  {journal} {\bibinfo  {journal} {Molecular Physics}\ }\textbf {\bibinfo {volume} {37}},\ \bibinfo {pages} {287} (\bibinfo {year} {1979})},\ \bibinfo {note} {publisher: Taylor \& Francis}\BibitemShut {NoStop}%
\bibitem [{\citenamefont {Bose}\ \emph {et~al.}(1980)\citenamefont {Bose}, \citenamefont {Chernovitz},\ and\ \citenamefont {Emptage}}]{bose_time_1980}%
  \BibitemOpen
  \bibfield  {author} {\bibinfo {author} {\bibfnamefont {S.~K.}\ \bibnamefont {Bose}}, \bibinfo {author} {\bibfnamefont {P.~A.}\ \bibnamefont {Chernovitz}},\ and\ \bibinfo {author} {\bibfnamefont {M.~R.}\ \bibnamefont {Emptage}},\ }\bibfield  {title} {\bibinfo {title} {Time correlation function theory of the dynamics of helix–coil transitions},\ }\href {https://doi.org/10.1063/1.440253} {\bibfield  {journal} {\bibinfo  {journal} {The Journal of Chemical Physics}\ }\textbf {\bibinfo {volume} {73}},\ \bibinfo {pages} {1368} (\bibinfo {year} {1980})}\BibitemShut {NoStop}%
\bibitem [{\citenamefont {Lanzafame}\ \emph {et~al.}(1992)\citenamefont {Lanzafame}, \citenamefont {Miller}, \citenamefont {Muenter},\ and\ \citenamefont {Parkinson}}]{lanzafame_ultrafast_1992}%
  \BibitemOpen
  \bibfield  {author} {\bibinfo {author} {\bibfnamefont {J.~M.}\ \bibnamefont {Lanzafame}}, \bibinfo {author} {\bibfnamefont {R.~J.~D.}\ \bibnamefont {Miller}}, \bibinfo {author} {\bibfnamefont {A.~A.}\ \bibnamefont {Muenter}},\ and\ \bibinfo {author} {\bibfnamefont {B.~A.}\ \bibnamefont {Parkinson}},\ }\bibfield  {title} {\bibinfo {title} {Ultrafast charge-transfer dynamics at tin disulfide surfaces},\ }\href {https://doi.org/10.1021/j100186a008} {\bibfield  {journal} {\bibinfo  {journal} {The Journal of Physical Chemistry}\ }\textbf {\bibinfo {volume} {96}},\ \bibinfo {pages} {2820} (\bibinfo {year} {1992})}\BibitemShut {NoStop}%
\bibitem [{\citenamefont {Schilling}\ and\ \citenamefont {Scheidsteger}(1997)}]{schilling_mode_1997}%
  \BibitemOpen
  \bibfield  {author} {\bibinfo {author} {\bibfnamefont {R.}~\bibnamefont {Schilling}}\ and\ \bibinfo {author} {\bibfnamefont {T.}~\bibnamefont {Scheidsteger}},\ }\bibfield  {title} {\bibinfo {title} {Mode coupling approach to the ideal glass transition of molecular liquids: {Linear} molecules},\ }\href {https://doi.org/10.1103/PhysRevE.56.2932} {\bibfield  {journal} {\bibinfo  {journal} {Physical Review E}\ }\textbf {\bibinfo {volume} {56}},\ \bibinfo {pages} {2932} (\bibinfo {year} {1997})},\ \bibinfo {note} {publisher: American Physical Society}\BibitemShut {NoStop}%
\bibitem [{\citenamefont {Chong}\ and\ \citenamefont {Hirata}(1998)}]{chong_mode-coupling_1998}%
  \BibitemOpen
  \bibfield  {author} {\bibinfo {author} {\bibfnamefont {S.-H.}\ \bibnamefont {Chong}}\ and\ \bibinfo {author} {\bibfnamefont {F.}~\bibnamefont {Hirata}},\ }\bibfield  {title} {\bibinfo {title} {Mode-coupling theory for molecular liquids based on the interaction-site model},\ }\href {https://doi.org/10.1103/PhysRevE.58.6188} {\bibfield  {journal} {\bibinfo  {journal} {Physical Review E}\ }\textbf {\bibinfo {volume} {58}},\ \bibinfo {pages} {6188} (\bibinfo {year} {1998})},\ \bibinfo {note} {publisher: American Physical Society}\BibitemShut {NoStop}%
\bibitem [{\citenamefont {Swinburne}\ and\ \citenamefont {Dudarev}(2015)}]{swinburne_phonon_2015}%
  \BibitemOpen
  \bibfield  {author} {\bibinfo {author} {\bibfnamefont {T.~D.}\ \bibnamefont {Swinburne}}\ and\ \bibinfo {author} {\bibfnamefont {S.~L.}\ \bibnamefont {Dudarev}},\ }\bibfield  {title} {\bibinfo {title} {Phonon drag force acting on a mobile crystal defect: {Full} treatment of discreteness and nonlinearity},\ }\href {https://doi.org/10.1103/PhysRevB.92.134302} {\bibfield  {journal} {\bibinfo  {journal} {Physical Review B}\ }\textbf {\bibinfo {volume} {92}},\ \bibinfo {pages} {134302} (\bibinfo {year} {2015})},\ \bibinfo {note} {publisher: American Physical Society}\BibitemShut {NoStop}%
\bibitem [{\citenamefont {Castellano}\ \emph {et~al.}(2023)\citenamefont {Castellano}, \citenamefont {Batista},\ and\ \citenamefont {Verstraete}}]{castellano_mode-coupling_2023}%
  \BibitemOpen
  \bibfield  {author} {\bibinfo {author} {\bibfnamefont {A.}~\bibnamefont {Castellano}}, \bibinfo {author} {\bibfnamefont {J.~P.~A.}\ \bibnamefont {Batista}},\ and\ \bibinfo {author} {\bibfnamefont {M.~J.}\ \bibnamefont {Verstraete}},\ }\bibfield  {title} {\bibinfo {title} {Mode-coupling theory of lattice dynamics for classical and quantum crystals},\ }\href {https://doi.org/10.1063/5.0174255} {\bibfield  {journal} {\bibinfo  {journal} {The Journal of Chemical Physics}\ }\textbf {\bibinfo {volume} {159}},\ \bibinfo {pages} {234501} (\bibinfo {year} {2023})}\BibitemShut {NoStop}%
\bibitem [{\citenamefont {Fulde}\ and\ \citenamefont {Lukas}(1982)}]{fulde_computation_1982}%
  \BibitemOpen
  \bibfield  {author} {\bibinfo {author} {\bibfnamefont {P.}~\bibnamefont {Fulde}}\ and\ \bibinfo {author} {\bibfnamefont {W.-D.}\ \bibnamefont {Lukas}},\ }\bibfield  {title} {\bibinfo {title} {On the computation of electronic excitations in solids},\ }\href {https://doi.org/10.1007/BF01362456} {\bibfield  {journal} {\bibinfo  {journal} {Zeitschrift für Physik B Condensed Matter}\ }\textbf {\bibinfo {volume} {48}},\ \bibinfo {pages} {113} (\bibinfo {year} {1982})}\BibitemShut {NoStop}%
\bibitem [{\citenamefont {Chorin}\ \emph {et~al.}(2000)\citenamefont {Chorin}, \citenamefont {Hald},\ and\ \citenamefont {Kupferman}}]{chorin_optimal_2000}%
  \BibitemOpen
  \bibfield  {author} {\bibinfo {author} {\bibfnamefont {A.~J.}\ \bibnamefont {Chorin}}, \bibinfo {author} {\bibfnamefont {O.~H.}\ \bibnamefont {Hald}},\ and\ \bibinfo {author} {\bibfnamefont {R.}~\bibnamefont {Kupferman}},\ }\bibfield  {title} {\bibinfo {title} {Optimal prediction and the {Mori}–{Zwanzig} representation of irreversible processes},\ }\href {https://doi.org/10.1073/pnas.97.7.2968} {\bibfield  {journal} {\bibinfo  {journal} {Proceedings of the National Academy of Sciences}\ }\textbf {\bibinfo {volume} {97}},\ \bibinfo {pages} {2968} (\bibinfo {year} {2000})},\ \bibinfo {note} {publisher: Proceedings of the National Academy of Sciences}\BibitemShut {NoStop}%
\bibitem [{\citenamefont {Li}\ and\ \citenamefont {E}(2007)}]{li_variational_2007}%
  \BibitemOpen
  \bibfield  {author} {\bibinfo {author} {\bibfnamefont {X.}~\bibnamefont {Li}}\ and\ \bibinfo {author} {\bibfnamefont {W.}~\bibnamefont {E}},\ }\bibfield  {title} {\bibinfo {title} {Variational boundary conditions for molecular dynamics simulations of crystalline solids at finite temperature: {Treatment} of the thermal bath},\ }\href {https://doi.org/10.1103/PhysRevB.76.104107} {\bibfield  {journal} {\bibinfo  {journal} {Physical Review B}\ }\textbf {\bibinfo {volume} {76}},\ \bibinfo {pages} {104107} (\bibinfo {year} {2007})},\ \bibinfo {note} {publisher: American Physical Society}\BibitemShut {NoStop}%
\bibitem [{\citenamefont {Silbermann}\ \emph {et~al.}(2008)\citenamefont {Silbermann}, \citenamefont {Schoen},\ and\ \citenamefont {Klapp}}]{silbermann_coarse-grained_2008}%
  \BibitemOpen
  \bibfield  {author} {\bibinfo {author} {\bibfnamefont {J.~R.}\ \bibnamefont {Silbermann}}, \bibinfo {author} {\bibfnamefont {M.}~\bibnamefont {Schoen}},\ and\ \bibinfo {author} {\bibfnamefont {S.~H.~L.}\ \bibnamefont {Klapp}},\ }\bibfield  {title} {\bibinfo {title} {Coarse-grained single-particle dynamics in two-dimensional solids and liquids},\ }\href {https://doi.org/10.1103/PhysRevE.78.011201} {\bibfield  {journal} {\bibinfo  {journal} {Physical Review E}\ }\textbf {\bibinfo {volume} {78}},\ \bibinfo {pages} {011201} (\bibinfo {year} {2008})},\ \bibinfo {note} {publisher: American Physical Society}\BibitemShut {NoStop}%
\bibitem [{\citenamefont {Darve}\ \emph {et~al.}(2009)\citenamefont {Darve}, \citenamefont {Solomon},\ and\ \citenamefont {Kia}}]{darve_computing_2009}%
  \BibitemOpen
  \bibfield  {author} {\bibinfo {author} {\bibfnamefont {E.}~\bibnamefont {Darve}}, \bibinfo {author} {\bibfnamefont {J.}~\bibnamefont {Solomon}},\ and\ \bibinfo {author} {\bibfnamefont {A.}~\bibnamefont {Kia}},\ }\bibfield  {title} {\bibinfo {title} {Computing generalized {Langevin} equations and generalized {Fokker}–{Planck} equations},\ }\href {https://doi.org/10.1073/pnas.0902633106} {\bibfield  {journal} {\bibinfo  {journal} {Proceedings of the National Academy of Sciences}\ }\textbf {\bibinfo {volume} {106}},\ \bibinfo {pages} {10884} (\bibinfo {year} {2009})}\BibitemShut {NoStop}%
\bibitem [{\citenamefont {Hijón}\ \emph {et~al.}(2010)\citenamefont {Hijón}, \citenamefont {Español}, \citenamefont {Vanden-Eijnden},\ and\ \citenamefont {Delgado-Buscalioni}}]{hijon_morizwanzig_2010}%
  \BibitemOpen
  \bibfield  {author} {\bibinfo {author} {\bibfnamefont {C.}~\bibnamefont {Hijón}}, \bibinfo {author} {\bibfnamefont {P.}~\bibnamefont {Español}}, \bibinfo {author} {\bibfnamefont {E.}~\bibnamefont {Vanden-Eijnden}},\ and\ \bibinfo {author} {\bibfnamefont {R.}~\bibnamefont {Delgado-Buscalioni}},\ }\bibfield  {title} {\bibinfo {title} {Mori–{Zwanzig} formalism as a practical computational tool},\ }\href {https://doi.org/10.1039/B902479B} {\bibfield  {journal} {\bibinfo  {journal} {Faraday Discussions}\ }\textbf {\bibinfo {volume} {144}},\ \bibinfo {pages} {301} (\bibinfo {year} {2010})},\ \bibinfo {note} {publisher: Royal Society of Chemistry}\BibitemShut {NoStop}%
\bibitem [{\citenamefont {Kauzlarić}\ \emph {et~al.}(2013)\citenamefont {Kauzlarić}, \citenamefont {Meier}, \citenamefont {Español}, \citenamefont {Greiner},\ and\ \citenamefont {Succi}}]{kauzlaric_markovian_2013}%
  \BibitemOpen
  \bibfield  {author} {\bibinfo {author} {\bibfnamefont {D.}~\bibnamefont {Kauzlarić}}, \bibinfo {author} {\bibfnamefont {J.~T.}\ \bibnamefont {Meier}}, \bibinfo {author} {\bibfnamefont {P.}~\bibnamefont {Español}}, \bibinfo {author} {\bibfnamefont {A.}~\bibnamefont {Greiner}},\ and\ \bibinfo {author} {\bibfnamefont {S.}~\bibnamefont {Succi}},\ }\bibfield  {title} {\bibinfo {title} {Markovian equations of motion for non-{Markovian} coarse-graining and properties for graphene blobs},\ }\href {https://doi.org/10.1088/1367-2630/15/12/125015} {\bibfield  {journal} {\bibinfo  {journal} {New Journal of Physics}\ }\textbf {\bibinfo {volume} {15}},\ \bibinfo {pages} {125015} (\bibinfo {year} {2013})},\ \bibinfo {note} {publisher: IOP Publishing}\BibitemShut {NoStop}%
\bibitem [{\citenamefont {Mukhopadhyay}\ and\ \citenamefont {Xing}(2013)}]{mukhopadhyay_numerical_2013}%
  \BibitemOpen
  \bibfield  {author} {\bibinfo {author} {\bibfnamefont {A.}~\bibnamefont {Mukhopadhyay}}\ and\ \bibinfo {author} {\bibfnamefont {J.}~\bibnamefont {Xing}},\ }\href {https://doi.org/10.48550/arXiv.1304.4908} {\bibinfo {title} {A numerical procedure for model reduction using the generalized {Langevin} equation formalism}} (\bibinfo {year} {2013}),\ \bibinfo {note} {arXiv:1304.4908 [cond-mat]}\BibitemShut {NoStop}%
\bibitem [{\citenamefont {Carof}\ \emph {et~al.}(2014{\natexlab{a}})\citenamefont {Carof}, \citenamefont {~}, \citenamefont {~}, \citenamefont {~}, \citenamefont {~}, ,\ and\ \citenamefont {Rotenberg}}]{carof_coarse_2014}%
  \BibitemOpen
  \bibfield  {author} {\bibinfo {author} {\bibfnamefont {A.}~\bibnamefont {Carof}}, \bibinfo {author} {\bibfnamefont {M.}~\bibnamefont {~}, \bibfnamefont {Virginie}}, \bibinfo {author} {\bibfnamefont {S.}~\bibnamefont {~}, \bibfnamefont {Mathieu}}, \bibinfo {author} {\bibfnamefont {H.}~\bibnamefont {~}, \bibfnamefont {Jean-Pierre}}, \bibinfo {author} {\bibfnamefont {T.}~\bibnamefont {~}, \bibfnamefont {Pierre}}, ,\ and\ \bibinfo {author} {\bibfnamefont {B.}~\bibnamefont {Rotenberg}},\ }\bibfield  {title} {\bibinfo {title} {Coarse graining the dynamics of nano-confined solutes: the case of ions in clays},\ }\href {https://doi.org/10.1080/08927022.2013.840894} {\bibfield  {journal} {\bibinfo  {journal} {Molecular Simulation}\ }\textbf {\bibinfo {volume} {40}},\ \bibinfo {pages} {237} (\bibinfo {year} {2014}{\natexlab{a}})},\ \bibinfo {note} {publisher: Taylor \& Francis}\BibitemShut {NoStop}%
\bibitem [{\citenamefont {Carof}\ \emph {et~al.}(2014{\natexlab{b}})\citenamefont {Carof}, \citenamefont {Vuilleumier},\ and\ \citenamefont {Rotenberg}}]{carof_two_2014}%
  \BibitemOpen
  \bibfield  {author} {\bibinfo {author} {\bibfnamefont {A.}~\bibnamefont {Carof}}, \bibinfo {author} {\bibfnamefont {R.}~\bibnamefont {Vuilleumier}},\ and\ \bibinfo {author} {\bibfnamefont {B.}~\bibnamefont {Rotenberg}},\ }\bibfield  {title} {\bibinfo {title} {Two algorithms to compute projected correlation functions in molecular dynamics simulations},\ }\href {https://doi.org/10.1063/1.4868653} {\bibfield  {journal} {\bibinfo  {journal} {The Journal of Chemical Physics}\ }\textbf {\bibinfo {volume} {140}},\ \bibinfo {pages} {124103} (\bibinfo {year} {2014}{\natexlab{b}})}\BibitemShut {NoStop}%
\bibitem [{\citenamefont {Lang}\ and\ \citenamefont {Franosch}(2014)}]{lang_tagged-particle_2014}%
  \BibitemOpen
  \bibfield  {author} {\bibinfo {author} {\bibfnamefont {S.}~\bibnamefont {Lang}}\ and\ \bibinfo {author} {\bibfnamefont {T.}~\bibnamefont {Franosch}},\ }\bibfield  {title} {\bibinfo {title} {Tagged-particle motion in a dense confined liquid},\ }\href {https://doi.org/10.1103/PhysRevE.89.062122} {\bibfield  {journal} {\bibinfo  {journal} {Physical Review E}\ }\textbf {\bibinfo {volume} {89}},\ \bibinfo {pages} {062122} (\bibinfo {year} {2014})},\ \bibinfo {note} {publisher: American Physical Society}\BibitemShut {NoStop}%
\bibitem [{\citenamefont {Parish}\ and\ \citenamefont {Duraisamy}(2017{\natexlab{a}})}]{parish_dynamic_2017}%
  \BibitemOpen
  \bibfield  {author} {\bibinfo {author} {\bibfnamefont {E.~J.}\ \bibnamefont {Parish}}\ and\ \bibinfo {author} {\bibfnamefont {K.}~\bibnamefont {Duraisamy}},\ }\bibfield  {title} {\bibinfo {title} {A dynamic subgrid scale model for {Large} {Eddy} {Simulations} based on the {Mori}–{Zwanzig} formalism},\ }\href {https://doi.org/10.1016/j.jcp.2017.07.053} {\bibfield  {journal} {\bibinfo  {journal} {Journal of Computational Physics}\ }\textbf {\bibinfo {volume} {349}},\ \bibinfo {pages} {154} (\bibinfo {year} {2017}{\natexlab{a}})}\BibitemShut {NoStop}%
\bibitem [{\citenamefont {Parish}\ and\ \citenamefont {Duraisamy}(2017{\natexlab{b}})}]{parish_non-markovian_2017}%
  \BibitemOpen
  \bibfield  {author} {\bibinfo {author} {\bibfnamefont {E.~J.}\ \bibnamefont {Parish}}\ and\ \bibinfo {author} {\bibfnamefont {K.}~\bibnamefont {Duraisamy}},\ }\bibfield  {title} {\bibinfo {title} {Non-{Markovian} closure models for large eddy simulations using the {Mori}-{Zwanzig} formalism},\ }\href {https://doi.org/10.1103/PhysRevFluids.2.014604} {\bibfield  {journal} {\bibinfo  {journal} {Physical Review Fluids}\ }\textbf {\bibinfo {volume} {2}},\ \bibinfo {pages} {014604} (\bibinfo {year} {2017}{\natexlab{b}})}\BibitemShut {NoStop}%
\bibitem [{\citenamefont {Maeyama}\ and\ \citenamefont {Watanabe}(2020)}]{maeyama_extracting_2020}%
  \BibitemOpen
  \bibfield  {author} {\bibinfo {author} {\bibfnamefont {S.}~\bibnamefont {Maeyama}}\ and\ \bibinfo {author} {\bibfnamefont {T.-H.}\ \bibnamefont {Watanabe}},\ }\bibfield  {title} {\bibinfo {title} {Extracting and {Modeling} the {Effects} of {Small}-{Scale} {Fluctuations} on {Large}-{Scale} {Fluctuations} by {Mori}–{Zwanzig} {Projection} {Operator} {Method}},\ }\href {https://doi.org/10.7566/JPSJ.89.024401} {\bibfield  {journal} {\bibinfo  {journal} {Journal of the Physical Society of Japan}\ }\textbf {\bibinfo {volume} {89}},\ \bibinfo {pages} {024401} (\bibinfo {year} {2020})},\ \bibinfo {note} {publisher: The Physical Society of Japan}\BibitemShut {NoStop}%
\bibitem [{\citenamefont {Curtis}\ and\ \citenamefont {Alford-Lago}(2021)}]{curtis_dynamic-mode_2021}%
  \BibitemOpen
  \bibfield  {author} {\bibinfo {author} {\bibfnamefont {C.~W.}\ \bibnamefont {Curtis}}\ and\ \bibinfo {author} {\bibfnamefont {D.~J.}\ \bibnamefont {Alford-Lago}},\ }\bibfield  {title} {\bibinfo {title} {Dynamic-mode decomposition and optimal prediction},\ }\href {https://doi.org/10.1103/PhysRevE.103.012201} {\bibfield  {journal} {\bibinfo  {journal} {Physical Review E}\ }\textbf {\bibinfo {volume} {103}},\ \bibinfo {pages} {012201} (\bibinfo {year} {2021})},\ \bibinfo {note} {publisher: American Physical Society}\BibitemShut {NoStop}%
\bibitem [{\citenamefont {Tian}\ \emph {et~al.}(2021)\citenamefont {Tian}, \citenamefont {Lin}, \citenamefont {Anghel},\ and\ \citenamefont {Livescu}}]{tian_data-driven_2021}%
  \BibitemOpen
  \bibfield  {author} {\bibinfo {author} {\bibfnamefont {Y.}~\bibnamefont {Tian}}, \bibinfo {author} {\bibfnamefont {Y.~T.}\ \bibnamefont {Lin}}, \bibinfo {author} {\bibfnamefont {M.}~\bibnamefont {Anghel}},\ and\ \bibinfo {author} {\bibfnamefont {D.}~\bibnamefont {Livescu}},\ }\bibfield  {title} {\bibinfo {title} {Data-driven learning of {Mori}–{Zwanzig} operators for isotropic turbulence},\ }\href {https://doi.org/10.1063/5.0070548} {\bibfield  {journal} {\bibinfo  {journal} {Physics of Fluids}\ }\textbf {\bibinfo {volume} {33}},\ \bibinfo {pages} {125118} (\bibinfo {year} {2021})}\BibitemShut {NoStop}%
\bibitem [{\citenamefont {Wulkow}\ \emph {et~al.}(2021)\citenamefont {Wulkow}, \citenamefont {Koltai},\ and\ \citenamefont {Schütte}}]{wulkow_memory-based_2021}%
  \BibitemOpen
  \bibfield  {author} {\bibinfo {author} {\bibfnamefont {N.}~\bibnamefont {Wulkow}}, \bibinfo {author} {\bibfnamefont {P.}~\bibnamefont {Koltai}},\ and\ \bibinfo {author} {\bibfnamefont {C.}~\bibnamefont {Schütte}},\ }\bibfield  {title} {\bibinfo {title} {Memory-{Based} {Reduced} {Modelling} and {Data}-{Based} {Estimation} of {Opinion} {Spreading}},\ }\href {https://doi.org/10.1007/s00332-020-09673-2} {\bibfield  {journal} {\bibinfo  {journal} {Journal of Nonlinear Science}\ }\textbf {\bibinfo {volume} {31}},\ \bibinfo {pages} {19} (\bibinfo {year} {2021})}\BibitemShut {NoStop}%
\bibitem [{\citenamefont {Brünig}\ \emph {et~al.}(2022{\natexlab{a}})\citenamefont {Brünig}, \citenamefont {Netz},\ and\ \citenamefont {Kappler}}]{brunig_barrier-crossing_2022}%
  \BibitemOpen
  \bibfield  {author} {\bibinfo {author} {\bibfnamefont {F.~N.}\ \bibnamefont {Brünig}}, \bibinfo {author} {\bibfnamefont {R.~R.}\ \bibnamefont {Netz}},\ and\ \bibinfo {author} {\bibfnamefont {J.}~\bibnamefont {Kappler}},\ }\bibfield  {title} {\bibinfo {title} {Barrier-crossing times for different non-{Markovian} friction in well and barrier: {A} numerical study},\ }\href {https://doi.org/10.1103/PhysRevE.106.044133} {\bibfield  {journal} {\bibinfo  {journal} {Physical Review E}\ }\textbf {\bibinfo {volume} {106}},\ \bibinfo {pages} {044133} (\bibinfo {year} {2022}{\natexlab{a}})},\ \bibinfo {note} {publisher: American Physical Society}\BibitemShut {NoStop}%
\bibitem [{\citenamefont {Brünig}\ \emph {et~al.}(2022{\natexlab{b}})\citenamefont {Brünig}, \citenamefont {Daldrop},\ and\ \citenamefont {Netz}}]{brunig_pair-reaction_2022}%
  \BibitemOpen
  \bibfield  {author} {\bibinfo {author} {\bibfnamefont {F.~N.}\ \bibnamefont {Brünig}}, \bibinfo {author} {\bibfnamefont {J.~O.}\ \bibnamefont {Daldrop}},\ and\ \bibinfo {author} {\bibfnamefont {R.~R.}\ \bibnamefont {Netz}},\ }\bibfield  {title} {\bibinfo {title} {Pair-{Reaction} {Dynamics} in {Water}: {Competition} of {Memory}, {Potential} {Shape}, and {Inertial} {Effects}},\ }\href {https://doi.org/10.1021/acs.jpcb.2c05923} {\bibfield  {journal} {\bibinfo  {journal} {The Journal of Physical Chemistry B}\ }\textbf {\bibinfo {volume} {126}},\ \bibinfo {pages} {10295} (\bibinfo {year} {2022}{\natexlab{b}})},\ \bibinfo {note} {publisher: American Chemical Society}\BibitemShut {NoStop}%
\bibitem [{\citenamefont {Brünig}\ \emph {et~al.}(2022{\natexlab{c}})\citenamefont {Brünig}, \citenamefont {Geburtig}, \citenamefont {Canal}, \citenamefont {Kappler},\ and\ \citenamefont {Netz}}]{brunig_time-dependent_2022}%
  \BibitemOpen
  \bibfield  {author} {\bibinfo {author} {\bibfnamefont {F.~N.}\ \bibnamefont {Brünig}}, \bibinfo {author} {\bibfnamefont {O.}~\bibnamefont {Geburtig}}, \bibinfo {author} {\bibfnamefont {A.~v.}\ \bibnamefont {Canal}}, \bibinfo {author} {\bibfnamefont {J.}~\bibnamefont {Kappler}},\ and\ \bibinfo {author} {\bibfnamefont {R.~R.}\ \bibnamefont {Netz}},\ }\bibfield  {title} {\bibinfo {title} {Time-{Dependent} {Friction} {Effects} on {Vibrational} {Infrared} {Frequencies} and {Line} {Shapes} of {Liquid} {Water}},\ }\href {https://doi.org/10.1021/acs.jpcb.1c09481} {\bibfield  {journal} {\bibinfo  {journal} {The Journal of Physical Chemistry B}\ }\textbf {\bibinfo {volume} {126}},\ \bibinfo {pages} {1579} (\bibinfo {year} {2022}{\natexlab{c}})},\ \bibinfo {note} {publisher: American Chemical Society}\BibitemShut {NoStop}%
\bibitem [{\citenamefont {Akcasu}\ \emph {et~al.}(1970)\citenamefont {Akcasu}, \citenamefont {Corngold},\ and\ \citenamefont {Duderstadt}}]{akcasu_theory_1970}%
  \BibitemOpen
  \bibfield  {author} {\bibinfo {author} {\bibfnamefont {A.~Z.}\ \bibnamefont {Akcasu}}, \bibinfo {author} {\bibfnamefont {N.}~\bibnamefont {Corngold}},\ and\ \bibinfo {author} {\bibfnamefont {J.~J.}\ \bibnamefont {Duderstadt}},\ }\bibfield  {title} {\bibinfo {title} {Theory of self‐diffusion in classical fluids: The van hove self‐correlation function g8(r, t)},\ }\href {https://doi.org/10.1063/1.1693227} {\bibfield  {journal} {\bibinfo  {journal} {The Physics of Fluids}\ }\textbf {\bibinfo {volume} {13}},\ \bibinfo {pages} {2213} (\bibinfo {year} {1970})}\BibitemShut {NoStop}%
\bibitem [{\citenamefont {Duderstadt}\ and\ \citenamefont {Akcasu}(1970)}]{duderstadt_calculation_1970}%
  \BibitemOpen
  \bibfield  {author} {\bibinfo {author} {\bibfnamefont {J.~J.}\ \bibnamefont {Duderstadt}}\ and\ \bibinfo {author} {\bibfnamefont {A.~Z.}\ \bibnamefont {Akcasu}},\ }\bibfield  {title} {\bibinfo {title} {Calculation of {Current} {Correlations} in {Classical} {Fluids} via {Modeled} {Kinetic} {Equations}},\ }\href {https://doi.org/10.1103/PhysRevA.1.905} {\bibfield  {journal} {\bibinfo  {journal} {Physical Review A}\ }\textbf {\bibinfo {volume} {1}},\ \bibinfo {pages} {905} (\bibinfo {year} {1970})},\ \bibinfo {note} {publisher: American Physical Society}\BibitemShut {NoStop}%
\bibitem [{\citenamefont {Mashiyama}\ and\ \citenamefont {Mori}(1978)}]{mashiyama_origin_1978}%
  \BibitemOpen
  \bibfield  {author} {\bibinfo {author} {\bibfnamefont {K.~T.}\ \bibnamefont {Mashiyama}}\ and\ \bibinfo {author} {\bibfnamefont {H.}~\bibnamefont {Mori}},\ }\bibfield  {title} {\bibinfo {title} {Origin of the {Landau}-{Lifshitz} hydrodynamic fluctuations in nonequilibrium systems and a new method for reducing the {Boltzmann} equation},\ }\href {https://doi.org/10.1007/BF01106730} {\bibfield  {journal} {\bibinfo  {journal} {Journal of Statistical Physics}\ }\textbf {\bibinfo {volume} {18}},\ \bibinfo {pages} {385} (\bibinfo {year} {1978})}\BibitemShut {NoStop}%
\bibitem [{\citenamefont {Grossmann}\ and\ \citenamefont {Thomae}(1982)}]{grossmann_correlation_1982}%
  \BibitemOpen
  \bibfield  {author} {\bibinfo {author} {\bibfnamefont {S.}~\bibnamefont {Grossmann}}\ and\ \bibinfo {author} {\bibfnamefont {S.}~\bibnamefont {Thomae}},\ }\bibfield  {title} {\bibinfo {title} {Correlation decay of {Lagrangian} velocity differences in locally isotropic turbulence},\ }\href {https://doi.org/10.1007/BF01313034} {\bibfield  {journal} {\bibinfo  {journal} {Zeitschrift für Physik B Condensed Matter}\ }\textbf {\bibinfo {volume} {49}},\ \bibinfo {pages} {253} (\bibinfo {year} {1982})}\BibitemShut {NoStop}%
\bibitem [{\citenamefont {Bixon}\ \emph {et~al.}(1989)\citenamefont {Bixon}, \citenamefont {Dorfman},\ and\ \citenamefont {Dufty}}]{bixon_hard_1989}%
  \BibitemOpen
  \bibfield  {author} {\bibinfo {author} {\bibfnamefont {M.}~\bibnamefont {Bixon}}, \bibinfo {author} {\bibfnamefont {J.~R.}\ \bibnamefont {Dorfman}},\ and\ \bibinfo {author} {\bibfnamefont {J.~W.}\ \bibnamefont {Dufty}},\ }\bibfield  {title} {\bibinfo {title} {Hard sphere {Langevin} equations},\ }\href {https://doi.org/10.1021/j100356a027} {\bibfield  {journal} {\bibinfo  {journal} {The Journal of Physical Chemistry}\ }\textbf {\bibinfo {volume} {93}},\ \bibinfo {pages} {7019} (\bibinfo {year} {1989})}\BibitemShut {NoStop}%
\bibitem [{\citenamefont {Koide}\ and\ \citenamefont {Kodama}(2008)}]{koide_transport_2008}%
  \BibitemOpen
  \bibfield  {author} {\bibinfo {author} {\bibfnamefont {T.}~\bibnamefont {Koide}}\ and\ \bibinfo {author} {\bibfnamefont {T.}~\bibnamefont {Kodama}},\ }\bibfield  {title} {\bibinfo {title} {Transport coefficients of non-{Newtonian} fluid and causal dissipative hydrodynamics},\ }\href {https://doi.org/10.1103/PhysRevE.78.051107} {\bibfield  {journal} {\bibinfo  {journal} {Physical Review E}\ }\textbf {\bibinfo {volume} {78}},\ \bibinfo {pages} {051107} (\bibinfo {year} {2008})},\ \bibinfo {note} {publisher: American Physical Society}\BibitemShut {NoStop}%
\bibitem [{\citenamefont {Gouasmi}\ \emph {et~al.}(2017)\citenamefont {Gouasmi}, \citenamefont {Parish},\ and\ \citenamefont {Duraisamy}}]{gouasmi_priori_2017}%
  \BibitemOpen
  \bibfield  {author} {\bibinfo {author} {\bibfnamefont {A.}~\bibnamefont {Gouasmi}}, \bibinfo {author} {\bibfnamefont {E.~J.}\ \bibnamefont {Parish}},\ and\ \bibinfo {author} {\bibfnamefont {K.}~\bibnamefont {Duraisamy}},\ }\bibfield  {title} {\bibinfo {title} {A priori estimation of memory effects in reduced-order models of nonlinear systems using the {Mori}–{Zwanzig} formalism},\ }\href {https://doi.org/10.1098/rspa.2017.0385} {\bibfield  {journal} {\bibinfo  {journal} {Proceedings of the Royal Society A: Mathematical, Physical and Engineering Sciences}\ }\textbf {\bibinfo {volume} {473}},\ \bibinfo {pages} {20170385} (\bibinfo {year} {2017})},\ \bibinfo {note} {publisher: Royal Society}\BibitemShut {NoStop}%
\bibitem [{\citenamefont {Bian}\ \emph {et~al.}(2018)\citenamefont {Bian}, \citenamefont {Li},\ and\ \citenamefont {Adams}}]{bian_note_2018}%
  \BibitemOpen
  \bibfield  {author} {\bibinfo {author} {\bibfnamefont {X.}~\bibnamefont {Bian}}, \bibinfo {author} {\bibfnamefont {Z.}~\bibnamefont {Li}},\ and\ \bibinfo {author} {\bibfnamefont {N.~A.}\ \bibnamefont {Adams}},\ }\bibfield  {title} {\bibinfo {title} {A note on hydrodynamics from dissipative particle dynamics},\ }\href {https://doi.org/10.1007/s10483-018-2257-9} {\bibfield  {journal} {\bibinfo  {journal} {Applied Mathematics and Mechanics}\ }\textbf {\bibinfo {volume} {39}},\ \bibinfo {pages} {63} (\bibinfo {year} {2018})}\BibitemShut {NoStop}%
\bibitem [{\citenamefont {Kadam}(2022)}]{kadam_dynamic_2022}%
  \BibitemOpen
  \bibfield  {author} {\bibinfo {author} {\bibfnamefont {G.}~\bibnamefont {Kadam}},\ }\bibfield  {title} {\bibinfo {title} {Dynamic density correlations in baryon rich fluid using {Mori}–{Zwanzig}–{Nakajima} projection operator method},\ }\href {https://doi.org/10.1140/epjc/s10052-022-10614-4} {\bibfield  {journal} {\bibinfo  {journal} {The European Physical Journal C}\ }\textbf {\bibinfo {volume} {82}},\ \bibinfo {pages} {644} (\bibinfo {year} {2022})}\BibitemShut {NoStop}%
\bibitem [{\citenamefont {Nordholm}\ and\ \citenamefont {Zwanzig}(1975)}]{nordholm_systematic_1975}%
  \BibitemOpen
  \bibfield  {author} {\bibinfo {author} {\bibfnamefont {S.}~\bibnamefont {Nordholm}}\ and\ \bibinfo {author} {\bibfnamefont {R.}~\bibnamefont {Zwanzig}},\ }\bibfield  {title} {\bibinfo {title} {A systematic derivation of exact generalized {Brownian} motion theory},\ }\href {https://doi.org/10.1007/BF01012013} {\bibfield  {journal} {\bibinfo  {journal} {Journal of Statistical Physics}\ }\textbf {\bibinfo {volume} {13}},\ \bibinfo {pages} {347} (\bibinfo {year} {1975})}\BibitemShut {NoStop}%
\bibitem [{\citenamefont {Tokuyama}\ and\ \citenamefont {Mori}(1976)}]{tokuyama_statistical-mechanical_1976}%
  \BibitemOpen
  \bibfield  {author} {\bibinfo {author} {\bibfnamefont {M.}~\bibnamefont {Tokuyama}}\ and\ \bibinfo {author} {\bibfnamefont {H.}~\bibnamefont {Mori}},\ }\bibfield  {title} {\bibinfo {title} {Statistical-{Mechanical} {Theory} of the {Boltzmann} {Equation} and {Fluctuations} in µ {Space}},\ }\href {https://doi.org/10.1143/PTP.56.1073} {\bibfield  {journal} {\bibinfo  {journal} {Progress of Theoretical Physics}\ }\textbf {\bibinfo {volume} {56}},\ \bibinfo {pages} {1073} (\bibinfo {year} {1976})}\BibitemShut {NoStop}%
\bibitem [{\citenamefont {Zwanzig}(1978)}]{zwanzig_nonlinear_1978}%
  \BibitemOpen
  \bibfield  {author} {\bibinfo {author} {\bibfnamefont {R.}~\bibnamefont {Zwanzig}},\ }\bibfield  {title} {\bibinfo {title} {Nonlinear {Transport} {Equations} from {Statistical} {Mechanics}*)},\ }\href {https://doi.org/10.1143/PTPS.64.74} {\bibfield  {journal} {\bibinfo  {journal} {Progress of Theoretical Physics Supplement}\ }\textbf {\bibinfo {volume} {64}},\ \bibinfo {pages} {74} (\bibinfo {year} {1978})}\BibitemShut {NoStop}%
\bibitem [{\citenamefont {Kawasaki}\ and\ \citenamefont {Kawakatsu}(1992)}]{kawasaki_projector_1992}%
  \BibitemOpen
  \bibfield  {author} {\bibinfo {author} {\bibfnamefont {K.}~\bibnamefont {Kawasaki}}\ and\ \bibinfo {author} {\bibfnamefont {T.}~\bibnamefont {Kawakatsu}},\ }\bibfield  {title} {\bibinfo {title} {Projector formalism of generalized {Brownian} motion theory applied to dissipative and noisy systems},\ }\href {https://doi.org/10.1007/BF01049727} {\bibfield  {journal} {\bibinfo  {journal} {Journal of Statistical Physics}\ }\textbf {\bibinfo {volume} {67}},\ \bibinfo {pages} {795} (\bibinfo {year} {1992})}\BibitemShut {NoStop}%
\bibitem [{\citenamefont {Vojta}\ and\ \citenamefont {Becker}(1998)}]{vojta_charge_1998}%
  \BibitemOpen
  \bibfield  {author} {\bibinfo {author} {\bibfnamefont {M.}~\bibnamefont {Vojta}}\ and\ \bibinfo {author} {\bibfnamefont {K.}~\bibnamefont {Becker}},\ }\bibfield  {title} {\bibinfo {title} {Charge correlations in the weakly doped t-{J} model calculated by projection technique},\ }\href {https://doi.org/10.1007/s100510050332} {\bibfield  {journal} {\bibinfo  {journal} {The European Physical Journal B - Condensed Matter and Complex Systems}\ }\textbf {\bibinfo {volume} {3}},\ \bibinfo {pages} {427} (\bibinfo {year} {1998})}\BibitemShut {NoStop}%
\bibitem [{\citenamefont {Lamba}\ and\ \citenamefont {Kumar}(1999)}]{lamba_variable-range_1999}%
  \BibitemOpen
  \bibfield  {author} {\bibinfo {author} {\bibfnamefont {S.}~\bibnamefont {Lamba}}\ and\ \bibinfo {author} {\bibfnamefont {D.}~\bibnamefont {Kumar}},\ }\bibfield  {title} {\bibinfo {title} {Variable-range hopping: {Role} of {Coulomb} interactions},\ }\href {https://doi.org/10.1103/PhysRevB.59.4752} {\bibfield  {journal} {\bibinfo  {journal} {Physical Review B}\ }\textbf {\bibinfo {volume} {59}},\ \bibinfo {pages} {4752} (\bibinfo {year} {1999})},\ \bibinfo {note} {publisher: American Physical Society}\BibitemShut {NoStop}%
\bibitem [{\citenamefont {Givon}\ \emph {et~al.}(2005)\citenamefont {Givon}, \citenamefont {Kupferman},\ and\ \citenamefont {Hald}}]{givon_existence_2005}%
  \BibitemOpen
  \bibfield  {author} {\bibinfo {author} {\bibfnamefont {D.}~\bibnamefont {Givon}}, \bibinfo {author} {\bibfnamefont {R.}~\bibnamefont {Kupferman}},\ and\ \bibinfo {author} {\bibfnamefont {O.~H.}\ \bibnamefont {Hald}},\ }\bibfield  {title} {\bibinfo {title} {Existence proof for orthogonal dynamics and the {Mori}-{Zwanzig} formalism},\ }\href {https://doi.org/10.1007/BF02786691} {\bibfield  {journal} {\bibinfo  {journal} {Israel Journal of Mathematics}\ }\textbf {\bibinfo {volume} {145}},\ \bibinfo {pages} {221} (\bibinfo {year} {2005})}\BibitemShut {NoStop}%
\bibitem [{\citenamefont {Lyubimov}\ and\ \citenamefont {Guenza}(2011)}]{lyubimov_first-principle_2011}%
  \BibitemOpen
  \bibfield  {author} {\bibinfo {author} {\bibfnamefont {I.}~\bibnamefont {Lyubimov}}\ and\ \bibinfo {author} {\bibfnamefont {M.~G.}\ \bibnamefont {Guenza}},\ }\bibfield  {title} {\bibinfo {title} {First-principle approach to rescale the dynamics of simulated coarse-grained macromolecular liquids},\ }\href {https://doi.org/10.1103/PhysRevE.84.031801} {\bibfield  {journal} {\bibinfo  {journal} {Physical Review E}\ }\textbf {\bibinfo {volume} {84}},\ \bibinfo {pages} {031801} (\bibinfo {year} {2011})},\ \bibinfo {note} {publisher: American Physical Society}\BibitemShut {NoStop}%
\bibitem [{\citenamefont {Venturi}\ \emph {et~al.}(2017)\citenamefont {Venturi}, \citenamefont {Cho},\ and\ \citenamefont {Karniadakis}}]{venturi_mori-zwanzig_2017}%
  \BibitemOpen
  \bibfield  {author} {\bibinfo {author} {\bibfnamefont {D.}~\bibnamefont {Venturi}}, \bibinfo {author} {\bibfnamefont {H.}~\bibnamefont {Cho}},\ and\ \bibinfo {author} {\bibfnamefont {G.~E.}\ \bibnamefont {Karniadakis}},\ }\bibfield  {title} {\bibinfo {title} {Mori-{Zwanzig} {Approach} to {Uncertainty} {Quantification}},\ }in\ \href {https://doi.org/10.1007/978-3-319-12385-1_28} {\emph {\bibinfo {booktitle} {Handbook of {Uncertainty} {Quantification}}}}\ (\bibinfo  {publisher} {Springer, Cham},\ \bibinfo {year} {2017})\ pp.\ \bibinfo {pages} {1037--1073}\BibitemShut {NoStop}%
\bibitem [{\citenamefont {Zhu}\ and\ \citenamefont {Venturi}(2020)}]{zhu_generalized_2020}%
  \BibitemOpen
  \bibfield  {author} {\bibinfo {author} {\bibfnamefont {Y.}~\bibnamefont {Zhu}}\ and\ \bibinfo {author} {\bibfnamefont {D.}~\bibnamefont {Venturi}},\ }\bibfield  {title} {\bibinfo {title} {Generalized {Langevin} {Equations} for {Systems} with {Local} {Interactions}},\ }\href {https://doi.org/10.1007/s10955-020-02499-y} {\bibfield  {journal} {\bibinfo  {journal} {Journal of Statistical Physics}\ }\textbf {\bibinfo {volume} {178}},\ \bibinfo {pages} {1217} (\bibinfo {year} {2020})}\BibitemShut {NoStop}%
\bibitem [{\citenamefont {Debets}\ \emph {et~al.}(2021)\citenamefont {Debets}, \citenamefont {Luo}, \citenamefont {Ciarella},\ and\ \citenamefont {Janssen}}]{debets_generalized_2021}%
  \BibitemOpen
  \bibfield  {author} {\bibinfo {author} {\bibfnamefont {V.~E.}\ \bibnamefont {Debets}}, \bibinfo {author} {\bibfnamefont {C.}~\bibnamefont {Luo}}, \bibinfo {author} {\bibfnamefont {S.}~\bibnamefont {Ciarella}},\ and\ \bibinfo {author} {\bibfnamefont {L.~M.~C.}\ \bibnamefont {Janssen}},\ }\bibfield  {title} {\bibinfo {title} {Generalized mode-coupling theory for mixtures of {Brownian} particles},\ }\href {https://doi.org/10.1103/PhysRevE.104.065302} {\bibfield  {journal} {\bibinfo  {journal} {Physical Review E}\ }\textbf {\bibinfo {volume} {104}},\ \bibinfo {pages} {065302} (\bibinfo {year} {2021})},\ \bibinfo {note} {publisher: American Physical Society}\BibitemShut {NoStop}%
\bibitem [{\citenamefont {te~Vrugt}(2022)}]{te_vrugt_understanding_2022}%
  \BibitemOpen
  \bibfield  {author} {\bibinfo {author} {\bibfnamefont {M.}~\bibnamefont {te~Vrugt}},\ }\bibfield  {title} {\bibinfo {title} {Understanding probability and irreversibility in the {Mori}-{Zwanzig} projection operator formalism},\ }\href {https://doi.org/10.1007/s13194-022-00466-w} {\bibfield  {journal} {\bibinfo  {journal} {European Journal for Philosophy of Science}\ }\textbf {\bibinfo {volume} {12}},\ \bibinfo {pages} {41} (\bibinfo {year} {2022})}\BibitemShut {NoStop}%
\bibitem [{\citenamefont {Fiorentino}\ and\ \citenamefont {Baroni}(2023)}]{fiorentino_green-kubo_2023}%
  \BibitemOpen
  \bibfield  {author} {\bibinfo {author} {\bibfnamefont {A.}~\bibnamefont {Fiorentino}}\ and\ \bibinfo {author} {\bibfnamefont {S.}~\bibnamefont {Baroni}},\ }\bibfield  {title} {\bibinfo {title} {From {Green}-{Kubo} to the full {Boltzmann} kinetic approach to heat transport in crystals and glasses},\ }\href {https://doi.org/10.1103/PhysRevB.107.054311} {\bibfield  {journal} {\bibinfo  {journal} {Physical Review B}\ }\textbf {\bibinfo {volume} {107}},\ \bibinfo {pages} {054311} (\bibinfo {year} {2023})},\ \bibinfo {note} {publisher: American Physical Society}\BibitemShut {NoStop}%
\bibitem [{\citenamefont {Kiefer}\ \emph {et~al.}(2025{\natexlab{a}})\citenamefont {Kiefer}, \citenamefont {Héry}, \citenamefont {Tepper}, \citenamefont {Dalton}, \citenamefont {Ayaz},\ and\ \citenamefont {Netz}}]{kiefer_analysis_2025}%
  \BibitemOpen
  \bibfield  {author} {\bibinfo {author} {\bibfnamefont {H.}~\bibnamefont {Kiefer}}, \bibinfo {author} {\bibfnamefont {B.~J.~A.}\ \bibnamefont {Héry}}, \bibinfo {author} {\bibfnamefont {L.}~\bibnamefont {Tepper}}, \bibinfo {author} {\bibfnamefont {B.~A.}\ \bibnamefont {Dalton}}, \bibinfo {author} {\bibfnamefont {C.}~\bibnamefont {Ayaz}},\ and\ \bibinfo {author} {\bibfnamefont {R.~R.}\ \bibnamefont {Netz}},\ }\href {https://doi.org/10.48550/arXiv.2505.15665} {\bibinfo {title} {Analysis and {Simulation} of {Generalized} {Langevin} {Equations} with {Non}-{Gaussian} {Orthogonal} {Forces}}} (\bibinfo {year} {2025}{\natexlab{a}})\BibitemShut {NoStop}%
\bibitem [{\citenamefont {Zon}\ and\ \citenamefont {Schofield}(2001)}]{zon_mode_2001}%
  \BibitemOpen
  \bibfield  {author} {\bibinfo {author} {\bibfnamefont {R.~v.}\ \bibnamefont {Zon}}\ and\ \bibinfo {author} {\bibfnamefont {J.}~\bibnamefont {Schofield}},\ }\bibfield  {title} {\bibinfo {title} {Mode coupling theory for multi-point and multi-time correlation functions},\ }\href {https://doi.org/10.1103/PhysRevE.65.011106} {\bibfield  {journal} {\bibinfo  {journal} {Physical Review E}\ }\textbf {\bibinfo {volume} {65}},\ \bibinfo {pages} {011106} (\bibinfo {year} {2001})}\BibitemShut {NoStop}%
\bibitem [{\citenamefont {Kiefer}\ \emph {et~al.}(2025{\natexlab{b}})\citenamefont {Kiefer}, \citenamefont {Ayaz}, \citenamefont {Dalton},\ and\ \citenamefont {Netz}}]{kiefer_uence_2025}%
  \BibitemOpen
  \bibfield  {author} {\bibinfo {author} {\bibfnamefont {H.}~\bibnamefont {Kiefer}}, \bibinfo {author} {\bibfnamefont {C.}~\bibnamefont {Ayaz}}, \bibinfo {author} {\bibfnamefont {B.~A.}\ \bibnamefont {Dalton}},\ and\ \bibinfo {author} {\bibfnamefont {R.~R.}\ \bibnamefont {Netz}},\ }\bibfield  {title} {\bibinfo {title} {The inﬂuence of multi-dimensionality and oﬀ-diagonal non-{Markovian} friction coupling on coarse-grained dynamics},\ }\bibfield  {journal} {\bibinfo  {journal} {New Journal of Physics}\ }\href {https://doi.org/10.1088/1367-2630/ae1d9f} {10.1088/1367-2630/ae1d9f} (\bibinfo {year} {2025}{\natexlab{b}})\BibitemShut {NoStop}%
\bibitem [{\citenamefont {Jangi}\ and\ \citenamefont {Bai}(2012)}]{Jangi01122012}%
  \BibitemOpen
  \bibfield  {author} {\bibinfo {author} {\bibfnamefont {M.}~\bibnamefont {Jangi}}\ and\ \bibinfo {author} {\bibfnamefont {X.-S.}\ \bibnamefont {Bai}},\ }\bibfield  {title} {\bibinfo {title} {Multidimensional chemistry coordinate mapping approach for combustion modelling with finite-rate chemistry},\ }\href@noop {} {\bibfield  {journal} {\bibinfo  {journal} {Combustion Theory and Modelling}\ }\textbf {\bibinfo {volume} {16}},\ \bibinfo {pages} {1109} (\bibinfo {year} {2012})}\BibitemShut {NoStop}%
\bibitem [{\citenamefont {Lai}\ \emph {et~al.}(2014)\citenamefont {Lai}, \citenamefont {Zhang},\ and\ \citenamefont {Wang}}]{C3CP54476A}%
  \BibitemOpen
  \bibfield  {author} {\bibinfo {author} {\bibfnamefont {Z.}~\bibnamefont {Lai}}, \bibinfo {author} {\bibfnamefont {K.}~\bibnamefont {Zhang}},\ and\ \bibinfo {author} {\bibfnamefont {J.}~\bibnamefont {Wang}},\ }\bibfield  {title} {\bibinfo {title} {Exploring multi-dimensional coordinate-dependent diffusion dynamics on the energy landscape of protein conformation change},\ }\href@noop {} {\bibfield  {journal} {\bibinfo  {journal} {Phys. Chem. Chem. Phys.}\ }\textbf {\bibinfo {volume} {16}},\ \bibinfo {pages} {6486} (\bibinfo {year} {2014})}\BibitemShut {NoStop}%
\bibitem [{\citenamefont {Baldovin}\ \emph {et~al.}(2020)\citenamefont {Baldovin}, \citenamefont {Cecconi},\ and\ \citenamefont {Vulpiani}}]{Baldovin_2020}%
  \BibitemOpen
  \bibfield  {author} {\bibinfo {author} {\bibfnamefont {M.}~\bibnamefont {Baldovin}}, \bibinfo {author} {\bibfnamefont {F.}~\bibnamefont {Cecconi}},\ and\ \bibinfo {author} {\bibfnamefont {A.}~\bibnamefont {Vulpiani}},\ }\bibfield  {title} {\bibinfo {title} {Effective equations for reaction coordinates in polymer transport},\ }\href@noop {} {\bibfield  {journal} {\bibinfo  {journal} {Journal of Statistical Mechanics: Theory and Experiment}\ }\textbf {\bibinfo {volume} {2020}},\ \bibinfo {pages} {013208} (\bibinfo {year} {2020})}\BibitemShut {NoStop}%
\bibitem [{\citenamefont {Liang}\ \emph {et~al.}(2020)\citenamefont {Liang}, \citenamefont {Díaz~Leines}, \citenamefont {Drautz},\ and\ \citenamefont {Rogal}}]{10.1063/5.0010074}%
  \BibitemOpen
  \bibfield  {author} {\bibinfo {author} {\bibfnamefont {Y.}~\bibnamefont {Liang}}, \bibinfo {author} {\bibfnamefont {G.}~\bibnamefont {Díaz~Leines}}, \bibinfo {author} {\bibfnamefont {R.}~\bibnamefont {Drautz}},\ and\ \bibinfo {author} {\bibfnamefont {J.}~\bibnamefont {Rogal}},\ }\bibfield  {title} {\bibinfo {title} {Identification of a multi-dimensional reaction coordinate for crystal nucleation in ni3al},\ }\href@noop {} {\bibfield  {journal} {\bibinfo  {journal} {The Journal of Chemical Physics}\ }\textbf {\bibinfo {volume} {152}},\ \bibinfo {pages} {224504} (\bibinfo {year} {2020})}\BibitemShut {NoStop}%
\bibitem [{\citenamefont {Kawasaki}(1973)}]{kawasaki_simple_1973}%
  \BibitemOpen
  \bibfield  {author} {\bibinfo {author} {\bibfnamefont {K.}~\bibnamefont {Kawasaki}},\ }\bibfield  {title} {\bibinfo {title} {Simple derivations of generalized linear and nonlinear {Langevin} equations},\ }\href {https://doi.org/10.1088/0305-4470/6/9/004} {\bibfield  {journal} {\bibinfo  {journal} {Journal of Physics A: Mathematical, Nuclear and General}\ }\textbf {\bibinfo {volume} {6}},\ \bibinfo {pages} {1289} (\bibinfo {year} {1973})}\BibitemShut {NoStop}%
\bibitem [{\citenamefont {Zwanzig}(2001)}]{zwanzig_nonequilibrium_2001}%
  \BibitemOpen
  \bibfield  {author} {\bibinfo {author} {\bibfnamefont {R.}~\bibnamefont {Zwanzig}},\ }\href@noop {} {\emph {\bibinfo {title} {Nonequilibrium statistical mechanics}}}\ (\bibinfo  {publisher} {Oxford University Press},\ \bibinfo {address} {Oxford ; New York},\ \bibinfo {year} {2001})\BibitemShut {NoStop}%
\bibitem [{\citenamefont {Wolf}\ \emph {et~al.}(2025)\citenamefont {Wolf}, \citenamefont {Klippenstein},\ and\ \citenamefont {van~der Vegt}}]{wolf_cross-correlations_2025}%
  \BibitemOpen
  \bibfield  {author} {\bibinfo {author} {\bibfnamefont {N.}~\bibnamefont {Wolf}}, \bibinfo {author} {\bibfnamefont {V.}~\bibnamefont {Klippenstein}},\ and\ \bibinfo {author} {\bibfnamefont {N.~F.~A.}\ \bibnamefont {van~der Vegt}},\ }\bibfield  {title} {\bibinfo {title} {Cross-correlations in the fluctuation–dissipation relation influence barrier-crossing dynamics},\ }\href {https://doi.org/10.1063/5.0246295} {\bibfield  {journal} {\bibinfo  {journal} {The Journal of Chemical Physics}\ }\textbf {\bibinfo {volume} {162}},\ \bibinfo {pages} {054113} (\bibinfo {year} {2025})}\BibitemShut {NoStop}%
\bibitem [{\citenamefont {Padrick}\ and\ \citenamefont {Miranker}(2002)}]{padrickIsletAmyloidPhase2002}%
  \BibitemOpen
  \bibfield  {author} {\bibinfo {author} {\bibfnamefont {S.~B.}\ \bibnamefont {Padrick}}\ and\ \bibinfo {author} {\bibfnamefont {A.~D.}\ \bibnamefont {Miranker}},\ }\bibfield  {title} {\bibinfo {title} {Islet {{Amyloid}}:\, {{Phase Partitioning}} and {{Secondary Nucleation Are Central}} to the {{Mechanism}} of {{Fibrillogenesis}}},\ }\href {https://doi.org/10.1021/bi0160462} {\bibfield  {journal} {\bibinfo  {journal} {Biochemistry}\ }\textbf {\bibinfo {volume} {41}},\ \bibinfo {pages} {4694} (\bibinfo {year} {2002})}\BibitemShut {NoStop}%
\bibitem [{\citenamefont {Westermark}\ \emph {et~al.}(2011)\citenamefont {Westermark}, \citenamefont {Andersson},\ and\ \citenamefont {Westermark}}]{doi:10.1152/physrev.00042.2009}%
  \BibitemOpen
  \bibfield  {author} {\bibinfo {author} {\bibfnamefont {P.}~\bibnamefont {Westermark}}, \bibinfo {author} {\bibfnamefont {A.}~\bibnamefont {Andersson}},\ and\ \bibinfo {author} {\bibfnamefont {G.~T.}\ \bibnamefont {Westermark}},\ }\bibfield  {title} {\bibinfo {title} {Islet amyloid polypeptide, islet amyloid, and diabetes mellitus},\ }\href@noop {} {\bibfield  {journal} {\bibinfo  {journal} {Physiological Reviews}\ }\textbf {\bibinfo {volume} {91}},\ \bibinfo {pages} {795} (\bibinfo {year} {2011})}\BibitemShut {NoStop}%
\bibitem [{\citenamefont {Patil}\ \emph {et~al.}(2011)\citenamefont {Patil}, \citenamefont {Mehta}, \citenamefont {Jha},\ and\ \citenamefont {Alexandrescu}}]{patilHeterogeneousAmylinFibril2011}%
  \BibitemOpen
  \bibfield  {author} {\bibinfo {author} {\bibfnamefont {S.~M.}\ \bibnamefont {Patil}}, \bibinfo {author} {\bibfnamefont {A.}~\bibnamefont {Mehta}}, \bibinfo {author} {\bibfnamefont {S.}~\bibnamefont {Jha}},\ and\ \bibinfo {author} {\bibfnamefont {A.~T.}\ \bibnamefont {Alexandrescu}},\ }\bibfield  {title} {\bibinfo {title} {Heterogeneous {{Amylin Fibril Growth Mechanisms Imaged}} by {{Total Internal Reflection Fluorescence Microscopy}}},\ }\href {https://doi.org/10.1021/bi101908m} {\bibfield  {journal} {\bibinfo  {journal} {Biochemistry}\ }\textbf {\bibinfo {volume} {50}},\ \bibinfo {pages} {2808} (\bibinfo {year} {2011})}\BibitemShut {NoStop}%
\bibitem [{\citenamefont {Buchanan}\ \emph {et~al.}(2013)\citenamefont {Buchanan}, \citenamefont {Dunkelberger}, \citenamefont {Tran}, \citenamefont {Cheng}, \citenamefont {Chiu}, \citenamefont {Cao}, \citenamefont {Raleigh}, \citenamefont {{de Pablo}}, \citenamefont {Nowick},\ and\ \citenamefont {Zanni}}]{buchananMechanismIAPPAmyloid2013}%
  \BibitemOpen
  \bibfield  {author} {\bibinfo {author} {\bibfnamefont {L.~E.}\ \bibnamefont {Buchanan}}, \bibinfo {author} {\bibfnamefont {E.~B.}\ \bibnamefont {Dunkelberger}}, \bibinfo {author} {\bibfnamefont {H.~Q.}\ \bibnamefont {Tran}}, \bibinfo {author} {\bibfnamefont {P.-N.}\ \bibnamefont {Cheng}}, \bibinfo {author} {\bibfnamefont {C.-C.}\ \bibnamefont {Chiu}}, \bibinfo {author} {\bibfnamefont {P.}~\bibnamefont {Cao}}, \bibinfo {author} {\bibfnamefont {D.~P.}\ \bibnamefont {Raleigh}}, \bibinfo {author} {\bibfnamefont {J.~J.}\ \bibnamefont {{de Pablo}}}, \bibinfo {author} {\bibfnamefont {J.~S.}\ \bibnamefont {Nowick}},\ and\ \bibinfo {author} {\bibfnamefont {M.~T.}\ \bibnamefont {Zanni}},\ }\bibfield  {title} {\bibinfo {title} {Mechanism of {{IAPP}} amyloid fibril formation involves an intermediate with a transient {$\beta$}-sheet},\ }\href {https://doi.org/10.1073/pnas.1314481110} {\bibfield  {journal} {\bibinfo  {journal} {Proceedings of the National Academy of Sciences}\ }\textbf {\bibinfo {volume} {110}},\
  \bibinfo {pages} {19285} (\bibinfo {year} {2013})}\BibitemShut {NoStop}%
\bibitem [{\citenamefont {Moore}\ \emph {et~al.}(2018)\citenamefont {Moore}, \citenamefont {Sonar}, \citenamefont {Bharadwaj}, \citenamefont {Deplazes},\ and\ \citenamefont {Mancera}}]{mooreCharacterisationStructureOligomerisation2018}%
  \BibitemOpen
  \bibfield  {author} {\bibinfo {author} {\bibfnamefont {S.~J.}\ \bibnamefont {Moore}}, \bibinfo {author} {\bibfnamefont {K.}~\bibnamefont {Sonar}}, \bibinfo {author} {\bibfnamefont {P.}~\bibnamefont {Bharadwaj}}, \bibinfo {author} {\bibfnamefont {E.}~\bibnamefont {Deplazes}},\ and\ \bibinfo {author} {\bibfnamefont {R.~L.}\ \bibnamefont {Mancera}},\ }\bibfield  {title} {\bibinfo {title} {Characterisation of the {{Structure}} and {{Oligomerisation}} of {{Islet Amyloid Polypeptides}} ({{IAPP}}): {{A Review}} of {{Molecular Dynamics Simulation Studies}}},\ }\href {https://doi.org/10.3390/molecules23092142} {\bibfield  {journal} {\bibinfo  {journal} {Molecules}\ }\textbf {\bibinfo {volume} {23}},\ \bibinfo {pages} {2142} (\bibinfo {year} {2018})}\BibitemShut {NoStop}%
\bibitem [{\citenamefont {Klimek}\ \emph {et~al.}(2026)\citenamefont {Klimek}, \citenamefont {Dalton}, \citenamefont {Tepper},\ and\ \citenamefont {Netz}}]{klimek_hierarchical_2026}%
  \BibitemOpen
  \bibfield  {author} {\bibinfo {author} {\bibfnamefont {A.}~\bibnamefont {Klimek}}, \bibinfo {author} {\bibfnamefont {B.~A.}\ \bibnamefont {Dalton}}, \bibinfo {author} {\bibfnamefont {L.}~\bibnamefont {Tepper}},\ and\ \bibinfo {author} {\bibfnamefont {R.~R.}\ \bibnamefont {Netz}},\ }\bibfield  {title} {\bibinfo {title} {Hierarchical friction memory leads to subdiffusive configurational dynamics of fast-folding proteins},\ }\href {https://doi.org/10.1073/pnas.2516506123} {\bibfield  {journal} {\bibinfo  {journal} {Proceedings of the National Academy of Sciences}\ }\textbf {\bibinfo {volume} {123}},\ \bibinfo {pages} {e2516506123} (\bibinfo {year} {2026})}\BibitemShut {NoStop}%
\bibitem [{\citenamefont {Milton}\ and\ \citenamefont {Harris}(2010)}]{https://doi.org/10.1100/tsw.2010.73}%
  \BibitemOpen
  \bibfield  {author} {\bibinfo {author} {\bibfnamefont {N.~G.~N.}\ \bibnamefont {Milton}}\ and\ \bibinfo {author} {\bibfnamefont {J.~R.}\ \bibnamefont {Harris}},\ }\bibfield  {title} {\bibinfo {title} {Human islet amyloid polypeptide fibril binding to catalase: A transmission electron microscopy and microplate study},\ }\href@noop {} {\bibfield  {journal} {\bibinfo  {journal} {The Scientific World Journal}\ }\textbf {\bibinfo {volume} {10}},\ \bibinfo {pages} {204535} (\bibinfo {year} {2010})}\BibitemShut {NoStop}%
\bibitem [{\citenamefont {Röder}\ \emph {et~al.}(2020)\citenamefont {Röder}, \citenamefont {Kupreichyk}, \citenamefont {Gremer}, \citenamefont {Schäfer}, \citenamefont {Pothula}, \citenamefont {Ravelli}, \citenamefont {Willbold}, \citenamefont {Hoyer},\ and\ \citenamefont {Schröder}}]{roder_cryo-em_2020}%
  \BibitemOpen
  \bibfield  {author} {\bibinfo {author} {\bibfnamefont {C.}~\bibnamefont {Röder}}, \bibinfo {author} {\bibfnamefont {T.}~\bibnamefont {Kupreichyk}}, \bibinfo {author} {\bibfnamefont {L.}~\bibnamefont {Gremer}}, \bibinfo {author} {\bibfnamefont {L.~U.}\ \bibnamefont {Schäfer}}, \bibinfo {author} {\bibfnamefont {K.~R.}\ \bibnamefont {Pothula}}, \bibinfo {author} {\bibfnamefont {R.~B.~G.}\ \bibnamefont {Ravelli}}, \bibinfo {author} {\bibfnamefont {D.}~\bibnamefont {Willbold}}, \bibinfo {author} {\bibfnamefont {W.}~\bibnamefont {Hoyer}},\ and\ \bibinfo {author} {\bibfnamefont {G.~F.}\ \bibnamefont {Schröder}},\ }\bibfield  {title} {\bibinfo {title} {Cryo-{EM} structure of islet amyloid polypeptide fibrils reveals similarities with amyloid-$\beta$ fibrils},\ }\href@noop {} {\bibfield  {journal} {\bibinfo  {journal} {Nature Structural \& Molecular Biology}\ }\textbf {\bibinfo {volume} {27}},\ \bibinfo {pages} {660} (\bibinfo {year} {2020})}\BibitemShut {NoStop}%
\bibitem [{\citenamefont {Abraham}\ \emph {et~al.}(2015)\citenamefont {Abraham}, \citenamefont {Murtola}, \citenamefont {Schulz}, \citenamefont {Páll}, \citenamefont {Smith}, \citenamefont {Hess},\ and\ \citenamefont {Lindahl}}]{abraham_gromacs_2015}%
  \BibitemOpen
  \bibfield  {author} {\bibinfo {author} {\bibfnamefont {M.~J.}\ \bibnamefont {Abraham}}, \bibinfo {author} {\bibfnamefont {T.}~\bibnamefont {Murtola}}, \bibinfo {author} {\bibfnamefont {R.}~\bibnamefont {Schulz}}, \bibinfo {author} {\bibfnamefont {S.}~\bibnamefont {Páll}}, \bibinfo {author} {\bibfnamefont {J.~C.}\ \bibnamefont {Smith}}, \bibinfo {author} {\bibfnamefont {B.}~\bibnamefont {Hess}},\ and\ \bibinfo {author} {\bibfnamefont {E.}~\bibnamefont {Lindahl}},\ }\bibfield  {title} {\bibinfo {title} {{GROMACS}: High performance molecular simulations through multi-level parallelism from laptops to supercomputers},\ }\href {https://doi.org/10.1016/j.softx.2015.06.001} {\bibfield  {journal} {\bibinfo  {journal} {{SoftwareX}}\ }\textbf {\bibinfo {volume} {1-2}},\ \bibinfo {pages} {19} (\bibinfo {year} {2015})}\BibitemShut {NoStop}%
\bibitem [{\citenamefont {Bussi}\ and\ \citenamefont {Tribello}(2019)}]{bussi_analyzing_2019}%
  \BibitemOpen
  \bibfield  {author} {\bibinfo {author} {\bibfnamefont {G.}~\bibnamefont {Bussi}}\ and\ \bibinfo {author} {\bibfnamefont {G.~A.}\ \bibnamefont {Tribello}},\ }\bibfield  {title} {\bibinfo {title} {Analyzing and biasing simulations with {PLUMED}},\ }in\ \href {https://doi.org/10.1007/978-1-4939-9608-7_21} {\emph {\bibinfo {booktitle} {Biomolecular Simulations: Methods and Protocols}}},\ \bibinfo {editor} {edited by\ \bibinfo {editor} {\bibfnamefont {M.}~\bibnamefont {Bonomi}}\ and\ \bibinfo {editor} {\bibfnamefont {C.}~\bibnamefont {Camilloni}}}\ (\bibinfo  {publisher} {Springer},\ \bibinfo {year} {2019})\ pp.\ \bibinfo {pages} {529--578}\BibitemShut {NoStop}%
\bibitem [{\citenamefont {Bussi}\ \emph {et~al.}(2007)\citenamefont {Bussi}, \citenamefont {Donadio},\ and\ \citenamefont {Parrinello}}]{bussi_canonical_2007}%
  \BibitemOpen
  \bibfield  {author} {\bibinfo {author} {\bibfnamefont {G.}~\bibnamefont {Bussi}}, \bibinfo {author} {\bibfnamefont {D.}~\bibnamefont {Donadio}},\ and\ \bibinfo {author} {\bibfnamefont {M.}~\bibnamefont {Parrinello}},\ }\bibfield  {title} {\bibinfo {title} {Canonical sampling through velocity rescaling},\ }\href {https://doi.org/10.1063/1.2408420} {\bibfield  {journal} {\bibinfo  {journal} {The Journal of Chemical Physics}\ }\textbf {\bibinfo {volume} {126}},\ \bibinfo {pages} {014101} (\bibinfo {year} {2007})}\BibitemShut {NoStop}%
\bibitem [{\citenamefont {Bernetti}\ and\ \citenamefont {Bussi}(2020)}]{bernetti_pressure_2020}%
  \BibitemOpen
  \bibfield  {author} {\bibinfo {author} {\bibfnamefont {M.}~\bibnamefont {Bernetti}}\ and\ \bibinfo {author} {\bibfnamefont {G.}~\bibnamefont {Bussi}},\ }\bibfield  {title} {\bibinfo {title} {Pressure control using stochastic cell rescaling},\ }\href {https://doi.org/10.1063/5.0020514} {\bibfield  {journal} {\bibinfo  {journal} {The Journal of Chemical Physics}\ }\textbf {\bibinfo {volume} {153}},\ \bibinfo {pages} {114107} (\bibinfo {year} {2020})}\BibitemShut {NoStop}%
\bibitem [{\citenamefont {Best}\ and\ \citenamefont {Hummer}(2009)}]{best_optimized_2009}%
  \BibitemOpen
  \bibfield  {author} {\bibinfo {author} {\bibfnamefont {R.~B.}\ \bibnamefont {Best}}\ and\ \bibinfo {author} {\bibfnamefont {G.}~\bibnamefont {Hummer}},\ }\bibfield  {title} {\bibinfo {title} {Optimized molecular dynamics force fields applied to the helix-coil transition of polypeptides},\ }\href {https://doi.org/10.1021/jp901540t} {\bibfield  {journal} {\bibinfo  {journal} {The Journal of Physical Chemistry B}\ }\textbf {\bibinfo {volume} {113}},\ \bibinfo {pages} {9004} (\bibinfo {year} {2009})}\BibitemShut {NoStop}%
\bibitem [{\citenamefont {Lindorff-Larsen}\ \emph {et~al.}(2010)\citenamefont {Lindorff-Larsen}, \citenamefont {Piana}, \citenamefont {Palmo}, \citenamefont {Maragakis}, \citenamefont {Klepeis}, \citenamefont {Dror},\ and\ \citenamefont {Shaw}}]{lindorff-larsen_improved_2010}%
  \BibitemOpen
  \bibfield  {author} {\bibinfo {author} {\bibfnamefont {K.}~\bibnamefont {Lindorff-Larsen}}, \bibinfo {author} {\bibfnamefont {S.}~\bibnamefont {Piana}}, \bibinfo {author} {\bibfnamefont {K.}~\bibnamefont {Palmo}}, \bibinfo {author} {\bibfnamefont {P.}~\bibnamefont {Maragakis}}, \bibinfo {author} {\bibfnamefont {J.~L.}\ \bibnamefont {Klepeis}}, \bibinfo {author} {\bibfnamefont {R.~O.}\ \bibnamefont {Dror}},\ and\ \bibinfo {author} {\bibfnamefont {D.~E.}\ \bibnamefont {Shaw}},\ }\bibfield  {title} {\bibinfo {title} {Improved side-chain torsion potentials for the amber ff99sb protein force field},\ }\href {https://doi.org/10.1002/prot.22711} {\bibfield  {journal} {\bibinfo  {journal} {Proteins: Structure, Function, and Bioinformatics}\ }\textbf {\bibinfo {volume} {78}},\ \bibinfo {pages} {1950} (\bibinfo {year} {2010})}\BibitemShut {NoStop}%
\bibitem [{\citenamefont {Piana}\ \emph {et~al.}(2011)\citenamefont {Piana}, \citenamefont {Lindorff-Larsen},\ and\ \citenamefont {Shaw}}]{piana_how_2011}%
  \BibitemOpen
  \bibfield  {author} {\bibinfo {author} {\bibfnamefont {S.}~\bibnamefont {Piana}}, \bibinfo {author} {\bibfnamefont {K.}~\bibnamefont {Lindorff-Larsen}},\ and\ \bibinfo {author} {\bibfnamefont {D.~E.}\ \bibnamefont {Shaw}},\ }\bibfield  {title} {\bibinfo {title} {How robust are protein folding simulations with respect to force field parameterization?},\ }\href {https://doi.org/10.1016/j.bpj.2011.03.051} {\bibfield  {journal} {\bibinfo  {journal} {Biophysical Journal}\ }\textbf {\bibinfo {volume} {100}},\ \bibinfo {pages} {L47} (\bibinfo {year} {2011})}\BibitemShut {NoStop}%
\bibitem [{\citenamefont {Hoffmann}\ \emph {et~al.}(2015)\citenamefont {Hoffmann}, \citenamefont {{McGovern}}, \citenamefont {Chiu},\ and\ \citenamefont {Pablo}}]{hoffmann_secondary_2015}%
  \BibitemOpen
  \bibfield  {author} {\bibinfo {author} {\bibfnamefont {K.~Q.}\ \bibnamefont {Hoffmann}}, \bibinfo {author} {\bibfnamefont {M.}~\bibnamefont {{McGovern}}}, \bibinfo {author} {\bibfnamefont {C.-c.}\ \bibnamefont {Chiu}},\ and\ \bibinfo {author} {\bibfnamefont {J.~J.~d.}\ \bibnamefont {Pablo}},\ }\bibfield  {title} {\bibinfo {title} {Secondary structure of rat and human amylin across force fields},\ }\href {https://doi.org/10.1371/journal.pone.0134091} {\bibfield  {journal} {\bibinfo  {journal} {{PLOS} {ONE}}\ }\textbf {\bibinfo {volume} {10}},\ \bibinfo {pages} {e0134091} (\bibinfo {year} {2015})}\BibitemShut {NoStop}%
\end{thebibliography}%

\appendix

\clearpage

\renewcommand{\thesection}{\arabic{section}} % Restore section numbering
\setcounter{section}{0} % Reset the section counter
\renewcommand{\appendixname}{} % Clear "Appendix" label

\newpage

%%%%%%%%%% Merge with Supplementary materials %%%%%%%%%%
\pagebreak
\newpage
\widetext
\begin{center}
    \textbf{\large Non-equilibrium generalized Langevin equation for multi-dimensional observables} \\
    \vspace{5mm}
    \textbf{Benjamin J. A. Hery$^{1}$, Lucas Tepper$^{1}$, Andrea Guljas$^{1}$, Artem Pavlov$^{2}$, Beate Koksch$^{2}$, Cecilia Clementi$^{1}$, Roland R. Netz$^{1}$} \\
    \vspace{1mm}
    \textit{$^{1}$ Freie  Universität  Berlin,  Department  of  Physics, 14195  Berlin,  Germany} \\
    \vspace{1mm}
    \textit{$^{2}$ Freie Universität Berlin, Institut für Chemie und Biochemie, 14195 Berlin, Germany}
\end{center}

%%%%%%%%%% Merge with Supplementary materials %%%%%%%%%%
%%%%%%%%%% Prefix a "S" to all equations, figures, tables and reset the counter %%%%%%%%%%
\setcounter{equation}{0}
\setcounter{figure}{0}
\setcounter{table}{0}
\setcounter{page}{1}
\makeatletter
\renewcommand{\thepage}{S\arabic{page}}
\renewcommand{\thesection}{S\arabic{section}}
\renewcommand{\thefigure}{S\arabic{figure}}
\renewcommand{\theequation}{S\arabic{section}.\arabic{equation}}

%\renewcommand{\bibnumfmt}[1]{[S#1]}
%\renewcommand{\citenumfont}[1]{S#1}
%%%%%%%%%% Prefix a "S" to all equations, figures, tables and reset the counter %%%%%%%%%%
%\section*{Supplementary Material}

\section{\label{si_1:solve}Solution of the Liouville equation}

We derive an expression for the probability density $\rho(t, \vec{w})$. First, we integrate eq.~\ref{s2_ss1_eq:liouville_equation} over the time variable from the initial time $t_{0}$ to the current time $t$. We obtain the integral equation
\begin{align}
    \label{si_1_eq:solution_1}
    & \rho(t, \vec{w}) = \rho(t_{0}, \vec{w}) - \int_{t_{0}}^{t} dt_{1} \mathcal{L}(t_{1}) \rho(t_{1}, \vec{w})
\end{align}
that we solve recursively, and by which we recover the expression of $\rho(t, \vec{w})$ in eq.~\ref{s2_ss1_eq:probability_density} alongside the definition of the Schrödinger-type propagation operator in eq.~\ref{s2_ss1_def:shcrödinger_evolution_operator}.

\section{\label{si_2:generic_observable}Derivation of the Heisenberg generic observable}

We derive the expression for the Heisenberg observable $O(t_{0}, t, \vec{w})$ in eq.~\ref{s2_ss2_def:generic_observable}. First, we compute
\begin{align}
    \label{si_2_eq:ensemble_expectation}
    & o(t) = \int_{\mathbb{W}} d\hat{\vec{w}} \: O_{S}(\hat{\vec{w}}) \left\lbrack \left( \mathcal{I} + \sum_{n \geq 1} \left( \prod_{k = 1}^{n} (-1)^{n} \int_{t_{0}}^{\delta_{k,1}t + (1-\delta_{k,1})t_{k-1}} dt_{k} \prod_{j = 1}^{n} \mathcal{L}(t_{j}) \right) \right) \rho(t_{0}, \hat{\vec{w}}) \right\rbrack
\end{align}
where we substituted for $\rho(t, \vec{w})$ the  expression in eq.~\ref{s2_ss1_eq:probability_density}. We recall that $\mathcal{L}(t_{j})$ defines an anti-self-adjoint operator and rewrite eq.~\ref{si_2_eq:ensemble_expectation} as
\begin{align}
    \label{si_2_eq:ensemble_expectation_2}
    & o(t) = \int_{\mathbb{W}} d\hat{\vec{w}} \: \rho(t_{0}, \hat{\vec{w}}) \left\lbrack \left( \mathcal{I} + \sum_{n \geq 1} \left( \prod_{k = 1}^{n} \int_{t_{0}}^{\delta_{k,1}t + (1-\delta_{k,1})t_{k-1}} dt_{k} \prod_{n = 1}^{1} \mathcal{L}(t_{j}) \right) \right) O_{S}(\hat{\vec{w}}) \right\rbrack,
\end{align}
from which we recover the definition of the Heisenberg-type propagator in eq.~\ref{s2_ss2_def:heisenberg_propagator}, the definition of the generic observable in eq.~\ref{s2_ss2_def:generic_observable}, and thus eq.~\ref{s2_ss2_eq:ensemble_expectation}.

\section{\label{si_3:projection_formalism}Projection operator formalism}

We review the general theory of projection operator formalism presented by H. Vroylandt~\cite{vroylandt_position-dependent_2022} and adapt its main results. First, we consider the projection set
\begin{align}
    \label{si_3_def:projection_set_discrete}
    & \varepsilon \equiv \bigcup_{k \in I} \{ E_{k}(\vec{w}) \}
\end{align}
where each $E_{k}(\vec{w})$ is a phase space observables and $I$ is a set of indices. The projection operator $\mathcal{P}_{\varepsilon}$ associated to $\varepsilon$ is defined by
\begin{align}
    \label{si_3_def:projection_operator_discrete}
    & \mathcal{P}_{\varepsilon} O(t_{0}, t, \vec{w}) \equiv \sum_{k, k' \in I} \langle O(t_{0}, t, \hat{\vec{w}}) E_{k}(\hat{\vec{w}}) \rangle (\hat{G}_{\varepsilon}^{-1})_{k k'} E_{k'}(w),
\end{align}
where $\hat{G}_{\varepsilon}^{-1}$ is the inverse of the Gram matrix $\hat{G}_{\varepsilon}$ defined component-wise by
\begin{align}
    \label{si_3_def:gram_matrix}
    & (\hat{G}_{\varepsilon})_{k k'} \equiv \langle E_{k}(\hat{\vec{w}}) E_{k'}(\hat{\vec{w}}) \rangle,
\end{align}
and which satisfies the relation
\begin{align}
    \label{si_3_eq:matricial_relation}
    & \sum_{k" \in I} (\hat{G}_{\varepsilon}^{-1})_{k k"} (\hat{G}_{\varepsilon})_{k" k'} = \delta_{k k'}.
\end{align}
Second, we show that $\mathcal{P}_{\varepsilon}$ defines a linear operator, i.e.
\begin{align}
    \label{si_3_eq:linearity}
    & \mathcal{P}_{\varepsilon} (\lambda_{1} X(t_{0}, t, \vec{w}) + \lambda_{2} Y(t_{0}, t', \vec{w})) = \lambda_{1} \mathcal{P}_{\varepsilon} X(t_{0}, t, \vec{w}) + \lambda_{2} \mathcal{P}_{\varepsilon} Y(t_{0}, t', \vec{w})
\end{align}
for arbitrary observables $X(t_{0}, t, \vec{w})$ and $Y(t_{0}, t', \vec{w})$ and arbitrary real numbers $\lambda_{1}$ and $\lambda_{2}$. We use that the expectation $\langle \cdot \rangle$ defined in eq.~\ref{s2_ss2_eq:ensemble_expectation} and the summations over the indices $k$ and $k'$ define linear operators to recover eq.~\ref{si_3_eq:linearity}. Third, we show that each $E_{k}(\vec{w})$ is an invariant of $\mathcal{P}_{\varepsilon}$, i.e.
\begin{align}
    \label{si_3_eq:invariants}
    & \mathcal{P}_{\varepsilon} E_{l}(\vec{w}) = E_{l}(\vec{w}).
\end{align}
We use eq.~\ref{si_3_def:projection_operator_discrete} where we substituted $E_{l}(\vec{w})$ for $O(t_{0}, t, \vec{w})$ for an arbitrary $l \in I$ and obtain
\begin{align}
    \label{si_3_def:invariants_2}
    & \mathcal{P}_{\varepsilon} E_{l}(\vec{w}) \equiv \sum_{k, k' \in I} \langle E_{l}(\hat{\vec{w}}) E_{k}(\hat{\vec{w}}) \rangle (\hat{G}_{\varepsilon}^{-1})_{k k'} E_{k'}(\vec{w}),
\end{align}
 using eq.~\ref{si_3_def:gram_matrix} and eq.~\ref{si_3_eq:matricial_relation} we recover eq.~\ref{si_3_eq:invariants}. Fourth, we show that $\mathcal{P}_{\varepsilon}$ defines an idempotent operator, i.e.
\begin{align}
    \label{si_3_eq:idempotency}
    & \mathcal{P}_{\varepsilon}^{2} O(t_{0}, t, \vec{w}) = \mathcal{P}_{\varepsilon} O(t_{0}, t, \vec{w}).
\end{align}
We compute
\begin{align}
    \label{si_3_eq:idempotency_2}
    & \mathcal{P}_{\varepsilon}^{2} O(t_{0}, t, \vec{w}) = \mathcal{P}_{\varepsilon} \left( \sum_{k, k' \in I} \langle O(t_{0}, t, \hat{\vec{w}}) E_{k}(\hat{\vec{w}}) \rangle (\hat{G}_{\varepsilon}^{-1})_{k k'} E_{k'}(\vec{w}) \right)
\end{align}
and use eqs.~\ref{si_3_eq:linearity} and~\ref{si_3_eq:invariants} to recover eq.~\ref{si_3_eq:idempotency}. Fifth, we show that $\mathcal{P}_{\varepsilon}$ defines a self-adjoint operator, i.e.
\begin{align}
    \label{si_3_eq:self_adjointedness}
    & \langle X(t_{0}, t, \hat{\vec{w}}) \mathcal{P}_{\varepsilon} Y(t_{0}, t', \hat{\vec{w}}) \rangle = \langle Y(t_{0}, t', \hat{\vec{w}}) \mathcal{P}_{\varepsilon} X(t_{0}, t, \hat{\vec{w}}) \rangle.
\end{align}
We compute
\begin{align}
    \label{si_3_eq:self_adjointedness_2}
    & \langle X(t_{0}, t, \hat{\vec{w}}) \mathcal{P}_{\varepsilon} Y(t_{0}, t', \hat{\vec{w}}) \rangle = \sum_{k, k' \in I} \langle X(t_{0}, t, \hat{\vec{w}}) E_{k}(\hat{\vec{w}}) \rangle (\hat{G}_{\varepsilon}^{-1})_{k k'} \langle Y(t_{0}, t', \hat{\vec{w}}) E_{k'}(\hat{\vec{w}}) \rangle
\end{align}
and reorder the right-hand side using in eq.~\ref{si_3_def:gram_matrix} that $\hat{G}_{\varepsilon}$ and $\hat{G}_{\varepsilon}^{-1}$ define symmetric matrices to recover eq.~\ref{si_3_eq:self_adjointedness}. Sixth, we define the complementary projection operator
\begin{align}
    \label{si_3_eq:complementary_projection_operator}
    & \mathcal{Q}_{\varepsilon} O(t_{0}, t, \vec{w}) \equiv (\mathcal{I}  - \mathcal{P}_{\varepsilon}) O(t_{0}, t, \vec{w}).
\end{align}
$\mathcal{Q}_{\varepsilon}$ also defines a linear, idempotent, and self-adjoint operator. Therefore, we obtain the relation
\begin{align}
    \label{si_3_def:orthogonal_projection}
    & \langle \lbrack \mathcal{Q}_{\varepsilon} X(t_{0}, t, \hat{\vec{w}}) \rbrack \mathcal{P}_{\varepsilon} Y(t_{0}, t', \hat{\vec{w}}) \rangle = 0
\end{align}
where in particular
\begin{align}
    \label{si_3_def:volterra_scheme}
    & \langle E_{k}(\hat{\vec{w}}) \mathcal{Q}_{\varepsilon} O(t_{0}, t, \hat{\vec{w}}) \rangle = 0.
\end{align}
Seventh, we split $\varepsilon$ into the finite union of subsets
\begin{align}
    \label{si_3_eq:projection_set_discrete_1}
    & \varepsilon = \bigcup_{i \geq 1} \varepsilon_{i}
\end{align}
where each subset is defined by
\begin{align}
    \label{si_3_eq:projection_set_discrete_2}
    & \varepsilon_{i} = \bigcup_{k_{i} \in I_{i}} \{ E_{k_{i}}(\vec{w}) \}
\end{align}
with $I_{i}$ being a subset $I$. Hence, eq.~\ref{si_3_def:projection_operator_discrete} becomes
\begin{align}
    \label{si_3_eq:projection_operator_discrete_2}
    & \mathcal{P}_{\varepsilon} O(t_{0}, t, \vec{w}) \equiv \sum_{i, i' \geq 1} \sum_{k_{i}, k_{i'}^{'} \in I_{i}, I_{i'}} \langle O(t_{0}, t, \hat{\vec{w}}) E_{k_{i}}(\hat{\vec{w}}) \rangle (\hat{G}_{\varepsilon}^{-1})_{k_{i} k_{i'}'} E_{k_{i'}'}(\vec{w})
\end{align}
and eq.~\ref{si_3_def:gram_matrix}
\begin{align}
    \label{si_3_eq:gram_matrix_1}
    & (\hat{G}_{\varepsilon})_{k_{i} k_{i'}'} = \langle E_{k_{i}}(\hat{\vec{w}}) E_{k_{i'}'}(\hat{\vec{w}}) \rangle,
\end{align}
where the original Gram matrix eq.~\ref{si_3_def:gram_matrix} is divided in blocks. We use that $(\hat{G}_{\varepsilon})_{k_{i}, k_{i'}'}$ still define a symmetric matrix to assert that there exists a unitary transfer matrix $\hat{T}_{\varepsilon \tilde{\varepsilon}}$ such that each $E_{k_{i}}(\vec{w})$ can be expressed as
\begin{align}
    \label{si_3_eq:transformation_1}
    & E_{k_{i}}(\vec{w}) = \sum_{j \geq 1} \sum_{l_{j} \in I_{j}} ( \hat{T}_{\varepsilon \tilde{\varepsilon}} )_{k_{i} l_{j}} \tilde{E}_{l_{j}}(\vec{w})
\end{align}
that is a linear combination of observables from the modified projection set $\varepsilon$ for which the modified subsets are orthogonal, i.e. $\langle \tilde{E}_{l_{j}}(\hat{\vec{w}}) \tilde{E}_{l_{j'}'}(\hat{\vec{w}}) \rangle = \delta_{j j'} \langle \tilde{E}_{l_{j'}}(\hat{\vec{w}}) \tilde{E}_{l_{j'}'}(\hat{\vec{w}}) \rangle$. Therefore, eq.~\ref{si_3_eq:projection_operator_discrete_2} reads
\begin{align}
    \label{si_3_eq:projection_operator_discrete_3}
    & \mathcal{P}_{\varepsilon} O(t_{0}, t, \vec{w}) \equiv \sum_{j, j' \geq 1} \sum_{l_{j}, l_{j'}' \in I_{j}, I_{j'}} \langle O(t_{0}, t, \hat{\vec{w}}) \tilde{E}_{l_{j}}(\hat{\vec{w}}) \rangle (\hat{G}_{\tilde{\varepsilon}}^{-1})_{l_{j} l_{j'}'} \tilde{E}_{l_{j'}'}(\vec{w}),
\end{align}
where
\begin{align}
    \label{si_3_eq:gram_matrix_2}
    & (\hat{G}_{\tilde{\varepsilon}})_{l_{j} l_{j'}^{'}} = \sum_{i, i' \geq 1} \sum_{k_{i}, k_{i'}' \in I_{i}, I_{i'}} ( \hat{T}_{\varepsilon \tilde{\varepsilon}}^{-1} )_{l_{j} k_{i}} (\hat{G}_{\varepsilon})_{k_{i} k_{i'}'} ( \hat{T}_{\varepsilon \tilde{\varepsilon}} )_{k_{i'}' l_{j'}'} = (\hat{G}_{\tilde{\varepsilon}})_{l_{j} l_{j}'} \delta_{j j'}.
\end{align}
Thus, eq.~\ref{si_3_eq:projection_operator_discrete_3} splits into
\begin{align}
    \label{si_3_eq:projection_operator_discrete_total}
    & \mathcal{P}_{\varepsilon} O(t_{0}, t, \vec{w}) \equiv \sum_{j \geq 1} \mathcal{P}_{\tilde{\varepsilon}_{j}} O(t_{0}, t, \vec{w})
\end{align}
where
\begin{align}
    \label{si_3_def:sub_projection_operator_discrete}
    & \mathcal{P}_{\tilde{\varepsilon}_{j}} O(t_{0}, t, \vec{w}) \equiv \sum_{l_{j}, l_{j}^{'} \in I_{j}} \langle O(t_{0}, t, \hat{\vec{w}}) \tilde{E}_{l_{j}}(\hat{\vec{w}}) \rangle (\hat{G}_{\tilde{\varepsilon}}^{-1})_{l_{j} l_{j}^{'}} \tilde{E}_{l_{j}^{'}}(\vec{w}).
\end{align}
We say that the projection subsets $\tilde{\varepsilon}_{j}$ are orthogonal with respect to each other, where
\begin{align}
    \label{si_3_eq:sub_idempotency}
    & \mathcal{P}_{\tilde{\varepsilon}_{j}} \mathcal{P}_{\tilde{\varepsilon}_{j'}} O(t_{0}, t, \vec{w}) = \delta_{j j'} \mathcal{P}_{\tilde{\varepsilon}_{j}} O(t_{0}, t, \vec{w})
\end{align}
and
\begin{align}
    \label{si_3_eq:sub_orthogonal_projection}
    & \langle \lbrack \mathcal{P}_{\tilde{\varepsilon}_{j}} X(t_{0}, t, \hat{\vec{w}}) \rbrack \mathcal{P}_{\tilde{\varepsilon}_{j'}} Y(t_{0}, t', \hat{\vec{w}}) \rangle = \delta_{j j'} \langle X(t_{0}, t, \hat{\vec{w}}) \mathcal{P}_{\tilde{\varepsilon}_{j}} Y(t_{0}, t', \hat{\vec{w}}) \rangle.
\end{align}
In practice, $\mathcal{P}_{\varepsilon}$ is split into  $\mathcal{P}_{\tilde{\varepsilon}_{j}}$ by applying a Gram-Schmidt algorithm on the projection set $\varepsilon$, which we do in sec.~\ref{si_4:mori_projection} for the multi-dimensional Mori projection operator.

\section{\label{si_4:mori_projection}Constructing the multi-dimensional Mori projection operator}

In this section, we construct $\mathcal{P}_{M}$ in eq.~\ref{s2_ss4_def:mori_projection_operator}. First, we consider the projection set
\begin{align}
    \label{si_4_def:projection_set}
    & \varepsilon_{M} \equiv \varepsilon_{1} \cup \varepsilon_{A} \cup \varepsilon_{\dot{A}}
\end{align}
where
\begin{subequations}
    \label{si_4_def:projection_subsets}
    \begin{align}
        \label{si_4_def:projection_subset_1}
        & \varepsilon_{1} \equiv \{ 1 \}, \\
        \label{si_4_def:projection_subset_A}
        & \varepsilon_{A} \equiv \bigcup_{k \in 1}^{d} \{ A_{S,k}(\vec{r}) \}, \\
        \label{si_4_def:projection_subset_dotA}
        & \varepsilon_{\dot{A}} = \bigcup_{k \in 1}^{d} \{ \dot{A}_{S,k}(\vec{w}) \}.
    \end{align}
\end{subequations}
Second, we consider the modified projection set $\tilde{\varepsilon}_{M}$
\begin{subequations}
    \label{si_4_def:modified_projection_subsets}
    \begin{align}
        \label{si_4_def:modified_projection_subset_1}
        & \tilde{\varepsilon}_{1} = \varepsilon_{1}, \\
        \label{si_4_def:modified_projection_subset_A}
        & \tilde{\varepsilon}_{A} \equiv \bigcup_{k \in 1}^{d} \{ A_{S,k}(\vec{r}) - \lambda_{k} \langle A_{S,k}(\hat{\vec{r}}) \rangle  \}, \\
        \label{si_4_def:modified_projection_subset_dotA}
        & \tilde{\varepsilon}_{\dot{A}} \equiv \bigcup_{k \in 1}^{d} \left\{ \dot{A}_{S,k}(\vec{w}) - \mu_{k} \langle \dot{A}_{S,k}(\hat{\vec{w}}) \rangle - \sum_{l}^{d} \hat{T}_{k l} ( A_{S,l}(\vec{r}) - \lambda_{l} \langle A_{S,l'}(\hat{\vec{r}}) \rangle ) \right\}
    \end{align}
\end{subequations}
and chose each $\lambda_{k}$, each $\mu_{k}$, and each component of the $d \times d$ matrix $\hat{T}$ such that $\tilde{\varepsilon}_{1}$, $\tilde{\varepsilon}_{A}$, and $\tilde{\varepsilon}_{\dot{A}}$ are pair-wise orthogonal, i.e.
\begin{subequations}
    \label{si_4_eq:system_parameters}
    \begin{align}
        \label{si_4_eq:parameters_subset_A}
        & \langle A_{S,k}(\hat{\vec{r}}) \rangle (1 - \lambda_{k}) = 0, \\
        \label{si_4_eq:parameters_subset_dotA_1}
        & \langle \dot{A}_{S,k}(\hat{\vec{w}}) \rangle (1 - \mu_{k}) - \sum_{l, l'} \hat{T}_{l l'} \langle A_{S,l'}(\hat{\vec{r}}) \rangle (1 - \lambda_{l'}) = 0, \\
        \label{si_4_eq:parameters_subset_dotA_2}
        & \left\langle \dot{A}_{S,k}(\hat{\vec{w}}) (A_{S,k}(\hat{\vec{r}}) - \lambda_{k} \langle A_{S,k}(\tilde{\vec{r}})) \right\rangle - \sum_{l}^{d} \hat{T}_{k l} \left\langle ( A_{S,l}(\hat{\vec{r}}) - \lambda_{l} \langle A_{S,l}(\tilde{\vec{r}}) \rangle ) ( A_{S,k}(\hat{\vec{r}}) - \lambda_{k} \langle A_{S,k}(\tilde{\vec{r}}) ) \right\rangle = 0.
    \end{align}
\end{subequations}
We obtain
\begin{subequations}
    \label{si_4_eq:system_parameters_2}
    \begin{align}
        \label{si_4_eq:parameters_2_subset_A}
        & \lambda_{k} = 1, \\
        \label{si_4_eq:parameters_2_subset_dotA_1}
        & \mu_{k} = 1, \\
        \label{si_4_eq:parameters_2_subset_dotA_2}
        & \hat{T}_{k k"} = \sum_{k' = 1}^{d} \left\langle \dot{A}_{S,k}(\hat{\vec{w}}) (A_{S,k'}(\hat{\vec{r}}) - \langle A_{S,k'} (\tilde{\vec{r}})) \right\rangle (\hat{G}_{A})_{k' k"},
    \end{align}
\end{subequations}
where we defined  the Gram matrix
\begin{align}
    \label{si_4_eq:gram_matrix_A}
    & (\hat{G}_{A})_{k k'} \equiv \langle ( A_{S, k}(\hat{\vec{r}}) - \langle A_{S, k}(\tilde{\vec{r}}) \rangle ) ( A_{S, k'}(\hat{\vec{r}}) - \langle A_{S, k'}(\tilde{\vec{r}}) \rangle ) \rangle.
\end{align}
Moreover, we recall from eq.~\ref{s2_ss4_eq:initial_distribution} that $\rho(t_{0}, \vec{w})$ is defined as the Boltzmann distribution determined by $H_{0}(\vec{w})$ and from eq.~\ref{s2_ss2_eq:observable_derivatives} that $\dot{A}_{S, k}(\vec{w}) = \mathcal{L}_{0} A_{S, k}(\vec{r})$. Thus, since $\mathcal{L}_{0}$ is the Liouville operator associated with $H_{0}(\vec{w})$, we obtain
\begin{align}
    \label{si_4_def:simplification}
    & \langle \dot{A}_{S, k}(\hat{\vec{w}}) \rangle = 0.
\end{align}
Furthermore, we compute 
\begin{align}
    \label{si_4_def:cross_correlation}
    & \langle ( A_{S, k}(\hat{\vec{r}}) - \langle A_{S, k}(\tilde{\vec{r}}) \rangle ) \dot{A}_{S, k'}(\hat{\vec{w}}) \rangle = \langle A_{S, k}(\hat{\vec{r}}) \dot{A}_{S, k'}(\hat{\vec{w}}) \rangle
\end{align}
which  we rewrite as
\begin{align}
    \label{si_4_def:cross_correlation_2}
    \langle A_{S, k}(\hat{\vec{r}}) \dot{A}_{S, k'}(\hat{\vec{w}}) \rangle = & \frac{1}{Z(\beta)} \sum_{j = 1}^{3N} \frac{1}{m_{j}} \prod_{i = 1}^{3N} \int_{- \infty}^{+ \infty} d\hat{r}_{i} \: A_{S, k}(\hat{\vec{r}}) \frac{\partial A_{S, k'}(\hat{\vec{r}})}{\partial \hat{r}_{j}} \exp \left( - \beta V(\vec{r}) \right) \\
    & \prod_{k = 1}^{3N} \int_{- \infty}^{+ \infty} d\hat{p}_{k} \: \hat{p}_{j} \exp \left( - \frac{\beta \hat{p}_{k}^{2}}{2 m_{k}} \right).
\end{align}
Since the $3N$ integrals over the components of $\vec{p}$ are the first moments of Gaussian integrals, $\langle A_{S, k}(\hat{\vec{r}}) \dot{A}_{S, k'}(\hat{\vec{w}}) \rangle = 0$, $\hat{T}_{k, k"} = 0$, and 
\begin{subequations}
    \label{si_4_eq:modified_projection_subsets_2}
    \begin{align}
        \label{si_4_eq:modified_projection_subset_1_2}
        & \tilde{\varepsilon}_{1} = \varepsilon_{1}, \\
        \label{si_4_eq:modified_projection_subset_A_2}
        & \tilde{\varepsilon}_{A} = \bigcup_{k \in 1}^{d} \{ A_{S,k}(\vec{r}) - \langle A_{S,k}(\hat{\vec{r}}) \rangle \}, \\
        \label{si_4_eq:modified_projection_subset_dotA_2}
        & \tilde{\varepsilon}_{\dot{A}} = \varepsilon_{\dot{A}}.
    \end{align}
\end{subequations}
Consequently, we split $\mathcal{P}_{M}$ into
\begin{align}
    \label{si_4_eq:mori_projection_split}
    \mathcal{P}_{M} O(t', t_{0}, t, \vec{w}) = \mathcal{P}_{1} O(t', t_{0}, t, \vec{w}) + \mathcal{P}_{A} O(t', t_{0}, t, \vec{w}) + \mathcal{P}_{\dot{A}} O(t', t_{0}, t, \vec{w})
\end{align}
where
\begin{subequations}
    \label{si_4_eq:orthogonal_projection_operators}
    \begin{align}
        \label{si_4_def:projection_operator_1}
        & \mathcal{P}_{1} O(t', t_{0}, t, \vec{w}) \equiv \langle O(t', t_{0}, t, \hat{\vec{w}}) \rangle, \\
        \label{si_4_def:projection_operator_A}
        & \mathcal{P}_{A} O(t', t_{0}, t, \vec{w}) \equiv \sum_{k = 1}^{d} \langle O(t', t_{0}, t, \hat{\vec{w}}) (A_{S, k}(\hat{\vec{r}}) - \langle A_{S, k}(\tilde{\vec{r}}) \rangle ) \rangle \: (\hat{G}_{A}^{-1})_{k k'} ( A_{S, k'}(\vec{r}) - \langle A_{S, k'}(\hat{\vec{r}}) \rangle, \\
        \label{si_4_def:projection_operator_dotA}
        & \mathcal{P}_{\dot{A}} O(t', t_{0}, t, \vec{w}) \equiv \sum_{k = 1}^{d} \langle O(t', t_{0}, t, \hat{\vec{w}}) \dot{A}_{S, k}(\hat{\vec{w}}) \rangle (\hat{G}_{\dot{A}}^{-1})_{k k'} \dot{A}_{S, k'}(\vec{w}),
    \end{align}
\end{subequations}
with $\hat{G}_{\dot{A}}$ being the Gram matrix associated to $\tilde{\varepsilon}_{\dot{A}}$ and defined component-wise by
\begin{align}
    \label{si_4_def:gram_matrix_dotA}
    & ( \hat{G}_{\dot{A}} )_{k k'} \equiv \langle \dot{A}_{S, k}(\hat{\vec{w}}) \dot{A}_{S, k'}(\hat{\vec{w}}) \rangle.
\end{align}
Finally, we recover from eqs.~\ref{si_4_eq:mori_projection_split} and~\ref{si_4_eq:orthogonal_projection_operators} the expression of $\mathcal{P}_{M}$ in eq.~\ref{s2_ss4_def:mori_projection_operator}. We know from section~\ref{si_3:projection_formalism} that $\mathcal{P}_{M}$ defines a linear, idempotent, and self-adjoint operator. Moreover, $1$, all components of $\vec{A}_{S}(\vec{r})$, and all components of $\dot{\vec{A}}_{S}(\vec{w})$ are its invariants.

\section{\label{si_5:prerequisite}Derivation of eq.~\ref{si_5:prerequisite_4}}

We compute the expression of $\langle X_{S}(\hat{\vec{w}}) \mathcal{L}(t) Y_{S}(\hat{\vec{w}}) \rangle$. First, we write
\begin{align}
    \label{si_5:prerequisite_1}
    & \langle X_{S}(\hat{\vec{w}}) \mathcal{L}(t) Y_{S}(\hat{\vec{w}}) \rangle = \int_{\mathbb{W}} d^{6N}\hat{\vec{w}} \: \rho(t_{0}, \hat{\vec{w}}) \: X_{S}(\hat{\vec{w}}) \: \mathcal{L}(t) \: Y_{S}(\hat{\vec{w}})
\end{align}
and use that $\mathcal{L}(t, \vec{w})$ defines an anti-self-adjoint operator to obtain
\begin{align}
    \label{si_5:prerequisite_2}
    & \langle X_{S}(\hat{\vec{w}}) \mathcal{L}(t) Y_{S}(\hat{\vec{w}}) \rangle = - \langle Y_{S}(\hat{\vec{w}}) \mathcal{L}(t) X_{S}(\hat{\vec{w}}) \rangle - \langle X_{S}(\hat{\vec{w}}) Y_{S}(\hat{\vec{w}}) \mathcal{L}(t) \log(\rho(t_{0}, \vec{w})) \rangle.
\end{align}
Second, we substitute for $\rho(t_{0}, \vec{w})$ its expression in eq.~\ref{s2_ss4_eq:initial_distribution} to obtain
\begin{align}
    \label{si_5:prerequisite_3}
    & \langle X_{S}(\hat{\vec{w}}) \mathcal{L}(t) Y_{S}(\hat{\vec{w}}) \rangle = - \langle Y_{S}(\hat{\vec{w}}) \mathcal{L}(t) X_{S}(\hat{\vec{w}}) \rangle + \beta \langle X_{S}(\hat{\vec{w}}) Y_{S}(\hat{\vec{w}}) \mathcal{L}(t) H_{0}(\hat{\vec{w}}) \rangle.
\end{align}
Third, we use that $\mathcal{L}(t) H_{0}(\vec{w}) = \mathcal{L}_{0} H_{0}(\vec{w}) - \mathcal{L}_{1}(t) H_{0}(\vec{w}) = \mathcal{L}_{0} H_{1}(t, \vec{r})$ and obtain
\begin{align}
    \label{si_5:prerequisite_4}
    & \langle X_{S}(\hat{\vec{w}}) \mathcal{L}(t) Y_{S}(\hat{\vec{w}}) \rangle = - \langle Y_{S}(\hat{\vec{w}}) \mathcal{L}(t) X_{S}(\hat{\vec{w}}) \rangle + \beta \langle X_{S}(\hat{\vec{w}}) \lbrack \mathcal{L}_{0} H_{1}(t, \hat{\vec{r}}) \rbrack Y_{S}(\hat{\vec{w}}) \rangle.
\end{align}

\section{\label{si_6:relevant_force}Computation of the effective force}

We compute the expression of $F_{\text{eff},i}(t_{0}, t, \vec{w})$ in eq.~\ref{s2_ss3_def:effective_force}. First, we substitute $\mathcal{P}_{M}$ for $\mathcal{P}$ in eq.~\ref{s2_ss3_def:effective_force} and obtain
\begin{align}
    \label{si_6_eq:effective_force_1}
    & F_{\text{eff},i}(t_{0}, t, \vec{w}) = \exp_{H} \left( \int_{t_{0}}^{t} ds \: \mathcal{L}(t) \right) \mathcal{P}_{M} \ddot{A}_{S, i}(t, \vec{w})
\end{align}
where
\begin{align}
    \label{si_6_eq:initial_effective_force_1}
    \mathcal{P}_{M} \ddot{A}_{S, i}(t, \vec{w}) = & \langle \ddot{A}_{S, i}(t, \hat{\vec{w}}) \rangle + \sum_{k, k' = 1}^{d} \langle \ddot{A}_{S, i}(t, \hat{\vec{w}}) (A_{S, k}(\hat{\vec{r}}) - \langle A_{S, k}(\tilde{\vec{r}}) \rangle ) \rangle (\hat{G}_{A}^{-1})_{k k'} ( A_{S, k'}(\vec{r}) - \langle A_{S, k'}(\hat{\vec{r}}) \rangle ) \\
    & + \sum_{k, k' = 1}^{d} \langle \ddot{A}_{S, i}(t, \hat{\vec{w}}) \dot{A}_{S, k}(\hat{\vec{w}}) \rangle (\hat{G}_{\dot{A}}^{-1})_{k k'} \dot{A}_{S, k'}(\vec{w}). \nonumber
\end{align}
Second, we use eq.~\ref{si_5:prerequisite_4} where we substituted $1$ for $O_{S,1}(t', \vec{w})$ and $\ddot{A}_{S, i}(t, \vec{w})$ for $O_{S,2}(t", \vec{w})$ to recover eq.~\ref{s2_ss5_def:neq_force} from $\langle \ddot{A}_{S, i}(t, \hat{\vec{w}}) \rangle$. Third, we use eq.~\ref{si_5:prerequisite_4} where we substituted $A_{S, k}(\vec{r}) - \langle A_{S, k}(\hat{\vec{r}}) \rangle$ for $O_{S,1}(t', \vec{w})$ and $\ddot{A}_{S, i}(t, \vec{w})$ for $O_{S,2}(t", \vec{w})$ to obtain
\begin{align}
    \label{si_6_eq:harmonic_force_1}
    & \langle \ddot{A}_{S, i}(t, \hat{\vec{w}}) (A_{S, k}(\hat{\vec{r}}) - \langle A_{S, k}(\tilde{\vec{r}}) \rangle ) \rangle = - \langle \dot{A}_{S, i}(\hat{\vec{w}}) \dot{A}_{S, k}(\hat{\vec{w}}) \rangle + \beta \langle \dot{A}_{S, i}(\hat{\vec{w}}) \lbrack \mathcal{L}_{0} H_{1}(t, \hat{\vec{r}}) \rbrack ( A_{S, k}(\hat{\vec{r}}) - \langle A_{S, k}(\tilde{\vec{r}}) \rangle ) \rangle.
\end{align}
Hence, the second second contribution to $\mathcal{P}_{M} \ddot{A}_{S, i}(t, \vec{w})$ in the right-hand-side of eq.~\ref{si_6_eq:initial_effective_force_1} reads
\begin{align}
    \label{si_6_eq:harmonic_force_2}
    & \sum_{k, k' = 1}^{d} \langle \ddot{A}_{S, i}(t, \hat{\vec{w}}) (A_{S, k}(\hat{\vec{r}}) - \langle A_{S, k}(\tilde{\vec{r}}) \rangle ) \rangle (\hat{G}_{A}^{-1})_{k k'} ( A_{S, k'}(\vec{r}) - \langle A_{S, k'}(\hat{\vec{r}}) \rangle ) = - \sum_{k' = 1}^{d} (\hat{K}(t))_{i k'} ( A_{S, k'}(\vec{r}) - \langle A_{S, k'}(\hat{\vec{r}}) \rangle )
\end{align}
from which we recover eq.~\ref{s2_ss5_def:stiffness_matrix}. Fourth, we compute
\begin{align}
    \label{si_6_eq:initial_velocity}
    \ddot{A}_{S, i}(\vec{w}) = \sum_{j, l = 1}^{3N} \frac{p_{j} p_{l}}{m_{j}m_{l}} \frac{\partial^{2} A_{S}(\vec{r})}{\partial r_{j} \partial r_{l}} - \sum_{k = 1}^{3N} \frac{1}{m_{k}} \left\lbrack \frac{\partial V(\vec{r})}{\partial r_{k}} - \frac{\partial H_{1}(t, \vec{r})}{\partial r_{k}} \right\rbrack \frac{\partial A_{S}(\vec{r})}{\partial r_{k}}
\end{align}
and
\begin{align}
    \label{si_6_eq:initial_acceleration}
    \dot{A}_{S, k}(\vec{w}) = \sum_{n = 1}^{3N} \frac{p_{n}}{m_{n}} \frac{\partial A_{S, k}(\vec{r})}{\partial r_{n}},
\end{align}
and use eq.~\ref{si_5:prerequisite_4} where we substituted $\dot{A}_{S, k}(\vec{w})$ for $O_{S,1}(t', \vec{w})$ and $\ddot{A}_{S, i}(t, \vec{w})$ for $O_{S,2}(t", \vec{w})$ to obtain
\begin{align}
    \label{si_6_eq:friction_force_1}
    \langle \ddot{A}_{S, i}(t, \hat{\vec{w}}) \dot{A}_{S, k}(\hat{\vec{w}}) \rangle = & \left\langle \sum_{j, l, n = 1}^{3N} \frac{\hat{p}_{j} \hat{p}_{l} \hat{p}_{n}}{m_{j}m_{l} m_{n}} \frac{\partial^{2} A_{S}(\hat{\vec{r}})}{\partial \hat{r}_{j} \partial \hat{r}_{l}} \frac{\partial A_{S, k}(\hat{\vec{r}})}{\partial \hat{r}_{n}} \right\rangle \\
    & - \left\langle \sum_{k, n = 1}^{3N} \frac{\hat{p}_{n}}{m_{k} m_{n}} \left\lbrack \frac{\partial V(\hat{\vec{r}})}{\partial \hat{r}_{k}} - \frac{\partial H_{1}(t, \hat{\vec{r}})}{\partial \hat{r}_{k}} \right\rbrack \frac{\partial A_{S}(\hat{\vec{r}})}{\partial \hat{r}_{k}} \frac{\partial A_{S}(\hat{\vec{r}})}{\partial \hat{r}_{n}} \right\rangle
\end{align}
where
\begin{align}
    \label{si_6_eq:gaussian_integral_third_moment}
    \left\langle \sum_{j, l, n = 1}^{3N} \frac{\hat{p}_{j} \hat{p}_{l} \hat{p}_{n}}{m_{j}m_{l} m_{n}} \frac{\partial^{2} A_{S}(\hat{\vec{r}})}{\partial \hat{r}_{j} \partial \hat{r}_{l}} \frac{\partial A_{S, k}(\hat{\vec{r}})}{\partial \hat{r}_{n}} \right\rangle = & \frac{1}{Z(\beta)} \sum_{j, l, n} \frac{1}{m_{j} m_{l} m_{n}} \prod_{k = 1}^{3N} \int_{- \infty}^{+ \infty} d\hat{r}_{k} \: \frac{\partial^{2} A_{S}(\hat{\vec{r}})}{\partial \hat{r}_{j} \partial \hat{r}_{l}} \frac{\partial A_{S, k}(\hat{\vec{r}})}{\partial \hat{r}_{n}} \exp \left( - \beta V(\hat{\vec{r}}) \right) \\
    & \prod_{k = 1}^{3N} \int_{- \infty}^{+ \infty} d\hat{p}_{k} \: \hat{p}_{j} \hat{p}_{l} \hat{p}_{n} \prod_{k = 1}^{3N} \exp \left( - \frac{\beta p_{k}^{2}}{2 m_{k}} \right) \nonumber
\end{align}
and
\begin{align}
    \label{si_6_eq:gaussian_integral_first_moment}
    & \left\langle \sum_{k, n = 1}^{3N} \frac{\hat{p}_{n}}{m_{k} m_{n}} \left\lbrack \frac{\partial V(\hat{\vec{r}})}{\partial \hat{r}_{k}} - \frac{\partial H_{1}(t, \hat{\vec{r}})}{\partial \hat{r}_{k}} \right\rbrack \frac{\partial A_{S}(\hat{\vec{r}})}{\partial \hat{r}_{k}} \frac{\partial A_{S}(\hat{\vec{r}})}{\partial \hat{r}_{n}} \right\rangle = \\
    & \frac{1}{Z(\beta)} \sum_{k, n} \frac{1}{m_{k} m_{n}} \prod_{j = 1}^{3N} \int_{- \infty}^{+ \infty} d\hat{p}_{j} \: \hat{p}_{n} \prod_{j = 1}^{3N} \exp \left( - \frac{\beta p_{j}^{2}}{2 m_{j}} \right) \prod_{j = 1}^{3N} \int_{- \infty}^{+ \infty} d\hat{r}_{j} \: \left\lbrack \frac{\partial V(\hat{\vec{r}})}{\partial \hat{r}_{k}} - \frac{\partial H_{1}(t, \hat{\vec{r}})}{\partial \hat{r}_{k}} \right\rbrack \frac{\partial A_{S}(\hat{\vec{r}})}{\partial \hat{r}_{k}} \frac{\partial A_{S}(\hat{\vec{r}})}{\partial \hat{r}_{n}} \exp \left( - \beta V(\hat{\vec{r}}) \right) \nonumber
\end{align}
vanish since the integrations over the components of $\vec{p}$ are  the odd  moments of a Gaussian integral. Consequently, the right-hand-side of eq.~\ref{si_6_eq:effective_force_1} reads
\begin{align}
    \label{si_6_eq:effective_force_2}
    & F_{\text{eff}, i}(t_{0}, t, \vec{w}) = D_{i}(t) - \sum_{k = 1}^{d} (\hat{K}(t))_{i k} \cdot ( A_{k}(t_{0}, t, \vec{w}) - \langle A_{S, k}(\hat{\vec{r}}) \rangle ).
\end{align}

\section{\label{si_7:non_markovian_force}Derivation of the properties of $F_{M, i}(t_{0}, t, \vec{w})$ and computation of $F_{\text{nM,i}}(t_{0}, t, \vec{w})$}

We compute the properties of $F_{M, i}(t_{0}, t, \vec{w})$ in eq.~\ref{s2_ss5_def:properties_orthogonal_force} and derive the expression for $F_{\text{nM,i}}(t_{0}, t, \vec{w})$ in eq.~\ref{s2_ss3_def:non_markovian_force}. First, we substitute $\mathcal{P}_{M}$ for $\mathcal{P}$ in eq.~\ref{s2_ss3_def:orthogonal_force} and obtain
\begin{align}
    \label{si_7_eq:orthogonal_force}
    & F_{M, i}(t_{0}, t, \vec{w}) = \exp_{H}\left( \mathcal{Q}_{M} \int_{t_{0}}^{t} du \: \mathcal{L}(u) \right) \mathcal{Q}_{M} \ddot{A}_{S, i}(t, \vec{w}).
\end{align}
We recall from sec.~\ref{si_4:mori_projection} that $\mathcal{P}_{M}$ defines an orthogonal projection, and that $1$, each component of $\vec{A}_{S}(\vec{r})$, and each component of $\dot{\vec{A}}_{S}(\vec{w})$ are invariants of $\mathcal{P}_{M}$. Therefore, we recover eq.~\ref{s2_ss5_def:properties_orthogonal_force}. Second, we substitute $\mathcal{P}_{M}$ for $\mathcal{P}$ in eq.~\ref{s2_ss3_def:non_markovian_force} and obtain
\begin{align}
    \label{si_7_eq:non_markovian_force}
    & F_{\text{nM,i}}(t_{0}, t, \vec{w}) = \int_{t_{0}}^{t} ds \: \exp_{H}\left( \int_{t_{0}}^{s} du \: \mathcal{L}(u) \right) \mathcal{P}_{M} \mathcal{L}(s) F_{M, i}(s, t, \vec{w})
\end{align}
where
\begin{align}
    \label{si_7_eq:initial_non_markovian_force}
    \mathcal{P}_{M} \mathcal{L}(s) F_{M, i}(s, t, \vec{w}) = & \langle \mathcal{L}(s) F_{M, i}(s, t, \hat{\vec{w}}) \rangle + \sum_{k, k' = 1}^{d} \langle \lbrack \mathcal{L}(s) F_{M, i}(s, t, \hat{\vec{w}}) \rbrack \dot{A}_{S, k}(\hat{\vec{w}}) \rangle (\hat{G}_{\dot{A}}^{-1})_{k k'} \dot{A}_{S, k'}(\vec{w}) \\
    & + \sum_{k, k' = 1}^{d} \langle \lbrack \mathcal{L}(s) F_{M, i}(s, t, \hat{\vec{w}}) \rbrack (A_{S, k}(\hat{\vec{r}}) - \langle A_{S, k}(\tilde{\vec{r}}) \rangle ) \rangle (\hat{G}_{A}^{-1})_{k k'} ( A_{S, k'}(\vec{r}) - \langle A_{S, k'}(\hat{\vec{r}}) \rangle ). \nonumber
\end{align}
Third, we use eq.~\ref{si_5:prerequisite_4} where we substituted $1$ and $F_{M, i}(s, t, \vec{w})$ for $O_{S, 1}(\vec{w})$ and $O_{S, 2}(\vec{w})$ and recover eq.~\ref{s2_ss5_def:constant_kernel}. Fourth, we use eq.~\ref{si_5:prerequisite_4} where we substituted $A_{S, k}(\vec{r}) - \langle A_{S, k}(\hat{\vec{r}}) \rangle$ and $F_{M, i}(s, t, \vec{w})$ for $O_{S, 1}(\vec{w})$ and $O_{S, 2}(\vec{w})$ and obtain
\begin{align}
    \label{si_7_eq:positional_memory_kernel}
    & \langle \lbrack \mathcal{L}(s) F_{M, i}(s, t, \hat{\vec{w}}) \rbrack (A_{S, k}(\hat{\vec{r}}) - \langle A_{S, k}(\tilde{\vec{r}}) \rangle ) \rangle = \beta \langle \lbrack \mathcal{L}_{0} H_{1}(s, \hat{\vec{r}}) \rbrack F_{M, i}(s, t, \hat{\vec{w}}) (A_{S, k}(\hat{\vec{r}}) - \langle A_{S, k}(\tilde{\vec{r}}) \rangle ) \rangle
\end{align}
from which we recover eq.~\ref{s2_ss5_def:positional_matrix_kernel}. Fifth, we use eq.~\ref{si_5:prerequisite_4} where we substituted $\dot{A}_{S, k}(\vec{w})$ and $F_{M, i}(s, t, \vec{w})$ for $O_{S, 1}(\vec{w})$ and $O_{S, 2}(\vec{w})$ and obtain
\begin{align}
    \label{si_7_eq:friction_memory_kernel}
    & \langle \lbrack \mathcal{L}(s) F_{M, i}(s, t, \hat{\vec{w}}) \rbrack \dot{A}_{S, k}(\hat{\vec{w}}) \rangle = - \langle F_{M, i}(s, t, \hat{\vec{w}}) F_{M, k}(s, s, \hat{\vec{w}}) \rangle + \beta \langle \lbrack \mathcal{L}_{0} H_{1}(s, \hat{\vec{r}}) \rbrack F_{M, i}(s, t, \hat{\vec{w}}) \dot{A}_{S, k}(\hat{\vec{w}}) \rangle.
\end{align}
from which we recover eq.~\ref{s2_ss5_def:friction_matrix_kernel}. Finally, eq.~\ref{si_7_eq:non_markovian_force} reads
\begin{align}
    \label{si_7_eq:non_markovian_force_2}
    F_{\text{nM,i}}(t_{0}, t, \vec{w}) = & \int_{t_{0}}^{t} ds \: (\vec{\Gamma}_{1}(s, t))_{i} + \int_{t_{0}}^{t} ds \: \sum_{k' = 1}^{d} (\hat{\Gamma}_{A}(s, t))_{i k'} ( A_{k'}(t_{0}, t, \vec{w}) - \langle A_{S, k'}(\hat{\vec{r}}) \rangle ) \\
    & - \int_{t_{0}}^{t} ds \: \sum_{k' = 1}^{d} ( \hat{\Gamma}_{\dot{A}}(s, t) )_{i k'} \dot{A}_{k'}(t_{0}, t, \vec{w}). \nonumber
\end{align}

\section{\label{si_8:uncorrelated}Properties of the multi-dimensional non-equilibrium Mori GLE in the uncorrelated limit}

We derive the parameters of the MD-NEQ Mori GLE in the uncorrelated limit. First, we relate the two Gram matrices from eqs.~\ref{si_3_eq:gram_matrix_1} and~\ref{si_3_eq:gram_matrix_2} to the covariance matrix in eq.~\ref{s3_ss1_def:correlation_matrix}. We obtain
\begin{align}
    \label{si_8_eq:gram_matrices}
    \left\{
    \begin{array}{c}
        ( \hat{G}_{A} )_{i k} = \left( \hat{C}(t, t, t) \right)_{i k} \\
        ( \hat{G}_{\dot{A}} )_{i k} = \left( \left. \frac{\partial^{2} \hat{C}(t_{0}, t, t')}{\partial t \partial t'} \right\vert_{t_{0} = t' = t} \right)_{i k}
    \end{array}
    \right..
\end{align}
Hence, in the uncorrelated limit, they become
\begin{align}
    \label{si_8_eq:gram_matrices_2}
    \left\{
    \begin{array}{c}
        ( \hat{G}_{A} )_{i k} = \delta_{i k} \langle ( A_{S, k}(\hat{\vec{r}}) - \langle A_{S, k}(\tilde{\vec{r}}) \rangle )^{2} \rangle \\
        ( \hat{G}_{\dot{A}} )_{i k} = \delta_{i k} \langle ( \dot{A}_{S, k}(\hat{\vec{w}}) )^{2} \rangle
    \end{array}
    \right..
\end{align}
Second, we use eqs.~\ref{si_8_eq:gram_matrices},~\ref{si_8_eq:gram_matrices_2}, ~\ref{s2_ss5_def:stiffness_matrix},~\ref{s2_ss5_def:positional_matrix_kernel}, and eq.~\ref{s2_ss5_def:friction_matrix_kernel} to recover eqs.~\ref{s3_ss2_eq:stiffness_matrix},~\ref{s3_ss2_eq:positional_kernel_matrix}, and~\ref{s3_ss2_eq:friction_kernel_matrix}.

\section{\label{si_9:equilibrium_observables}Stationary Hamiltonian: consequences for the observables}

We compute the respective expressions of $\vec{A}(t_{0}, t, \vec{w})$ and $\vec{F}_{M}(t_{0}, t, \vec{w})$ when $H_{1}(t, \vec{r})$ is either time-independent or vanishes. First, we consider the right-to-left time-ordered exponential-type operator
\begin{align}
    \label{si_9_def:exponential_operator}
    & \exp_{H}\left( \int_{t_{0}}^{t} du \: \mathcal{A}(u) \right) \equiv \mathcal{I} + \sum_{n \geq 1} \left( \prod_{k = 1}^{n} \int_{t_{0}}^{\delta_{k,1}t + (1-\delta_{k,1})t_{k-1}} dt_{k} \prod_{j = n}^{1} \mathcal{A}(t_{j}) \right)
\end{align}
where $\mathcal{A}(t)$ is an arbitrary time-dependent operator that  becomes time-independent when $H(t, \vec{w})$ in eq.~\ref{s2_ss1_def:hamiltonian} becomes time-independent as well, which implies that $\mathcal{A}(t) \Rightarrow \mathcal{A}$. In this limit, $\mathcal{A}$  commutes with itself, and we obtain
\begin{align}
    \label{si_9_eq:product}
    & \prod_{j = n}^{1} \mathcal{A}(t_{j}) \Rightarrow \mathcal{A}^{n}
\end{align}
with
\begin{align}
    \label{si_9_eq:nested_integral}
    & \int_{t_{0}}^{\delta_{k,1}t + (1-\delta_{k,1})t_{k-1}} dt_{k} \Rightarrow \frac{(t - t_{0})^{n}}{n!}
\end{align}
where $0 \leq t_{i}$ and $0 \leq \sum_{i = 1}^{n} t_{i} \leq t$. We use the convention $\mathcal{A}^{0} \equiv \mathcal{I}$ and obtain the expression of the right-hand-side of eq.~\ref{si_9_def:exponential_operator} in the equilibrium limit
\begin{align}
    \label{si_9_eq:exponential_operator}
    & \exp_{H}\left( \int_{t_{0}}^{t} du \: \mathcal{A}(u) \right) \Rightarrow \exp \left( (t - t_{0}) \mathcal{A} \right) \equiv \sum_{n \geq 0} \frac{(t - t_{0} )^{n}}{n!} \mathcal{A}^{n},
\end{align}
which  is the standard  operator exponential of $\mathcal{A}$.

\section{\label{si_10:gle_extraction_parameters}Extracting the parameters of the MD-NEQ Mori GLE}

In this section, we explain how to extract the parameters of the MD-NEQ Mori GLE for the case
 when the system evolves under a stationary perturbation $H_{1}(\vec{r})$. First, we use the results of  sec.~\ref{si_9:equilibrium_observables} in conjunction 
with eqs.~\ref{s2_ss5_def:neq_force}, ~\ref{s2_ss5_def:stiffness_matrix}, ~\ref{s2_ss5_def:constant_kernel}, ~\ref{s2_ss5_def:positional_matrix_kernel}, and~\ref{s2_ss5_def:friction_matrix_kernel} to derive from eq.~\ref{s2_ss5_eq:mori_gle} the modified expression 
\begin{align}
    \label{si_10_eq:mori_gle}
    \ddot{A}_{i}(t - t_{0}, \vec{w}) = & D_{i} - \sum_{j = 1}^{d} (\hat{K})_{i j} ( A_{j}(t - t_{0}, \vec{w}) - \langle A_{S, j}(\hat{\vec{r}}) \rangle ) + \int_{t_{0}}^{t} ds \: (\vec{\Gamma}_{1}(t - s))_{i} - \int_{t_{0}}^{t} ds \: \sum_{j = 1}^{d} (\hat{\Gamma}_{\dot{A}}(t - s))_{i j} \dot{A}_{j}(s - t_{0}, \vec{w}) \nonumber \\
    & + \int_{t_{0}}^{t} ds \: \sum_{j = 1}^{d} (\hat{\Gamma}_{A}(t - s))_{i j} ( A_{j}(s - t_{0}, \vec{w}) - \langle A_{S, j}(\hat{\vec{r}}) \rangle ) + F_{M, i}(t - t_{0}, \vec{w}).
\end{align}
To  extract the parameters of the effective force, we consider the time $t = t_{0}$, for  which eq.~\ref{si_10_eq:mori_gle} reads
\begin{align}
    \label{si_10_eq:mori_gle_initial_time}
    \ddot{A}_{S, i}(\vec{w}) = & D_{i} - \sum_{j = 1}^{d} (\hat{K})_{i j} ( A_{S, j}(\vec{r}) - \langle A_{S, j}(\hat{\vec{r}}) \rangle ) + F_{M, i}(0, \vec{w}),
\end{align}
and use eqs.~\ref{s2_ss5_def:properties_orthogonal_force} and~\ref{si_4_eq:gram_matrix_A} to obtain
\begin{subequations}
    \label{si_10_eq:voltera_equation_effective_force}
    \begin{align}
        \label{si_10_eq:voltera_equation_neq_force}
        & D_{i} = \langle \ddot{A}_{S, i}(\hat{\vec{w}}) \rangle \\
        \label{si_10_eq:voltera_equation_harmonic_force}
        & (\hat{K})_{i j} = - \sum_{k = 1}^{3N} \langle \ddot{A}_{S, i}(\hat{\vec{w}}) ( A_{S, j}(\hat{\vec{r}}) - \langle A_{S, j}(\tilde{\vec{r}}) \rangle ) \rangle (\hat{G}_{A}^{-1})_{k j},
    \end{align}
\end{subequations}
which correspond exactly to eqs.~\ref{si_6_eq:initial_effective_force_1} and~\ref{si_6_eq:harmonic_force_2} in sec.~\ref{si_6:relevant_force}. 
To extract the parameters of the non-Markovian force, we use sec.~\ref{si_9:equilibrium_observables} to modify  the expression of eq.~\ref{s3_ss1_def:correlation_matrix} as 
\begin{align}
    \label{si_10_eq:correlation_matrix}
    & ( \hat{C}(t - t_{0}, t' - t_{0}) )_{i j} = \langle ( A_{i}(t  - t_{0}, \hat{\vec{w}}) - \langle A_{i}(t - t_{0}, \vec{\tilde{w}}) \rangle ) ( A_{j}(t' - t_{0}, \hat{\vec{w}}) - \langle A_{j}(t' - t_{0}, \vec{\tilde{w}}) \rangle ) \rangle,
\end{align}
which together with eqs.~\ref{s2_ss5_def:properties_orthogonal_force} and~\ref{si_10_eq:voltera_equation_effective_force} leads to 
\begin{subequations}
    \label{si_10_eq:voltera_equation_non_markovian_force}
    \begin{align}
        \label{si_10_eq:voltera_equation_1}
        \langle \ddot{A}_{i}(t - t_{0}, \hat{\vec{w}}) \rangle = & D_{i} - \sum_{j = 1}^{d} (\hat{K})_{i j} ( \langle A_{j}(t - t_{0}, \hat{\vec{w}}) \rangle - \langle A_{S, j}(\hat{\vec{r}}) \rangle ) + \int_{t_{0}}^{t} ds \: (\vec{\Gamma}_{1}(t - s))_{i} \\
        & + \int_{t_{0}}^{t} ds \: \sum_{j = 1}^{d} (\hat{\Gamma}_{A}(t - s))_{i j} ( \langle A_{j}(s - t_{0}, \hat{\vec{w}}) \rangle - \langle A_{S, j}(\hat{\vec{r}}) \rangle ) - \int_{t_{0}}^{t} ds \: \sum_{j = 1}^{d} (\hat{\Gamma}_{\dot{A}}(t - s))_{i j} \langle \dot{A}_{j}(s - t_{0}, \hat{\vec{w}}) \rangle, \nonumber
    \end{align}
    \begin{align}
        \label{si_10_eq:voltera_equation_A}
        \langle \ddot{A}_{i}(t - t_{0}, \hat{\vec{w}}) ( A_{S, k}(\hat{\vec{r}}) - \langle A_{S, k}(\tilde{\vec{r}}) \rangle ) \rangle = & - \sum_{j = 1}^{d} (\hat{K})_{i j} (\hat{C}(t - t_{0}, 0))_{j k} + \int_{t_{0}}^{t} ds \: \sum_{j = 1}^{d} (\hat{\Gamma}_{A}(t - s))_{i j} (\hat{C}(s - t_{0}, 0))_{j k} \nonumber \\
        & - \int_{t_{0}}^{t} ds \: \sum_{j = 1}^{d} (\hat{\Gamma}_{\dot{A}}(t - s))_{i j} \left( \frac{\partial \hat{C}(s - t_{0}, 0)}{\partial s} \right)_{j k},
    \end{align}
    \begin{align}
        \label{si_10_eq:voltera_equation_dotA}
        \langle \ddot{A}_{i}(t - t_{0}, \hat{\vec{w}}) \dot{A}_{S, k}(\hat{\vec{w}}) \rangle = & - \sum_{j = 1}^{d} (\hat{K})_{i j} \left( \frac{\partial \hat{C}(t - t_{0}, 0)}{\partial t'} \right)_{j k} + \int_{t_{0}}^{t} ds \: \sum_{j = 1}^{d} (\hat{\Gamma}_{A}(t - s))_{i j} \left( \frac{\partial \hat{C}(s - t_{0}, 0)}{\partial s'} \right)_{j k} \nonumber \\
        & - \int_{t_{0}}^{t} ds \: \sum_{j = 1}^{d} (\hat{\Gamma}_{\dot{A}}(t - s))_{i j} \left( \frac{\partial^{2} \hat{C}(s - t_{0}, 0)}{\partial s \partial s'} \right)_{j k}.
    \end{align}
\end{subequations}
To proceed, we integrate eq.~\ref{si_10_eq:voltera_equation_non_markovian_force} over $t$ from $t_{0}$ to $\tau \geq t_{0}$. Using eqs.\ref{si_4_def:simplification} and~\ref{si_4_def:cross_correlation} with changes of variables inside the integral terms, we obtain
\begin{subequations}
    \label{si_10_eq:integrated_voltera_equation_non_markovian_force}
    \begin{align}
        \label{si_10_eq:integrated_voltera_equation_1}
        \langle \dot{A}_{i}(\tau - t_{0}, \hat{\vec{w}}) \rangle = & D_{i}(\tau - t_{0}) - \sum_{j = 1}^{d} (\hat{K})_{i j} \int_{0}^{\tau - t_{0}} ds \: ( \langle A_{j}(s, \hat{\vec{w}}) \rangle - \langle A_{S, j}(\hat{\vec{r}}) \rangle ) + \int_{0}^{\tau - t_{0}} ds \: (\vec{J}_{1}(s))_{i} \\
        & + \int_{0}^{\tau - t_{0}} ds \: \sum_{j = 1}^{d} (\hat{J}_{A}(s))_{i j} ( \langle A_{j}(\tau - t_{0} - s, \hat{\vec{w}}) \rangle - \langle A_{S, j}(\hat{\vec{r}}) \rangle ) \nonumber \\
        & - \int_{0}^{\tau - t_{0}} ds \: \sum_{j = 1}^{d} (\hat{J}_{\dot{A}}(s))_{i j} \langle \dot{A}_{j}(\tau - t_{0} - s, \hat{\vec{w}}) \rangle, \nonumber
    \end{align}
    \begin{align}
        \label{si_10_eq:integrated_voltera_equation_A}
        \langle \dot{A}_{i}(\tau - t_{0}, \hat{\vec{w}}) ( A_{S, k}(\hat{\vec{r}}) - \langle A_{S, k}(\tilde{\vec{r}}) \rangle ) \rangle = & - \sum_{j = 1}^{d} (\hat{K})_{i j} \int_{0}^{\tau - t_{0}} (\hat{C}(s, 0))_{j k} + \int_{0}^{\tau - t_{0}} ds \: \sum_{j = 1}^{d} (\hat{J}_{A}(s))_{i j} (\hat{C}(\tau - t_{0} - s, 0))_{j k} \nonumber \\
        & - \int_{0}^{\tau - t_{0}} ds \: \sum_{j = 1}^{d} (\hat{J}_{\dot{A}}(s))_{i j} \left( \frac{\partial \hat{C}(\tau - t_{0} - s, 0)}{\partial s} \right)_{j k},
    \end{align}
    \begin{align}
        \label{si_10_eq:integrated_voltera_equation_dotA}
        \langle \dot{A}_{i}(t - t_{0}, \hat{\vec{w}}) \dot{A}_{S, k}(\hat{\vec{w}}) \rangle = & \langle \dot{A}_{S, i}(\hat{\vec{w}}) \dot{A}_{S, k}(\hat{\vec{w}}) \rangle - \sum_{j = 1}^{d} (\hat{K})_{i j} \int_{0}^{\tau - t_{0}} ds \: \left( \frac{\partial \hat{C}(s, 0)}{\partial s'} \right)_{j k} \\
        & + \int_{0}^{\tau - t_{0}} ds \: \sum_{j = 1}^{d} (\hat{J}_{A}(s))_{i j} \left( \frac{\partial \hat{C}(\tau - t_{0} - s, 0)}{\partial s'} \right)_{j k} \nonumber \\
        & - \int_{0}^{\tau - t_{0}} ds \: \sum_{j = 1}^{d} (\hat{J}_{\dot{A}}(s))_{i j} \left( \frac{\partial^{2} \hat{C}(\tau - t_{0} - s, 0)}{\partial s \partial s'} \right)_{j k}. \nonumber
    \end{align}
\end{subequations}
where we defined the integrals
\begin{subequations}
    \label{si_10_def:integrated_kernels}
    \begin{align}
        \label{si_10_def:integrated_constant_kernel_vector}
        & \vec{J}_{1}(t) \equiv \int_{0}^{t} ds \: \vec{\Gamma}_{1}(s),
    \end{align}
    \begin{align}
        \label{si_10_eq:integrated_positional_memory_kernel_matrix}
        & \hat{J}_{A}(t) \equiv \int_{0}^{t} ds \: \hat{\Gamma}_{A}(s),
    \end{align}
    \begin{align}
        \label{si_10_eq:integrated_friction_memory_kernel_matrix}
        & \hat{J}_{\dot{A}}(t) \equiv \int_{0}^{t} ds \: \hat{\Gamma}_{\dot{A}}(s)
    \end{align}
\end{subequations}
of $\vec{\Gamma}_{1}(t)$, $\hat{\Gamma}_{A}(t)$, and $\hat{\Gamma}_{\dot{A}}(t)$. Next, we consider an interval of time $[t_{0}, t_{\text{end}}]$ that we slice in $N_{\text{steps}}$ intervals of length $\Delta t \equiv \frac{(t_{\text{end}} - t_{0})}{N_{\text{steps}}}$ such that $t_{m} = t_{0} + m \Delta t$ $\forall m$ 
and  $0 \leq m \leq N_{\text{steps}}$. Using the rectangular rule for discretization of the integral terms and eqs.\ref{si_4_def:simplification} and~\ref{si_4_def:cross_correlation}, eq.~\ref{si_10_eq:integrated_voltera_equation_non_markovian_force} reads
\begin{subequations}
    \label{si_10_eq:discretized_voltera_equation_non_markovian_force}
    \begin{align}
        \label{si_10_eq:discretized_voltera_equation_1}
        \langle \dot{A}_{i}(m \Delta t, \hat{\vec{w}}) \rangle = & D_{i} m \Delta t - \sum_{j = 1}^{d} (\hat{K})_{i j} \sum_{n = 0}^{m - 1} ( \langle A_{j}(n \Delta t, \hat{\vec{w}}) \rangle - \langle A_{S, j}(\hat{\vec{r}}) \rangle ) + \sum_{n = 0}^{m} \: (\vec{J}_{1}(n \Delta t))_{i} \\
        & + \sum_{n = 0}^{m - 1} \sum_{j = 1}^{d} (\hat{J}_{A}(n \Delta t))_{i j} ( \langle A_{j}( (m - n) \Delta t, \hat{\vec{w}}) \rangle - \langle A_{S, j}(\hat{\vec{r}}) \rangle ) - \sum_{n = 0}^{m - 1} \sum_{j = 1}^{d} (\hat{J}_{\dot{A}}(n \Delta t))_{i j} \langle \dot{A}_{j}( (m - n) \Delta t, \hat{\vec{w}}) \rangle, \nonumber
    \end{align}
    \begin{align}
        \label{si_10_eq:discretized_voltera_equation_A}
        \langle \dot{A}_{i}(m \Delta t, \hat{\vec{w}}) ( A_{S, k}(\hat{\vec{r}}) - \langle A_{S, k}(\tilde{\vec{r}}) \rangle ) \rangle = & - \sum_{j = 1}^{d} (\hat{K})_{i j} \sum_{n = 0}^{m} (\hat{C}(n \Delta t, 0))_{j k} + \sum_{n = 0}^{m} \sum_{j = 1}^{d} (\hat{J}_{A}(n \Delta t))_{i j} (\hat{C}( (m - n) \Delta t, 0))_{j k} \nonumber \\
        & - \sum_{n = 0}^{m - 1} \sum_{j = 1}^{d} (\hat{J}_{\dot{A}}(n \Delta t))_{i j} \left( \frac{\partial \hat{C}( (m - n) \Delta t, 0)}{\partial s} \right)_{j k},
    \end{align}
    \begin{align}
        \label{si_10_eq:discretized_voltera_equation_dotA}
        \langle \dot{A}_{i}(m \Delta t, \hat{\vec{w}}) \dot{A}_{S, k}(\hat{\vec{w}}) \rangle = & \langle \dot{A}_{S, i}(\hat{\vec{w}}) \dot{A}_{S, k}(\hat{\vec{w}}) \rangle - \sum_{j = 1}^{d} (\hat{K})_{i j} \sum_{n = 0}^{m - 1} \left( \frac{\partial \hat{C}(n \Delta t, 0)}{\partial s'} \right)_{j k} \\
        & + \sum_{n = 0}^{m - 1} \sum_{j = 1}^{d} (\hat{J}_{A}(n \Delta t))_{i j} \left( \frac{\partial \hat{C}( (m - n) \Delta t, 0)}{\partial s'} \right)_{j k} - \sum_{n = 0}^{m} \sum_{j = 1}^{d} (\hat{J}_{\dot{A}}(n \Delta t))_{i j} \left( \frac{\partial^{2} \hat{C}( (m - n) \Delta t, 0)}{\partial s \partial s'} \right)_{j k}, \nonumber
    \end{align}
\end{subequations}
from which we extract step by step $\vec{J}_{1}(m \Delta t)$, $\hat{J}_{A}(m \Delta t)$, and $\hat{J}_{\dot{A}}(m \Delta t)$ $\forall m \in [0, N_{\text{steps}}]$. Finally, we compute a numerical derivative of these three quantities to obtain $\Vec{\Gamma}_{1}(t)$, $\hat{\Gamma}_{A}(t)$, and $\hat{\Gamma}_{\dot{A}}(t)$.

\section{\label{si_11:simulation_details}Molecular dynamics simulation details}

The molecular dynamics (MD) simulations are initialized with the published Cryo$-$EM structure of human IAPP~\cite{roder_cryo-em_2020}, from  which we construct a stack of six layers of two  IAPP chains that are periodically repeated along the $z-$axis within an orthogonal simulation box of dimensions $11.44 \: nm \times 11.44 \: nm \times 2.85 \: nm$. Because of the flexibility of residues $1–12$ (LYS1, CYS2, ASN3, THR4, ALA5, THR6, CYS7, ALA8, THR9, GLN10, ARG11, LEU12), these are unresolved in the experimental structure~\cite{roder_cryo-em_2020} and are therefore appended in an extended conformation.\newline

Second, the system is subsequently energy-minimized, maintaining residues $13-38$ fixed. In order to introduce the disulfide bridge between CYS2 and CYS7, the sulfur atoms are repositioned to facilitate automatic bond detection by GROMACS.~\cite{abraham_gromacs_2015} This is achieved by applying a  harmonic interaction potential via PLUMED~\cite{bussi_analyzing_2019}, incrementally increasing the interaction strength until all S-S distances are less than approximately $0.19 \: nm$, while keeping the fibril core fixed.  Solvent molecules are added using gmx solvate, followed by charge neutralization by the addition of $36$ chloride ions using gmx genion. Water molecules erroneously placed inside the amyloid fibril core by gmx solvate are removed via a custom script, eliminating a total of $90$ water molecules.\newline

Third, a gradual heating protocol is applied, increasing the temperature from $5 \: K$ to $310 \: K$ over $100 \: ps$. 
Based on this initial configuration, an unbiased $10 \: \mu s$ molecular dynamics (MD) simulation is conducted starting with
an equilibration phase of $100 \: ns$ at $310 \: K$, with a semi-isotropic pressure of $1 \: bar$.\newline

All simulations, except for energy minimizations, are performed using a modified Berendsen thermostat~\cite{bussi_canonical_2007} at $310 \: K$ and a semi-isotropic barostat~\cite{bernetti_pressure_2020} with a pressure of $1$ bar in all directions, where the $x/y$ positions (which span the fibrillar plane) and the $z$ positions (along the fibrillar axis) are independently rescaled. The Amberff99sb-ILDN force field~\cite{best_optimized_2009, lindorff-larsen_improved_2010, piana_how_2011} is employed, as it demonstrated good agreement with experimental data for IAPP in a comprehensive force-field survey~\cite{hoffmann_secondary_2015}. A time step of 2  fs is used for all simulations, which were carried out using GROMACS $2022.5$

\section{\label{si_12:two-dimensional}Application to a two-dimensional observable of interest}

\begin{figure}
    \includegraphics[width=0.3\textwidth]{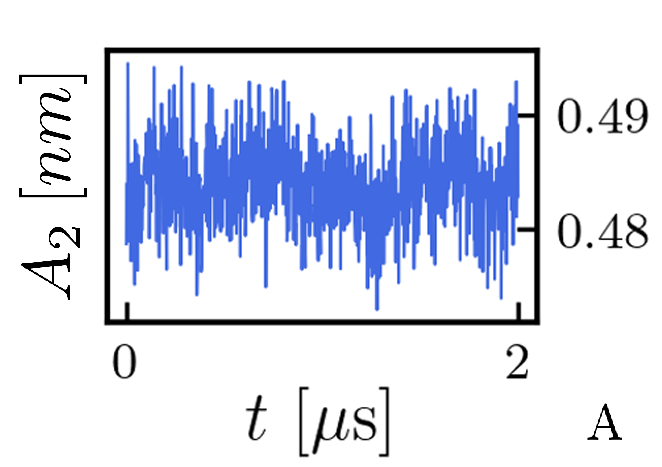}
    \includegraphics[width=0.3\textwidth]{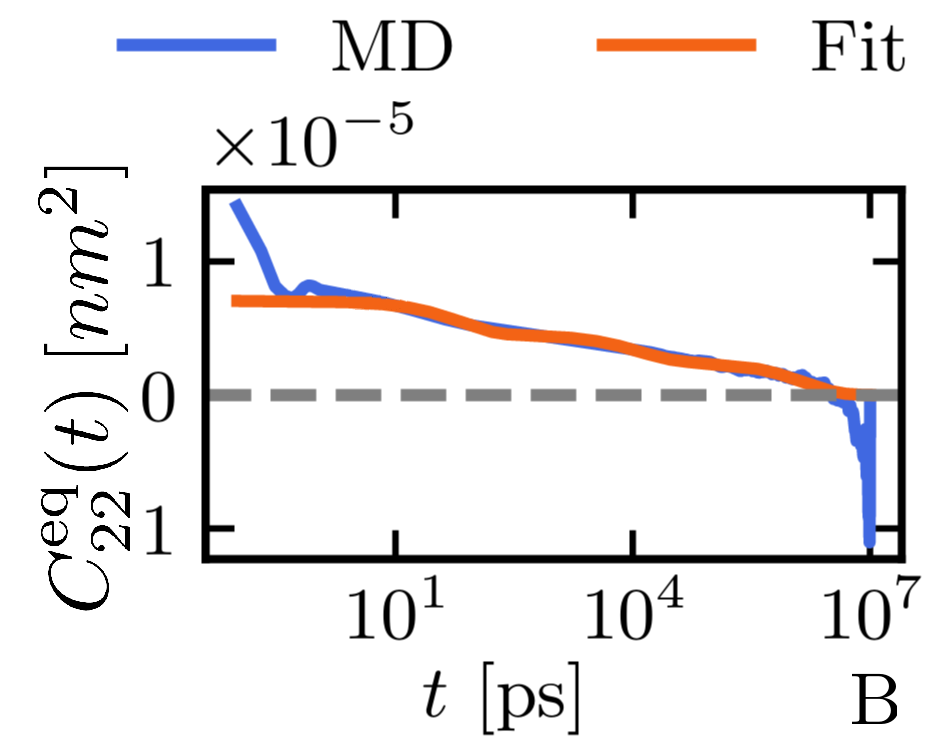}
    \includegraphics[width=0.3\textwidth]{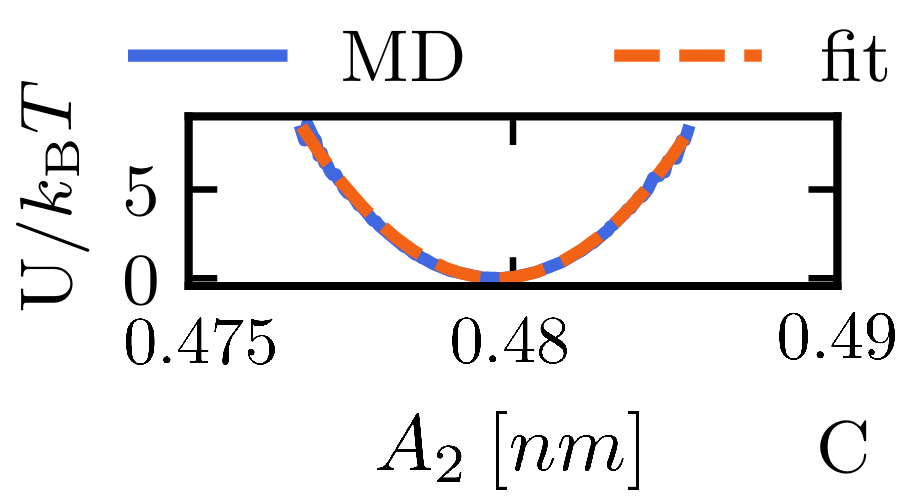}
    \caption{Results from a 10 microsecond long   molecular dynamics equilibrium simulation of a periodically replicated fibril formed
    from 12 human islet amyloid polypeptides (IAPPs) in explicit water. \textbf{A} Trajectory  for the inter-layer distance $A_{2}$ corresponding to  a $2-\mu s$ section of the full $10-\mu s$ unconstrained molecular dynamics simulation. (\textbf{B}) Inter-layer distance correlation function 
    $C^{eq}_{22}(t) \equiv \langle (A_{2}(t) - \langle A_{2}\rangle) (A_{2}(0) - \langle A_{2} \rangle) \rangle$ (blue line) 
    and corresponding fit using 4 exponentials (orange line). 
     \textbf{C}  Free energy $U(A_{2})$ from  MD simulations (blue line) together with a harmonic fit (orange  broken line).}
    \label{fig:supplementary}
\end{figure}

In this section, we specialize  the results of secs.~\ref{s2:derivation} and~\ref{s3:limits} to the two-dimensional  ($d = 2$) case,
which is relevant for  the fibril dynamics as discussed in the main text.
First, we consider the two-dimensional observable $\vec{A}(t_{0}, t, \vec{w}) \equiv (A_{1}(t_{0}, t, \vec{w}), A_{2}(t_{0}, t, \vec{w}))^{T}$ and write down the expressions of the inverse of the two Gram matrices defined in eqs.~\ref{si_4_eq:gram_matrix_A} and~\ref{si_4_def:gram_matrix_dotA}, i.e.
\begin{align}
    \label{si_12_eq:inverse_gram_matrix_A}
    \hat{G}_{A}^{-1} = & \frac{1}{\langle ( A_{S, 1}(\hat{\vec{r}}) - \langle A_{S, 1}(\Tilde{\vec{r}}) \rangle )^{2} \rangle \langle ( A_{S, 2}(\hat{\vec{r}}) - \langle A_{S, 2}(\Tilde{\vec{r}}) \rangle )^{2} \rangle - \langle ( A_{S, 1}(\hat{\vec{r}}) - \langle A_{S, 1}(\Tilde{\vec{r}}) \rangle )( A_{S, 2}(\hat{\vec{r}}) - \langle A_{S, 2}(\Tilde{\vec{r}}) \rangle ) \rangle^{2} } \nonumber \\
    & 
    \begin{pmatrix}
        \langle ( A_{S, 2}(\hat{\vec{r}}) - \langle A_{S, 2}(\Tilde{\vec{r}}) \rangle )^{2} \rangle & - \langle ( A_{S, 1}(\hat{\vec{r}}) - \langle A_{S, 1}(\Tilde{\vec{r}}) \rangle )( A_{S, 2}(\hat{\vec{r}}) - \langle A_{S, 2}(\Tilde{\vec{r}}) \rangle ) \rangle \\
        - \langle ( A_{S, 1}(\hat{\vec{r}}) - \langle A_{S, 1}(\Tilde{\vec{r}}) \rangle )( A_{S, 2}(\hat{\vec{r}}) - \langle A_{S, 2}(\Tilde{\vec{r}}) \rangle ) \rangle & \langle ( A_{S, 1}(\hat{\vec{r}}) - \langle A_{S, 1}(\Tilde{\vec{r}}) \rangle )^{2} \rangle
    \end{pmatrix}
\end{align}
and
\begin{align}
    \label{si_12_eq:inverse_gram_matrix_dotA}
    & \hat{G}_{\dot{A}}^{-1} = \frac{1}{\langle ( \dot{A}_{S, 1}(\hat{\vec{w}}) )^{2} \rangle \langle ( \dot{A}_{S, 2}(\hat{\vec{w}}) )^{2} \rangle - \langle \dot{A}_{S, 1}(\hat{\vec{w}}) \dot{A}_{S, 2}(\hat{\vec{w}}) \rangle^{2}}
    \begin{pmatrix}
        \langle ( \dot{A}_{S, 2}(\hat{\vec{w}}) )^{2} \rangle & -\langle \dot{A}_{S, 1}(\hat{\vec{w}}) \dot{A}_{S, 2}(\hat{\vec{w}}) \rangle \\
        - \langle \dot{A}_{S, 1}(\hat{\vec{w}}) \dot{A}_{S, 2}(\hat{\vec{w}}) \rangle & \langle ( \dot{A}_{S, 1}(\hat{\vec{w}}) )^{2} \rangle
    \end{pmatrix} .
\end{align}
Second, we write  $\mathcal{P}_{M} O(t_{0}, t, \vec{w})$ as 
\begin{align}
    \label{si_12_eq:mori_projection_operator}
    \mathcal{P}_{M} O(t_{0}, t, \vec{w}) = & \langle O(t_{0}, t, \hat{\vec{w}}) \rangle + 
    \begin{pmatrix}
        \langle O(t_{0}, t, \hat{\vec{w}}) \dot{A}_{S, 1}(\hat{\vec{w}}) ) \rangle & \langle O(t_{0}, t, \hat{\vec{w}}) \dot{A}_{S, 2}(\hat{\vec{w}}) \rangle
    \end{pmatrix}
    \cdot \hat{G}_{\dot{A}}^{-1} \cdot
    \begin{pmatrix}
        \dot{A}_{S, 1}(\vec{w}) \\
        \dot{A}_{S, 2}(\vec{w})
    \end{pmatrix} \\
    & + 
    \begin{pmatrix}
        \langle O(t_{0}, t, \hat{\vec{w}}) ( A_{S, 1}(\hat{\Vec{r}}) - \langle A_{S, 1}(\Tilde{\Vec{r}}) \rangle ) \rangle & \langle O(t_{0}, t, \hat{\vec{w}}) ( A_{S, 2}(\hat{\Vec{r}}) - \langle A_{S, 2}(\Tilde{\Vec{r}}) \rangle ) \rangle
    \end{pmatrix}
    \cdot \hat{G}_{A}^{-1} \cdot
    \begin{pmatrix}
        A_{S, 1}(\Vec{r}) - \langle A_{S, 1}(\hat{\Vec{r}}) \rangle \\
        A_{S, 2}(\Vec{r}) - \langle A_{S, 2}(\hat{\Vec{r}}) \rangle
    \end{pmatrix}. \nonumber
\end{align}
Third, we write down the expression of the 2D-NEQ Mori GLE in the two-dimensional case that reads
\begin{align}
    \label{si_12_eq:mori_gle}
    \begin{pmatrix}
        \ddot{A}_{1}(t_{0}, t, \vec{w}) \\
        \ddot{A}_{2}(t_{0}, t, \vec{w})
    \end{pmatrix}
    = & \begin{pmatrix}
        D_{1}(t) \\
        D_{2}(t)
    \end{pmatrix}
    - \hat{K}(t) \cdot
    \begin{pmatrix}
        A_{1}(t_{0}, t, \vec{w}) - \langle A_{S, 1}(\hat{\vec{r}}) \rangle \\
        A_{2}(t_{0}, t, \vec{w}) - \langle A_{S, 2}(\hat{\vec{r}}) \rangle
    \end{pmatrix}
    + \int_{t_{0}}^{t} ds
    \begin{pmatrix}
        \Gamma_{1, 1}(s, t) \\
        \Gamma_{1, 2}(s, t)
    \end{pmatrix} \\
    & + \int_{t_{0}}^{t} ds \: \hat{\Gamma}_{A}(s, t) \cdot
    \begin{pmatrix}
        A_{1}(t_{0}, s, \vec{w}) - \langle A_{S, 1}(\hat{\vec{r}}) \rangle \\
        A_{2}(t_{0}, s, \vec{w}) - \langle A_{S, 2}(\hat{\vec{r}}) \rangle
    \end{pmatrix}
     - \int_{t_{0}}^{t} ds \: \hat{\Gamma}_{\dot{A}}(s, t) \cdot
     \begin{pmatrix}
         \dot{A}_{1}(t_{0}, s, \vec{w}) \\
         \dot{A}_{2}(t_{0}, s, \vec{w})
     \end{pmatrix}
     +
     \begin{pmatrix}
        F_{M, 1}(t_{0}, t, \vec{w}) \\
        F_{M, 2}(t_{0}, t, \vec{w})
     \end{pmatrix}, \nonumber
\end{align}
and the expression for the  parameters appearing in the GLE. Namely, the two-dimensional non-equilibrium force
\begin{align}
    \label{si_12_eq:neq_force}
    &
    \begin{pmatrix}
        D_{1}(t) \\
        D_{2}(t)
    \end{pmatrix}
    = \beta 
    \begin{pmatrix}
        \langle \lbrack \mathcal{L}_{0} H_{1}(t, \hat{\vec{r}}) \rbrack \dot{A}_{S, 1}(\hat{\vec{w}}) \rangle \\
        \langle \lbrack \mathcal{L}_{0} H_{1}(t, \hat{\vec{r}}) \rbrack \dot{A}_{S, 2}(\hat{\vec{w}}) \rangle
    \end{pmatrix},
\end{align}
the $2 \times 2$ time-dependent stiffness matrix
\begin{align}
    \label{si_12_eq:stiffness_matrix}
    \hat{K}(t) = & \frac{1}{\langle ( A_{S, 1}(\hat{\vec{r}}) - \langle A_{S, 1}(\Tilde{\vec{r}}) \rangle )^{2} \rangle \langle ( A_{S, 2}(\hat{\vec{r}}) - \langle A_{S, 2}(\Tilde{\vec{r}}) \rangle )^{2} \rangle - \langle ( A_{S, 1}(\hat{\vec{r}}) - \langle A_{S, 1}(\Tilde{\vec{r}}) \rangle )( A_{S, 2}(\hat{\vec{r}}) - \langle A_{S, 2}(\Tilde{\vec{r}}) \rangle ) \rangle^{2} } \\
    & \left\lbrack
    \begin{pmatrix}
        \langle ( \dot{A}_{S, 1}(\hat{\vec{w}}) )^{2} \rangle  & \langle \dot{A}_{S, 1}(\hat{\vec{w}}) \dot{A}_{S, 2}(\hat{\vec{w}}) \rangle \\
        \langle \dot{A}_{S, 1}(\hat{\vec{w}}) \dot{A}_{S, 2}(\hat{\vec{w}}) \rangle & \langle ( \dot{A}_{S, 2}(\hat{\vec{w}}) )^{2} \rangle
    \end{pmatrix}
    \right.
    \nonumber \\
    & \left. - \beta
    \begin{pmatrix}
        \langle \dot{A}_{S, 1}(\hat{\vec{w}}) \lbrack \mathcal{L}_{0} H_{1}(t, \hat{\vec{r}}) \rbrack ( A_{S, 1}(\hat{\vec{r}}) - \langle A_{S, 1}(\Tilde{\vec{r}}) \rangle ) \rangle & \langle \dot{A}_{S, 1}(\hat{\vec{w}}) \lbrack \mathcal{L}_{0} H_{1}(t, \hat{\vec{r}}) \rbrack  ( A_{S, 2}(\hat{\vec{r}}) - \langle A_{S, 2}(\Tilde{\vec{r}}) \rangle ) \rangle \\
        \langle \dot{A}_{S, 2}(\hat{\vec{w}}) \lbrack \mathcal{L}_{0} H_{1}(t, \hat{\vec{r}}) \rbrack  ( A_{S, 1}(\hat{\vec{r}}) - \langle A_{S, 1}(\Tilde{\vec{r}}) \rangle ) \rangle & \langle \dot{A}_{S, 2}(\hat{\vec{w}}) \lbrack \mathcal{L}_{0} H_{1}(t, \hat{\vec{r}}) \rbrack ( A_{S, 2}(\hat{\vec{r}}) - \langle A_{S, 2}(\Tilde{\vec{r}}) \rangle ) \rangle
    \end{pmatrix}
    \right\rbrack
    \nonumber \\
    & \cdot
    \begin{pmatrix}
        \langle ( A_{S, 2}(\hat{\vec{r}}) - \langle A_{S, 2}(\Tilde{\vec{r}}) \rangle )^{2} \rangle & - \langle ( A_{S, 1}(\hat{\vec{r}}) - \langle A_{S, 1}(\Tilde{\vec{r}}) \rangle )( A_{S, 2}(\hat{\vec{r}}) - \langle A_{S, 2}(\Tilde{\vec{r}}) \rangle ) \rangle \\
        - \langle ( A_{S, 1}(\hat{\vec{r}}) - \langle A_{S, 1}(\Tilde{\vec{r}}) \rangle )( A_{S, 2}(\hat{\vec{r}}) - \langle A_{S, 2}(\Tilde{\vec{r}}) \rangle ) \rangle & \langle ( A_{S, 1}(\hat{\vec{r}}) - \langle A_{S, 1}(\Tilde{\vec{r}}) \rangle )^{2} \rangle
    \end{pmatrix}, \nonumber
\end{align}
the constant memory kernel
\begin{align}
    \label{si_12_eq:constant_kernel}
    &
    \begin{pmatrix}
        \Gamma_{1, 1}(t) \\
        \Gamma_{1, 2}(t)
    \end{pmatrix}
    = \beta 
    \begin{pmatrix}
        \langle \lbrack \mathcal{L}_{0} H_{1}(s, \hat{\vec{r}}) \rbrack F_{M, 1}(s, t, \hat{\vec{w}}) \rangle \\
        \langle \lbrack \mathcal{L}_{0} H_{1}(s, \hat{\vec{r}}) \rbrack F_{M, 2}(s, t, \hat{\vec{w}}) \rangle
    \end{pmatrix},
\end{align}
the $2 \times 2$ positional kernel matrix
\begin{align}
    \hat{\Gamma}_{A}(s, t) =
    & \frac{\beta}{\langle ( A_{S, 1}(\hat{\vec{r}}) - \langle A_{S, 1}(\Tilde{\vec{r}}) \rangle )^{2} \rangle \langle ( A_{S, 2}(\hat{\vec{r}}) - \langle A_{S, 2}(\Tilde{\vec{r}}) \rangle )^{2} \rangle - \langle ( A_{S, 1}(\hat{\vec{r}}) - \langle A_{S, 1}(\Tilde{\vec{r}}) \rangle )( A_{S, 2}(\hat{\vec{r}}) - \langle A_{S, 2}(\Tilde{\vec{r}}) \rangle ) \rangle^{2} } \\
    &
    \begin{pmatrix}
        \langle F_{M, 1}(s, t, \hat{\vec{w}}) \lbrack \mathcal{L}_{0} H_{1}(s, \hat{\vec{r}}) \rbrack (A_{S, 1}(\hat{\vec{r}}) - \langle A_{S, 1}(\Tilde{\vec{r}}) \rangle ) \rangle & \langle F_{M, 1}(s, t, \hat{\vec{w}}) \lbrack \mathcal{L}_{0} H_{1}(s, \hat{\vec{r}}) \rbrack (A_{S, 2}(\hat{\vec{r}}) - \langle A_{S, 2}(\Tilde{\vec{r}}) \rangle ) \rangle \\
        \langle F_{M, 2}(s, t, \hat{\vec{w}}) \lbrack \mathcal{L}_{0} H_{1}(s, \hat{\vec{r}}) \rbrack (A_{S, 1}\hat{\vec{r}}) - \langle A_{S, 1}(\Tilde{\vec{r}}) \rangle ) \rangle & \langle F_{M, 2}(s, t, \hat{\vec{w}}) \lbrack \mathcal{L}_{0} H_{1}(s, \hat{\vec{r}}) \rbrack (A_{S, 2}(\hat{\vec{r}}) - \langle A_{S, 2}(\Tilde{\vec{r}}) \rangle ) \rangle
    \end{pmatrix}
    \nonumber \\
    & \cdot
    \begin{pmatrix}
        \langle ( A_{S, 2}(\hat{\vec{r}}) - \langle A_{S, 2}(\Tilde{\vec{r}}) \rangle )^{2} \rangle & - \langle ( A_{S, 1}(\hat{\vec{r}}) - \langle A_{S, 1}(\Tilde{\vec{r}}) \rangle )( A_{S, 2}(\hat{\vec{r}}) - \langle A_{S, 2}(\Tilde{\vec{r}}) \rangle ) \rangle \\
        - \langle ( A_{S, 1}(\hat{\vec{r}}) - \langle A_{S, 1}(\Tilde{\vec{r}}) \rangle )( A_{S, 2}(\hat{\vec{r}}) - \langle A_{S, 2}(\Tilde{\vec{r}}) \rangle ) \rangle & \langle ( A_{S, 1}(\hat{\vec{r}}) - \langle A_{S, 1}(\Tilde{\vec{r}}) \rangle )^{2} \rangle
    \end{pmatrix}, \nonumber
\end{align}
and the $2 \times 2$ friction kernel matrix
\begin{align}
    \label{si_12_eq:friction_matrix_kernel}
    \hat{\Gamma}_{\dot{A}}(s, t) = & \frac{1}{\langle ( \dot{A}_{S, 1}(\hat{\vec{w}}) )^{2} \rangle \langle ( \dot{A}_{S, 2}(\hat{\vec{w}}) )^{2} \rangle - \langle \dot{A}_{S, 1}(\hat{\vec{w}}) \dot{A}_{S, 2}(\hat{\vec{w}}) \rangle^{2}} \\
    & \left\lbrack
    \begin{pmatrix}
        \langle F_{M, 1}(s, t, \hat{\vec{w}}) F_{M, 1}(s, s, \hat{\vec{w}}) \rangle & \langle F_{M, 1}(s, t, \hat{\vec{w}}) F_{M, 2}(s, s, \hat{\vec{w}}) \rangle \\
        \langle F_{M, 2}(s, t, \hat{\vec{w}}) F_{M, 1}(s, s, \hat{\vec{w}}) \rangle & \langle F_{M, 2}(s, t, \hat{\vec{w}}) F_{M, 2}(s, s, \hat{\vec{w}}) \rangle
    \end{pmatrix}
    \right.
    \nonumber \\
    & \left. - \beta
    \begin{pmatrix}
        \langle F_{M, 1}(s, s, \hat{\vec{w}}) \lbrack \mathcal{L}_{0} H_{1}(s, \hat{\vec{r}}) \rbrack \dot{A}_{S, 1}(\hat{\vec{w}}) \rangle & \langle F_{M, 1}(s, s, \hat{\vec{w}}) \lbrack \mathcal{L}_{0} H_{1}(s, \hat{\vec{r}}) \rbrack \dot{A}_{S, 2}(\hat{\vec{w}}) \rangle \\
        \langle F_{M, 2}(s, s, \hat{\vec{w}}) \lbrack \mathcal{L}_{0} H_{1}(s, \hat{\vec{r}}) \rbrack \dot{A}_{S, 1}(\hat{\vec{w}}) \rangle & \langle F_{M, 2}(s, s, \hat{\vec{w}}) \lbrack \mathcal{L}_{0} H_{1}(s, \hat{\vec{r}}) \rbrack \dot{A}_{S, 2}(\hat{\vec{w}}) \rangle
    \end{pmatrix}
    \right\rbrack
    \nonumber \\
    & \cdot 
    \begin{pmatrix}
        \langle ( \dot{A}_{S, 2}(\hat{\vec{w}}) )^{2} \rangle & - \langle \dot{A}_{S, 1}(\hat{\vec{w}}) \dot{A}_{S, 2}(\hat{\vec{w}}) \rangle \\
        - \langle \dot{A}_{S, 1}(\hat{\vec{w}}) \dot{A}_{S, 2}(\hat{\vec{w}}) \rangle & \langle ( \dot{A}_{S, 1}(\hat{\vec{w}}) )^{2} \rangle
    \end{pmatrix}. \nonumber
\end{align}
Fourth, we compute the parameter of eq.~\ref{si_12_eq:mori_gle} in the uncorrelated limit, where the two Gram matrices read
\begin{align}
    \label{si_12_eq:uncorrelated_inverse_gram_matrix_A}
    & \hat{G}_{A}^{-1} =
    \begin{pmatrix}
        \langle ( A_{S, 1}(\hat{\vec{r}}) - \langle A_{S, 1}(\Tilde{\vec{r}}) \rangle )^{2} \rangle^{-1} & 0 \\
        0 & \langle ( A_{S, 2}(\hat{\vec{r}}) - \langle A_{S, 2}(\Tilde{\vec{r}}) \rangle )^{2} \rangle^{-1}
    \end{pmatrix}
\end{align}
and
\begin{align}
    \label{si_12_eq:uncorrelated_inverse_gram_matrix_dotA}
    & \hat{G}_{\dot{A}}^{-1} =
    \begin{pmatrix}
        \langle ( \dot{A}_{S, 1}(\hat{\vec{w}}) )^{2} \rangle^{-1} & 0 \\
        0 & \langle ( \dot{A}_{S, 2}(\hat{\vec{w}}) )^{2} \rangle^{-1}
    \end{pmatrix},
\end{align}
the $2 \times 2$ time-dependent stiffness matrix reads
\begin{align}
    \label{si_12_eq:uncorrelated_stiffness_matrix}
    \hat{K}(t) = &
    \begin{pmatrix}
        \frac{\langle ( \dot{A}_{S, 1}(\hat{\vec{w}}) )^{2} \rangle}{\langle ( A_{S, 1}(\hat{\vec{r}}) - \langle A_{S, 1}(\Tilde{\vec{r}}) \rangle )^{2}} & 0 \\
        0 & \frac{\langle ( \dot{A}_{S, 2}(\hat{\vec{w}}) )^{2} \rangle}{\langle ( A_{S, 2}(\hat{\vec{r}}) - \langle A_{S, 2}(\Tilde{\vec{r}}) \rangle )^{2} \rangle}
    \end{pmatrix}
    \nonumber \\
    & - \beta
    \begin{pmatrix}
        \frac{\langle \dot{A}_{S, 1}(\hat{\vec{w}}) \lbrack \mathcal{L}_{0} H_{1}(t, \vec{r}) \rbrack (A_{S, 1}(\hat{\vec{r}}) - \langle A_{S, 1}(\Tilde{\vec{r}}) \rangle ) \rangle}{\langle ( A_{S, 1}(\hat{\vec{r}}) - \langle A_{S, 1}(\Tilde{\vec{r}}) \rangle )^{2}}  & \frac{\langle \dot{A}_{S, 1}(\hat{\vec{w}}) \lbrack \mathcal{L}_{0} H_{1}(t, \vec{r}) \rbrack (A_{S, 2}(\hat{\vec{r}}) - \langle A_{S, 2}(\Tilde{\vec{r}}) \rangle ) \rangle}{\langle ( A_{S, 1}(\hat{\vec{r}}) - \langle A_{S, 1}(\Tilde{\vec{r}}) \rangle )^{2}} \\
        \frac{\langle \dot{A}_{S, 2}(\hat{\vec{w}}) \lbrack \mathcal{L}_{0} H_{1}(t, \vec{r}) \rbrack (A_{S, 1}(\hat{\vec{r}}) - \langle A_{S, 1}(\Tilde{\vec{r}}) \rangle ) \rangle}{\langle ( A_{S, 2}(\hat{\vec{r}}) - \langle A_{S, 2}(\Tilde{\vec{r}}) \rangle )^{2} \rangle} & \frac{\langle \dot{A}_{S, 2}(\hat{\vec{w}}) \lbrack \mathcal{L}_{0} H_{1}(t, \vec{r}) \rbrack (A_{S, 2}(\hat{\vec{r}}) - \langle A_{S, 2}(\Tilde{\vec{r}}) \rangle ) \rangle}{\langle ( A_{S, 2}(\hat{\vec{r}}) - \langle A_{S, 2}(\Tilde{\vec{r}}) \rangle )^{2} \rangle}
    \end{pmatrix},
\end{align}
the positional kernel matrix reads
\begin{align}
    \label{si_12_eq:uncorrelated_positional_kernel_matrix}
    & \hat{\Gamma}_{A}(s, t) = \beta
    \begin{pmatrix}
        \frac{\langle F_{M, 1}(s, t, \hat{\vec{w}}) \lbrack \mathcal{L}_{0} H_{1}(s, \hat{\vec{r}}) \rbrack (A_{S, 1}(\hat{\vec{r}}) - \langle A_{S, 1}(\Tilde{\vec{r}}) \rangle ) \rangle}{\langle ( A_{S, 1}(\hat{\vec{r}}) - \langle A_{S, 1}(\Tilde{\vec{r}}) \rangle )^{2} \rangle} & \frac{\langle F_{M, 1}(s, t, \hat{\vec{w}}) \lbrack \mathcal{L}_{0} H_{1}(s, \hat{\vec{r}}) \rbrack (A_{S, 2}(\hat{\vec{r}}) - \langle A_{S, 2}(\Tilde{\vec{r}}) \rangle ) \rangle}{\langle ( A_{S, 1}(\hat{\vec{r}}) - \langle A_{S, 1}(\Tilde{\vec{r}}) \rangle )^{2} \rangle} \\
        \frac{\langle F_{M, 2}(s, t, \hat{\vec{w}}) \lbrack \mathcal{L}_{0} H_{1}(s, \hat{\vec{r}}) \rbrack (A_{S, 1}(\hat{\vec{r}}) - \langle A_{S, 1}(\Tilde{\vec{r}}) \rangle ) \rangle}{\langle ( A_{S, 2}(\hat{\vec{r}}) - \langle A_{S, 2}(\Tilde{\vec{r}}) \rangle )^{2} \rangle} & \frac{\langle F_{M, 2}(s, t, \hat{\vec{w}}) \lbrack \mathcal{L}_{0} H_{1}(s, \hat{\vec{r}}) \rbrack (A_{S, 2}(\hat{\vec{r}}) - \langle A_{S, 2}(\Tilde{\vec{r}}) \rangle ) \rangle}{\langle ( A_{S, 2}(\hat{\vec{r}}) - \langle A_{S, 2}(\Tilde{\vec{r}}) \rangle )^{2} \rangle}
    \end{pmatrix},
\end{align}
and the friction kernel matrix reads
\begin{align}
    \label{si_12_eq:uncorrelated_friction_kernel_matrix}
    \hat{\Gamma}_{\dot{A}}(s, t) = &
    \begin{pmatrix}
        \frac{\langle F_{M, 1}(s, t, \hat{\vec{w}}) F_{M, 1}(s, s, \hat{\vec{w}}) \rangle}{\langle ( \dot{A}_{S, 1}(\hat{\vec{w}}) )^{2} \rangle} & \frac{\langle F_{M, 1}(s, t, \hat{\vec{w}}) F_{M, 2}(s, s, \hat{\vec{w}}) \rangle}{\langle ( \dot{A}_{S, 1}(\hat{\vec{w}}) )^{2} \rangle} \\
        \frac{\langle F_{M, 2}(s, t, \hat{\vec{w}}) F_{M, 1}(s, s, \hat{\vec{w}}) \rangle}{\langle ( \dot{A}_{S, 2}(\hat{\vec{w}}) )^{2} \rangle} & \frac{\langle F_{M, 2}(s, t, \hat{\vec{w}}) F_{M, 2}(s, s, \hat{\vec{w}}) \rangle}{\langle ( \dot{A}_{S, 2}(\hat{\vec{w}}) )^{2} \rangle}
    \end{pmatrix} \\
    & - \beta
    \begin{pmatrix}
        \frac{\langle F_{M, 1}(s, t, \hat{\vec{w}}) \lbrack \mathcal{L}_{0} H_{1}(s, \hat{\vec{r}}) \rbrack \dot{A}_{S, 1}(\hat{\vec{w}}) \rangle}{\langle ( \dot{A}_{S, 1}(\hat{\vec{w}}) )^{2} \rangle} & \frac{\langle F_{M, 1}(s, t, \hat{\vec{w}}) \lbrack \mathcal{L}_{0} H_{1}(s, \hat{\vec{r}}) \rbrack \dot{A}_{S, 2}(\hat{\vec{w}}) \rangle}{\langle ( \dot{A}_{S, 1}(\hat{\vec{w}}) )^{2} \rangle} \\
        \frac{\langle F_{M, 2}(s, t, \hat{\vec{w}}) \lbrack \mathcal{L}_{0} H_{1}(s, \hat{\vec{r}}) \rbrack \dot{A}_{S, 1}(\hat{\vec{w}}) \rangle}{\langle ( \dot{A}_{S, 2}(\hat{\vec{w}}) )^{2} \rangle} & \frac{\langle F_{M, 2}(s, t, \hat{\vec{w}}) \lbrack \mathcal{L}_{0} H_{1}(s, \hat{\vec{r}}) \rbrack \dot{A}_{S, 2}(\hat{\vec{w}}) \rangle}{\langle ( \dot{A}_{S, 2}(\hat{\vec{w}}) )^{2} \rangle}
    \end{pmatrix}. \nonumber
\end{align}
Fifth, the expression of the 2D-NEQ Mori GLE in the equilibrium limit is given by
\begin{align}
    \label{si_12_eq:equilibrium_mori_gle}
    \begin{pmatrix}
        \ddot{A}_{1}(t - t_{0}, \vec{w}) \\
        \ddot{A}_{2}(t - t_{0}, \vec{w})
    \end{pmatrix}
    = &
    - \hat{K} \cdot
    \begin{pmatrix}
        A_{1}(t - t_{0}, \vec{w}) - \langle A_{S, 1} \rangle \\
        A_{2}(t - t_{0}, \vec{w}) - \langle A_{S, 2} \rangle
    \end{pmatrix}
     - \int_{t_{0}}^{t} ds \: \hat{\Gamma}_{\dot{A}}(t-s) \cdot
     \begin{pmatrix}
         \dot{A}_{1}(s - t_{0}, \vec{w}) \\
         \dot{A}_{2}(s - t_{0}, \vec{w})
     \end{pmatrix} \\
     & +
     \begin{pmatrix}
        F_{M, 1}(t - t_{0}, \vec{w}) \\
        F_{M, 2}(t - t_{0}, \vec{w})
    \end{pmatrix}, \nonumber
\end{align}
where the $2 \times 2$ stiffness matrix reads
\begin{align}
    \label{si_12_eq:equilibrium_stiffness_matrix}
    \hat{K} = & \frac{1}{\langle ( A_{S, 1}(\hat{\vec{r}}) - \langle A_{S, 1}(\Tilde{\vec{r}}) \rangle )^{2} \rangle \langle ( A_{S, 2}(\hat{\vec{r}}) - \langle A_{S, 2}(\Tilde{\vec{r}}) \rangle )^{2} \rangle - \langle ( A_{S, 1}(\hat{\vec{r}}) - \langle A_{S, 1}(\Tilde{\vec{r}}) \rangle )( A_{S, 2}(\hat{\vec{r}}) - \langle A_{S, 2}(\Tilde{\vec{r}}) \rangle ) \rangle^{2} } \\
    &
    \begin{pmatrix}
        \langle ( \dot{A}_{S, 1}(\hat{\vec{w}}) )^{2} \rangle & \langle \dot{A}_{S, 1}(\hat{\vec{w}}) \dot{A}_{S, 2}(\hat{\vec{w}}) \rangle \\
        \langle \dot{A}_{S, 1}(\hat{\vec{w}}) \dot{A}_{S, 2}(\hat{\vec{w}}) \rangle & \langle ( \dot{A}_{S, 2}(\hat{\vec{w}}) )^{2} \rangle
    \end{pmatrix} \nonumber \\
    & \cdot 
    \begin{pmatrix}
        \langle ( A_{S, 2}(\hat{\vec{r}}) - \langle A_{S, 2}(\Tilde{\vec{r}}) \rangle )^{2} \rangle & - \langle ( A_{S, 1}(\hat{\vec{r}}) - \langle A_{S, 1}(\Tilde{\vec{r}}) \rangle )( A_{S, 2}(\hat{\vec{r}}) - \langle A_{S, 2}(\Tilde{\vec{r}}) \rangle ) \rangle \\
        - \langle ( A_{S, 1}(\hat{\vec{r}}) - \langle A_{S, 1}(\Tilde{\vec{r}}) \rangle )( A_{S, 2}(\hat{\vec{r}}) - \langle A_{S, 2}(\Tilde{\vec{r}}) \rangle ) \rangle & \langle ( A_{S, 1}(\hat{\vec{r}}) - \langle A_{S, 1}(\Tilde{\vec{r}}) \rangle )^{2} \rangle
    \end{pmatrix}, \nonumber
\end{align}
and the $2 \times 2$ friction kernel matrix reads 
\begin{align}
    \label{si_12_eq:equilibrium_friction_kernel_matrix}
    \hat{\Gamma}_{\dot{A}}(t) = & \frac{1}{\langle ( \dot{A}_{S, 1}(\hat{\vec{w}}) )^{2} \rangle \langle ( \dot{A}_{S, 2}(\hat{\vec{w}}) )^{2} \rangle - \langle \dot{A}_{S, 1}(\hat{\vec{w}}) \dot{A}_{S, 2}(\hat{\vec{w}}) \rangle^{2}}
    \begin{pmatrix}
        \langle F_{M, 1}(t, \hat{\vec{w}})) F_{M, 1}(0, \hat{\vec{w}}) \rangle  & \langle F_{M, 1}(t, \hat{\vec{w}})) F_{M, 2}(0, \hat{\vec{w}})) \rangle \\
        \langle F_{M, 2}(t, \hat{\vec{w}})) F_{M, 1}(0, \hat{\vec{w}})) \rangle & \langle F_{M, 2}(t, \hat{\vec{w}})) F_{M, 2}(0, \hat{\vec{w}})) \rangle
    \end{pmatrix} \nonumber \\
    & \cdot 
    \begin{pmatrix}
        \langle ( \dot{A}_{S, 2}(\hat{\vec{w}}) )^{2} \rangle & -\langle \dot{A}_{S, 1}(\hat{\vec{w}}) \dot{A}_{S, 2}(\hat{\vec{w}}) \rangle \\
        - \langle \dot{A}_{S, 1}(\hat{\vec{w}}) \dot{A}_{S, 2}(\hat{\vec{w}}) \rangle & \langle ( \dot{A}_{S, 1}(\hat{\vec{w}}) )^{2} \rangle
    \end{pmatrix}.
\end{align}
Finally, we compute the expression of the parameters of eq.~\ref{si_12_eq:equilibrium_mori_gle} in the uncorrelated equilibrium limit, where the expression of the $2 \times 2$ stiffness matrix reads
\begin{align}
    \label{si_12_eq:uncorrelated_equilibrium_stiffness_matrix}
    & \hat{K} = 
    \begin{pmatrix}
        \frac{\langle ( \dot{A}_{S, 1}(\hat{\vec{w}}) )^{2} \rangle}{\langle ( A_{S, 1}(\hat{\Vec{r}}) - \langle A_{S, 1}(\Tilde{\Vec{r}}) \rangle )^{2} \rangle} & 0 \\
        0 & \frac{\langle ( \dot{A}_{S, 2}(\hat{\vec{w}}) )^{2} \rangle}{\langle ( A_{S, 2}(\hat{\Vec{r}}) - \langle A_{S, 2}(\Tilde{\Vec{r}}) \rangle )^{2} \rangle}
    \end{pmatrix}
\end{align}
and the expression of the friction kernel matrix reads
\begin{align}
    \label{si_12_eq:uncorrelated_equilibrium_friction_kernel_matrix}
    & \hat{\Gamma}_{\dot{A}}(t) = 
    \begin{pmatrix}
        \frac{\langle F_{M, 1}(t, \hat{\vec{w}}) F_{M, 1}(0, \hat{\vec{w}}) \rangle}{\langle ( \dot{A}_{S, 1}(\hat{\vec{w}}) )^{2} \rangle}  & \frac{\langle F_{M, 1}(t, \hat{\vec{w}}) F_{M, 2}(0, \hat{\vec{w}}) \rangle}{\langle ( \dot{A}_{S, 1}(\hat{\vec{w}}) )^{2} \rangle} \\
        \frac{\langle F_{M, 2}(t, \hat{\vec{w}}) F_{M, 1}(0, \hat{\vec{w}}) \rangle}{\langle ( \dot{A}_{S, 2}(\hat{\vec{w}}) )^{2} \rangle} & \frac{\langle F_{M, 2}(t, \hat{\vec{w}}) F_{M, 2}(0, \hat{\vec{w}}) \rangle}{\langle ( \dot{A}_{S, 2}(\hat{\vec{w}}) )^{2} \rangle}
    \end{pmatrix}.
\end{align}

\end{document}